\documentclass[journal]{IEEEtran}
\usepackage{amsmath}
\usepackage{stmaryrd}
\usepackage{amsfonts}
\usepackage{mathrsfs}
\usepackage{amssymb,color,balance,cite}
\usepackage[dvips]{graphicx}
\usepackage{multicol}
\usepackage{mathtools}
\usepackage{caption}
\usepackage{subfigure}
\usepackage{bbm}
\usepackage{multirow}

\newtheorem{remark}{Remark}

 \allowdisplaybreaks[4]

\begin{document}

\title{ \LARGE Frequency Synchronization for Uplink Massive MIMO with Adaptive MUI Suppression in Angle-domain }
\author{\IEEEauthorblockN{Yinghao Ge, Weile Zhang, Feifei Gao, and Geoffrey Ye Li
\vspace{-1.5em}
}
\thanks{Y. Ge and W. Zhang are with the MOE Key Lab for Intelligent Networks and Network Security, School of Electronic and
Information Engineering, Xi'an Jiaotong University, Xi'an, Shaanxi, 710049, China. (email: ge\_yinghao\_jacques@163.com, wlzhang@mail.xjtu.edu.cn).

F. Gao is with the State Key Laboratory of Intelligent Technology and Systems, Tsinghua National Laboratory for Information Science and Technology, Department of Automation, Tsinghua University, Beijing, 100084, China (e-mail: feifeigao@ieee.org).

Geoffrey Ye Li is with the Department of Electrical and Computer Engineering, Georgia Institute of Technology, Atlanta, GA, 30332 USA (e-mail: liye@ece.gatech.edu).
}
}

 \maketitle

 \pagestyle{empty}
 \thispagestyle{empty}


  \begin{abstract}
In this paper, we develop a novel angle-domain adaptive filtering (ADAF)-based frequency synchronization method for the uplink of a massive multiple-input multiple-output (MIMO) multiuser network, which is applicable for users with either separate or overlapped angle-of-arrival (AoA) regions.
For each user, we first introduce the angle-constraining matrix (ACM), which consists of a set of selected match-filter (MF) beamformers pointing to the AoAs of the interested user. Then, the adaptive beamformer can be acquired by appropriately designing the ADAF vectors.
Such beamformer can achieve two-stage adaptive multiuser interference (MUI) suppression, i.e., inherently suppressing the MUI from non-overlapping users by ACM in the first stage and substantially mitigating the MUI from adjacent overlapping users by ADAF vector in the second stage.
For both separate and mutually overlapping users, the carrier frequency offset (CFO) estimation and subsequent data detection can be performed independently for each user, which significantly reduces computational complexity. Moreover, ADAF is rather robust to imperfect AoA knowledge and side cluster of scatterers, making itself promising for multiuser uplink transmission.
We also analyze the performance of CFO estimation and obtain an approximated closed-form expression of mean-squared error (MSE).
The effectiveness of the proposed method and its superiority over the existing methods are demonstrated by numerical results.
   \end{abstract}

\vspace{-2mm}


\vspace{-0.5em}
\section{Introduction}
Over the past few years, large-scale multiple-input multiple-output (MIMO) or massive MIMO systems have drawn exploding interests from both academia and industry~\cite{Marzetta10, LuJSTSP2014, GaoTVT16, GaoAccess16, GaoJSAC16, Larsson_CM2014}. Massive MIMO is considered as a promising technique for the next generation wireless systems due to its high spectral and energy efficiency as well as the ability to substantially relieve the multiuser interference (MUI) with simple linear transceivers. However, these potential gains depend heavily on perfect frequency synchronization, which is quite challenging owing to the existence of carrier frequency offsets (CFOs)~\cite{Morelli_TWC2012, Morelli_WCL2013} among the different users.

There are many conventional synchronization schemes for traditional multiuser MIMO~\cite{Tsai-TWC13, Chen_TSP2008, Wu_EUJWCN2011, Zhang_TWC2016}. In~\cite{Tsai-TWC13}, a robust multi-CFO estimation with iterative interference cancellation (RMCE-IIC) scheme was proposed for a coordinated multi-point (CoMP) orthogonal frequency division multiplexing (OFDM) system. Based on the Zadoff-Chu (ZC) sequences, the optimal set of training sequences is designed to minimize the mutual interference, which is also robust to different CFOs. By employing the optimal training sequence set, the CFO search and channel estimation for different users can be performed independently within each iteration.
For massive MIMO systems, however, only a few frequency synchronization methods have been designed.
The effect of CFOs on the sum-rate performance of massive MIMO has been studied in~\cite{Cheng13} for both collocated and distributed antennas.
The impact of frequency selectivity on the information rate of CFO impaired single-carrier multiuser MIMO uplink with the time-reversal maximum-ratio-combining (MRC) receiver and the impact of residual CFOs on the information rate of zero-forcing (ZF) receiver have been discussed in~\cite{Mukherjee_TVT2016-2} and~\cite{Mukherjee_TVT2016-1}, respectively.
The work in~\cite{Mukherjee_GLOBECOM2015} has addressed the multi-CFO estimation problem for the time division duplexed (TDD) massive MIMO system. However, it requires that the CFO estimated at the base station (BS) must be fed back to multiple users over an error-free control channel for CFO compensation prior to data transmission.
A blind frequency synchronization approach has been developed in~\cite{Zhang_TSP2016} for multiuser OFDM uplink with a large number of receive antennas. However, many blocks are necessary to obtain the satisfactory CFO estimation performance, which delays the decoding and cannot be directly applied to latency-sensitive scenarios.

More recently, a multiuser frequency synchronization scheme has been proposed in~\cite{ZhangGLO16} when a large-scale uniform linear antenna array (ULA) is configured at the BS. The CFO estimation therein is performed individually for each user through a joint spatial-frequency alignment (JSFA) procedure~\cite{ZhangGLO16},
which has been further modified in~\cite{GE_SPAWC2017} to address the multi-CFO estimation problem with antenna array imperfection.
The JSFA proposed in~\cite{ZhangGLO16} can be implemented efficiently via Fast Fourier Transform (FFT), and its computational burden increases linearly with the number of coexisting users.
However, both the JSFA in~\cite{ZhangGLO16} and its variant in~\cite{GE_SPAWC2017} require the angle-of-arrival (AoA) regions to be separated by certain guard zone. As a result, the JSFA scheme can only support the simultaneous transmission of very few spatially multiplexed users over the same time-frequency resources, and cannot fully exploit the abundant degrees of freedom provided by the massive ULA, especially in the case of relatively large angular spread (AS).
In view of this limitation, a user grouping (UG) technique has been further designed in~\cite{Zhang_TWC2018} to cope with the frequency synchronization of AoA-overlapping users. However, its computational complexity grows sharply with the number of intragroup users. Moreover, as the intragroup users exceed a certain number, the UG technique in~\cite{Zhang_TWC2018} exhibits a underdetermination problem and thereby suffers from severe performance deterioration.

Motivated by the aforementioned discussion, we design an angle-domain adaptive filtering (ADAF)-based frequency synchronization approach for multiuser MIMO uplink in this paper.
By introducing the angle-constraining matrix (ACM) and appropriately designing the ADAF vectors, we obtain the qualified adaptive beamformers for each user.
The ACM consists of a set of match-filter (MF) beamformers that point to the AoA region of the interested user. It constitutes the first stage MUI suppression to handle the MUI from the non-overlapping users.
The second stage MUI suppression is accomplished by judiciously designing the ADAF vectors such that the acquired beamformers could substantially relieve the severe interference from the adjacent overlapping users.
With the above two-stage MUI suppression, the interference from 
the other users can be successfully cancelled, resulting in an equivalent single-user transmission network and enabling separate CFO estimation and data detection.
Therefore, the proposed ADAF method can tolerate certain overlapping of AoA regions among coexisting users without consuming unbearable computational resources.
Both analytical and numerical results demonstrate the effectiveness of the proposed ADAF approach and its superiority over existing approaches.

The rest of this paper is organized as follows. The system model is described in Section II. Section III elaborates the proposed ADAF algorithm. The CFO estimation mean-squared error (MSE) of the proposed algorithm is analyzed in Section IV.
Simulation results are provided in Section V. Section VI concludes the paper.

\textit{Notations:} Superscripts $(\cdot)^*$, $(\cdot)^T$, $(\cdot)^H$, $(\cdot)^{-1}$ and $E\{\cdot\}$ represent conjugate, transpose, Hermitian, inverse and expectation, respectively; ${\mathrm j}=\sqrt{-1}$ is the imaginary unit;
$\|\cdot\|$ denotes the Frobenius norm operator; $\operatorname{tr}(\cdot)$ denotes the trace operation.
For a vector $\mathbf x$, $\operatorname{diag}(\mathbf x)$ is a diagonal matrix with $\mathbf x$ as the main diagonal;
$\otimes$ stands for the Kronecker product;
${\mathbb C}^{m\times n}$ denotes the vector space of all $m\times n$ complex matrices.
$\lambda_q(\mathbf{A})$ and $\boldsymbol{\nu}_q(\mathbf{A})$ denote the $q$th eigenvalue and the corresponding eigenvector of a positive semi-definite Hermitian matrix $\mathbf A$.
${\bf 0}$ represents an all-zero matrix with appropriate dimension.

\section{System Model}
We consider a multiuser OFDM uplink system that consists of $K$ distributed single-antenna users and one BS with $M \!\gg\! 1$ antennas in the form of ULA.
The total number of subcarriers is $N$. We assume perfect time synchronization among all users for the time being so that we could get clearer insights in CFO related issues. Denote the normalized CFO between the $k\rm{th}$ user and the BS as $\phi_k$, which is the ratio of the real CFO and the subcarrier spacing. We consider the fractional CFOs, i.e., $|\phi_k| \!<\! 0.5$, which should be sufficient for the multiuser uplink. The CFO will affect the uplink received OFDM signal by an $N \!\times\! N$ diagonal phase rotation matrix defined as
\begin{align}
{\bf E}(\phi_k) = \operatorname{diag}\big(1, {\rm e}^{{\mathrm j}\frac{2 {\mathrm{\pi}} \phi_k}{N}},\cdots\!, {\rm e}^{{\mathrm j}\frac{2 {\mathrm{\pi}} (N-1)\phi_k}{N}}  \big).
\end{align}

The classical one-ring channel propagation model~\cite{You15, Sun15} is adopted to characterize the multiuser OFDM uplink scenario. Each user is surrounded by a ring of $P \!\gg\! 1$ local scatterers.
The channel between the BS and the $k$th user consists of $L$ taps. Denote the average power of the $l$th tap as $\sigma_l^2$. Then, ${\sigma_{l}^{2}}, l \!=\! 1,2,\cdots\!, L$ models the power delay profile (PDP) of the channel with $\sum\nolimits_{l=1}^{L}{\sigma_{l}^{2}} \!=\! 1$ such that the total average channel gain per receive antenna is normalized. Each tap is composed of $P$ separable subpaths identifiable by its unique AoA $\theta_{l,p}^{(k)}$ and associated independent and identically distributed (i.i.d.) complex gain $g_{l,p}^{(k)} \sim \mathcal{CN}\left( 0, {\sigma_{l}^{2}}/P \right)$.

Let ${\boldsymbol a}(\theta_{l,p}^{(k)} ) \in\mathbb{C}^{M\times 1}$ denote the steering vector
\begin{align}
{\boldsymbol a}(\theta_{l,p}^{(k)} )  = \big [1, {\rm e}^{-{\mathrm j}2\chi\cos\theta_{l,p}^{(k)}},\cdots\!, {\rm e}^{-{\mathrm j}2\chi(M-1)\cos\theta_{l,p}^{(k)}} \big]^T,
\end{align}
where $\chi = {\mathrm \pi} {\tilde d} = {\mathrm{\pi}} \frac{ d}{\lambda}$, $d$ is the antenna spacing and $\lambda$ denotes the carrier wavelength. Then, the multi-tap channel matrix between the $k\rm{th}$ user and the BS can be expressed as
\begin{align}
{\bf H}_k = \big[ {\bf h}^{(k)}_1,{\bf h}^{(k)}_2,\cdots\!, {\bf h}^{(k)}_L    \big]^T \in\mathbb{C}^{L\times M},
\end{align}
where ${\bf h}^{(k)}_l \!=\! \sum_{p=1}^{P} g_{l,p}^{(k)} {\boldsymbol a}(\theta_{l,p}^{(k)} )$ corresponds to the $l$th tap channel vector. Similar to \cite{You15, Sun15}, we assume that the incident rays of each user can be constrained within a certain AS $\theta_{\rm as}$; namely, the incident angles of the $k\rm{th}$ user $\theta_{l,p}^{(k)}$ is limited in the AoA region
$(\theta_k - \theta_{\rm as}, \theta_k + \theta_{\rm as})$
with uniform distribution, where $\theta_k$ denotes the average AoA of the $k\rm{th}$ user.

Considering that the AoA regions of users vary slowly~\cite{GaoAccess16}, it is sufficient to send one time-division training sequence as in~\cite{Zhang_TWC2018} to estimate the average AoAs of all users, and the so obtained AoAs can be used for a bunch of upcoming frames. In this case, the AoA identification is independent of CFO estimation with tolerable expenditure, which significantly facilitates the receiver design. Thus, the accurate average AoAs are assumed available at the BS when addressing the CFO estimation. Nevertheless, we will provide numerical evidence in the simulation part that the proposed approach is quite robust to the AoA inaccuracy. Moreover, we have also verified that the proposed estimation is fairly insusceptible to the AS inaccuracy, though accurate AS is assumed known at the BS.

Define ${\bf F}$ as the $N\times N$ normalized discrete Fourier transform (DFT) matrix whose ($m,n$)-$\rm{th}$ entry is $\frac{1}{\sqrt{N}}{\rm e}^{-{\mathrm j}\frac{2 {\mathrm{\pi}} (m-1)(n-1)}{N}}$. Besides, ${\bf F}_L$ stands for the submatrix of ${\bf F}$ consisting of its first $L$ columns.
Each frame consists of $N_b$ blocks, including a whole training block at the beginning of each uplink frame and $N_b\!-\!1$ data blocks.
Denote ${\bf x}_i^{(k)} \!=\! \big[ x_i^{(k)}(0), \ \! x_i^{(k)}(1), \ \! \cdots\!, \\ x_i^{(k)}(N\!-\!1)\big]^T$ as the frequency domain data symbols transmitted from the $k\rm{th}$ user in the $i$th block. Then, the received time-domain signal in the $i$th block after cyclic-prefix (CP) removal is given by
\begin{align}{\label{ReceivedSignal}}
{\bf Y}_i = \sum_{k=1}^K \eta_i(\phi_k) {\bf E}(\phi_k) \mathbf{B}_i^{(k)} {\bf H}_k + {\bf N}_i,
\end{align}
where $\eta_i(\phi_k) \!=\! {\mathrm e}^{{\mathrm j} \frac {2{\mathrm \pi}(i-1)(N+N_{\mathrm{cp}})\phi_k} {N} }$ represents the accumulative phase rotation introduced by the CFO of the $k$th user with $N_{\mathrm{cp}}$ being the length of the CP, ${\bf B}_i^{(k)} \!=\! \sqrt{N} {\bf F}^H \operatorname{diag}({\bf x}_i^{(k)}){\bf F}_L \in\mathbb{C}^{N\times L}$, and ${\bf N}_i \in\mathbb{C}^{N\times M}$ denotes the corresponding additive white Gaussian noise (AWGN) matrix. We assume that each element of ${\bf N}_i$ follows i.i.d. complex Gaussian distribution with variance $\sigma_{\mathrm n}^2$, i.e., $E\{{\bf N}_i{\bf N}_i^H\} \!=\! M\sigma_{\mathrm n}^2 {\bf I}_N$. Note that $i\!=\!1$ denotes the training block, and ${\bf B}_1^{(k)}$ and $\mathbf{Y}_1$ will be simplified as ${\bf B}_k$ and $\mathbf{Y}$ hereinbelow for notation convenience.

\section{ADAF-Based Frequency Synchronization}
We first illustrate how the proposed ADAF approach can achieve two-stage MUI suppression in Section III-A. Then, the CFO and multiple ADAF vectors are jointly optimized for each user in Section III-B. In Section III-C, the multi-branch beamforming and CFO compensation are performed for each user before finally recovering the transmitted data symbols via MRC detection.

\vspace{-0.6em}
\subsection{Angle-Domain Adaptive Filtering for Two-stage MUI Suppression}
Owing to the large number of antennas, the steering vectors pointing to distinct AoAs would become asymptotically orthogonal to each other, i.e., ${\boldsymbol a}^H(\theta_1) {\boldsymbol a}(\theta_2) \!\simeq\! 0$ for $\theta_1 \!\ne\! \theta_2$. Thus, as long as the AoA regions of all users are mutually non-overlapping, we can simply employ the steering vector, ${\boldsymbol a}(\tilde{\theta}_k )$, pointing to the AoA region of the $k$th user as receive beamformer (MF beamformer) to substantially suppress the interference from the other users.
Unfortunately, in the presence of AoA intersection among different users, the steering vector, ${\boldsymbol a}(\tilde{\theta}_k )$, pointing to the $k$th user may allow the signals of both the $k$th user and the adjacent overlapping users to pass through. The signal after beamforming is thus not purely the desired signal of the $k$th user but intermingled with severe interference from other users sharing or partially sharing the same AoA regions. Consequently, regarding the $k$th user, ${\boldsymbol a}(\tilde{\theta}_k)$, $\tilde{\theta}_k \!\in\! (\theta_k \!-\! \theta_{\rm as}, \theta_k \!+\! \theta_{\rm as})$ may not be a qualified candidate receive beamformer to perform MUI suppression. Therefore, the JSFA scheme cannot be directly applied to the scenario with spatially overlapping users.

Nevertheless, any attempt to capture the information of the $k$th user from the spatial domain must resort to the directions satisfying $\tilde{\theta}_k \!\in\! (\theta_k \!-\!\theta_{\rm as}, \theta_k \!+\!\theta_{\rm as})$, though some directions may be more or less interfered with. Let $\vartheta_{k}^{(q)}$ denote the $q$th selected beamforming angle for the $k$th user drawn from the set $\mathcal{D}_{\mathrm{IDFT}} \!=\! \{ \mathcal{D}_1, \mathcal{D}_2, \cdots\!, \mathcal{D}_{M} \}$, where $\mathcal{D}_i \!=\! {\rm{acrcos}} \big( \frac{{\rm \pi}(i-\frac{M}{2})}{\chi M }  \big) $. Then, $ \mathcal{S}_{\theta_k} \!=\! \{\vartheta_{k}^{(q)} \!:\! | \vartheta_{k}^{(q)} \!-\! \theta_k | \!<\! \theta_{\rm as}  \}$ represents the set of selected beamforming directions falling into the AoA region of the $k$th user and ${Q}_{k} \!=\! | \mathcal{S}_{\theta_k} |$ denotes the cardinality of $\mathcal{S}_{\theta_k} $.

We introduce the following $M \!\times\! 1$ vector as the beamformer to effectively perform MUI cancellation for the $k\rm{th}$ user
\begin{align}
{\boldsymbol \omega}_k = {\mathcal{U}}_k^* {\boldsymbol \gamma}_k,
\end{align}
where
\begin{align}
{\mathcal{U}}_k= \frac{1}{\sqrt{M}} \left[ \begin{matrix}
   \mathbf{a} \big( \vartheta_{k}^{(1)} \big), & \mathbf{a}\big( \vartheta_{k}^{(2)} \big), & \cdots\!, & \mathbf{a} \big( \vartheta_{k}^{{({Q}_{k})}} \big)  \\
\end{matrix} \right],
\end{align}
is the ACM for the $k$th user with ${\mathcal{U}}_k^T {\mathcal{U}}_k^* \!=\! \mathbf{I}_{Q_k}$ and ${\boldsymbol \gamma}_k \!\in\! \mathbb{C}^{Q_k\times 1}$ is the ADAF vector for the $k\rm{th}$ user.

The ACM comprises all the MF beamformers pointing to the AoA region of the interested user. Considering that the information of the $k$th user is contained only within the AoA region $(\theta_k-\theta_{\rm as}, \theta_k+\theta_{\rm as})$, such ACM can well preserve the signal of the $k$th user and does not introduce excessive interference.
The ADAF vector, ${\boldsymbol \gamma}_k$, aims at performing angle-domain filtering on the set of MF beamformers contained in the ACM, such that the resulting beamformer, ${\boldsymbol \omega}_k$, can substantially suppress the interference from both the overlapping and non-overlapping users.
Note that any other user whose AoA region partially or completely overlaps with that of the $k$th user is deemed as overlapping user for the $k$th user, while any other user without AoA intersection with the $k$th user is deemed as non-overlapping user.

Different from the pure MF beamformer, the adaptive beamformer, ${\boldsymbol \omega}_k$, may not point to any realistic physical angles. However, such beamformer not only inherits the advantage of the MF beamformer but also proves to be more favorable for two-stage MUI suppression.
On the one hand, as each column of ${\mathcal{U}}_k$ is in fact an MF beamformer, the interference caused by non-overlapping users can be inherently cancelled by ${\mathcal{U}}_k$, irrespective of the ADAF vector. That is, the first stage MUI suppression is achieved by the ACM and targets at mitigating the MUI from non-overlapping users. On the other hand, the optimal ADAF vector, ${\boldsymbol \gamma}_k$, is adaptively determined by minimizing the interference incorporated by the resulting beamformer, ${\boldsymbol \omega}_k$. Owing to the first stage MUI suppression with the ACM, only the signals of the adjacent overlapping users and very weak residual signals from the non-overlapping users may enter into the corresponding beamforming branch and intermingle with the desired signal. Such interference could be effectively attenuated by appropriately designing the ADAF vector. Therefore, the design of the ADAF vector constitutes the key to the second stage MUI suppression, which mainly addresses the severe interference stemming from the adjacent overlapping users.

\begin{figure}[t]
\setlength{\abovecaptionskip}{-0.1cm}
\setlength{\belowcaptionskip}{-0.7cm}
\begin{center}
\includegraphics[width=90mm]{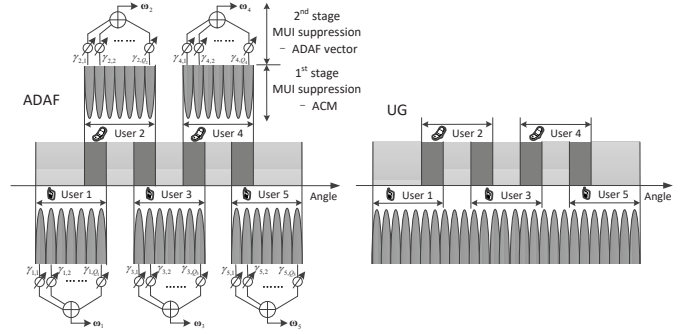}
\end{center}
\caption{ Illustration of the difference between ADAF and UG with $5$ mutually overlapping users. }
\end{figure}

\begin{remark}
It should be pointed out that we conceptually distinguish the ``overlapping users'' from ``non-overlapping users'' only for facilitating the presentation of the two-stage MUI suppression. Such distinction is, however, unnecessary for the derivation and implementation of the ADAF algorithm itself,
since the algorithm does not deal with the ``overlapping users'' and ``non-overlapping users'' discriminatingly. All the MUI, whether from ``overlapping users'' or ``non-overlapping users'', would be simultaneously suppressed by the final adaptive beamformer ${\boldsymbol \omega}_k$.
\end{remark}

\begin{remark}
The difference of the main idea between the ADAF and the UG is illustrated in Fig. 1 with $5$ mutually overlapping users. As already explained, the proposed ADAF approach can achieve two-stage MUI suppression. The ACM consisting of the selected MF beamformers constitutes the first stage, which can inherently suppress the MUI from non-overlapping users but maintain the desired signal from the expected user. Certainly, the UG can also accomplish such MUI suppression. The design of the ADAF vector mainly targets at cancelling the severe MUI caused by adjacent overlapping users and is the key to the second stage. Instead, the UG groups the users with AoA intersection and performs joint CFO estimation and data detection for each group, which circumvents the intractable problem of the severe mutual interference among the intragroup users.
The distinction between the ADAF and the UG in dealing with the MUI among the AoA-overlapping users accounts for the fact that the ADAF can estimate multiple CFOs separately with tremendously reduced computational burden, whereas the UG necessitates the joint multi-CFO estimation and hence is more likely to encounter the underdetermination problem that the number of parameters to be estimated exceeds the number of pilots.
\end{remark}

\vspace{-0.6em}
\subsection{CFO Estimation}
There may not be solely one qualified $\hat{\boldsymbol \gamma}_k$, which can be employed to perform MUI suppression for the $k$th user. Using only one ADAF vector may lead to the loss of space diversity. Hence, we propose hereinbelow to perform beamforming with multi-ADAF beamformer so as to improve CFO estimation accuracy through more complete MUI suppression and to achieve the potential multi-branch diversity gain~\cite{Zhang_DSP2017}.

We compensate the CFO with the trial CFO value $\tilde{\phi}_k$ and perform multi-branch beamforming with the trial beamformers $\tilde{\boldsymbol \omega}_k^{(q)} \!=\! {\mathcal{U}}_k^* \tilde{\boldsymbol \gamma}_k^{(q)} \!\in\! \mathbb{C}^{M \times 1}, q\!=\!1,2,\cdots\!,Q_k$, where $\tilde{\boldsymbol \gamma}_k^{(q)} \!\in\! \mathbb{C}^{Q_k \times 1}$ stands for the $q$th trial ADAF vector. As $\mathcal{U}_k$ has only $Q_k$ columns, there exist at most $Q_k$ linearly independent ADAF vectors~\cite{Meyer_2001}. Moreover, let $\tilde{\boldsymbol \gamma}(k)$ denote the set of $Q_k$ trial ADAF vectors, i.e., $\tilde{\boldsymbol \gamma}(k) \!=\! \left\{ \tilde{\boldsymbol \gamma}_k^{(q)}, q\!=\!1,2,\cdots\!,Q_k \right\}$. Then
\begin{align}
\rho_k^{(q)} =
\frac{ \left\| {\bf P}_{{\bf B}_k} {\bf E}^H\big(\tilde\phi_{k}\big) {\bf Y} \mathcal{U}_k^* \tilde{\boldsymbol \gamma}_k^{(q)} \right\|^2
}{  \left \| {\bf P}_{{\bf B}_k}^\bot {\bf E}^H\big(\tilde\phi_k\big) {\bf Y} \mathcal{U}_k^* \tilde{\boldsymbol \gamma}_k^{(q)} \right \|^2}
\end{align}
reflects the equivalent signal-to-interference-plus-noise ratio (SINR) after receive beamforming with the $q$th beamformer $\tilde{\boldsymbol \omega}_k^{(q)}$. Here, ${\bf P}_{{\bf B}_k} \!=\! {{\bf B}_k} ({{\bf B}_k^H}{{\bf B}_k})^{-1} {{\bf B}_k^H}$ represents the orthogonal projection operator onto the subspace spanned by the columns of ${{\bf B}_k}$ and ${\bf P}_{{\bf B}_k}^\bot \!=\! {\bf{I}}_N \!-\! {\bf P}_{{\bf B}_k}$.

The maximization of the overall SINR leads to
\begin{align}{\label{max_SINR_1}}
& \big\{ \hat\phi_k, \hat{\boldsymbol\gamma}(k) \big\}=\arg\max_{ \{\tilde\phi_k, \tilde{\boldsymbol\gamma}(k) \} } \sum_{q=1}^{Q_k} \rho_k^{(q)}, \nonumber \\
=& \arg\max_{ \{\tilde\phi_k, \tilde{\boldsymbol\gamma}(k) \} } \sum_{q=1}^{Q_k} \frac{ \left\| {\bf P}_{{\bf B}_k} {\bf E}^H\big(\tilde\phi_{k}\big) {\bf Y} \mathcal{U}_k^* \tilde{\boldsymbol \gamma}_k^{(q)} \right\|^2
}{  \left \| {\bf P}_{{\bf B}_k}^\bot {\bf E}^H\big(\tilde\phi_k\big) {\bf Y} \mathcal{U}_k^* \tilde{\boldsymbol \gamma}_k^{(q)} \right \|^2} \nonumber \\
 & \hspace{2em}  s.t.\ \left\| {\bf Y} \mathcal{U}_k^* \tilde{\boldsymbol \gamma}_k^{(q)} \right\|^2 = 1, \nonumber \\
=& \arg\max_{ \{\tilde\phi_k, \tilde{\boldsymbol\gamma}(k) \} } \sum_{q=1}^{Q_k} \frac{ \left\| {\bf Y} \mathcal{U}_k^* \tilde{\boldsymbol \gamma}_k^{(q)} \right\|^2
}{  \left \| {\bf P}_{{\bf B}_k}^\bot {\bf E}^H\big(\tilde\phi_k\big) {\bf Y} \mathcal{U}_k^* \tilde{\boldsymbol \gamma}_k^{(q)} \right \|^2} \nonumber \\
 & \hspace{2em}  s.t.\ \left\| {\bf Y} \mathcal{U}_k^* \tilde{\boldsymbol \gamma}_k^{(q)} \right\|^2 = 1, \nonumber \\
=& \arg\max_{ \{ \tilde\phi_k, \tilde{\boldsymbol\gamma}(k)  \} } \sum_{q=1}^{Q_k} \frac{ \tilde{\boldsymbol \gamma}_k^{(q)H}  {\bf\Psi}_k   \tilde{\boldsymbol \gamma}_k^{(q)}  }{ \tilde{\boldsymbol \gamma}_k^{(q)H} {\bf\Xi}_k\big(\tilde\phi_k\big)  \tilde{\boldsymbol \gamma}_k^{(q)} }, \ \! s.t.\ \! \tilde{\boldsymbol \gamma}_k^{(q)H}  {\bf\Psi}_k   \tilde{\boldsymbol \gamma}_k^{(q)} \!=\! 1,
\end{align}
where
\begin{align}
{\bf\Psi}_k & =  \mathcal{U}_k^T {\bf Y}^H {\bf Y} \mathcal{U}_k^*, \\
{\bf\Xi}_k\big(\tilde\phi_k\big) & = \mathcal{U}_k^T {\bf Y}^H  {\bf E}\big(\tilde\phi_k\big)  {\bf P}_{{\bf B}_k}^\bot {\bf E}^H\big(\tilde\phi_k\big) {\bf Y} \mathcal{U}_k^*, \label{Xi_BDAF}
\end{align}
and the constraints on $\tilde{\boldsymbol \gamma}_k^{(q)}, q\!=\!1,2,\cdots\!,Q_k$ are added to avoid degenerate solutions of $\hat{\boldsymbol \gamma}_k^{(q)} \!=\! \mathbf{0}$.

By decomposing ${\bf\Psi}_k$ as ${\bf\Psi}_k \!=\! {\bf Q}_k{\bf Q}_k^H$ and defining $\tilde{\boldsymbol \beta}_k^{(q)} \!=\! {\bf Q}_k^H \tilde{\boldsymbol \gamma}_k^{(q)}$, the maximization problem in (\ref{max_SINR_1}) can be equivalently transformed into
\begin{align}{\label{max_SINR}}
\big\{ \hat\phi_k, \hat{\boldsymbol\beta}(k) \big\}
=& \arg\max_{ \{\tilde\phi_k, \tilde{\boldsymbol\beta}(k) \} } \sum_{q=1}^{Q_k} \frac{1}{ \tilde{\boldsymbol \beta}_k^{(q)H}  {\bf Q}_k^{-1} {\bf\Xi}_k \big(\tilde\phi_k\big) {\bf Q}_k^{-H} \tilde{\boldsymbol \beta}_k^{(q)} }, \nonumber \\
& \hspace{2em} s.t.\ \| \tilde{\boldsymbol \beta}_k^{(q)} \|^2 = 1,
\end{align}
where $\tilde{\boldsymbol \beta}(k) \!=\! \left\{ \tilde{\boldsymbol \beta}_k^{(q)}, q\!=\!1,2,\cdots\!,Q_k \right\}$ denotes the set of $\tilde{\boldsymbol \beta}_k^{(q)}$
and the constraints $\| \tilde{\boldsymbol \beta}_k^{(q)} \|^2 \!=\! 1$ avoid degenerate solutions of $\hat{\boldsymbol \beta}_k^{(q)} \!=\! \mathbf{0}$. Besides, $\tilde{\boldsymbol \beta}_k^{(q)}, q \!=\! 1, 2,\cdots\!, Q_k$ should be designed as mutually uncorrelated, i.e., $\tilde{\boldsymbol \beta}_k^{(p)H} \! \tilde{\boldsymbol \beta}_k^{(q)} \!=\! \delta_{p,q}$ in order to ensure that the equivalent noise inside each branch is also mutually uncorrelated. (Note that $\Big[ \tilde{\boldsymbol \beta}_k^{(1)}, \tilde{\boldsymbol \beta}_k^{(2)}, \cdots\!,  \tilde{\boldsymbol \beta}_k^{(Q_k)} \Big]$ is unitary).

The following Lemma is essential for further derivation of the ADAF algorithm.

\newtheorem{lemma}{Lemma}
\begin{lemma}
For a positive semi-definite full-rank Hermitian matrix ${\mathbf{A}} \!\in\! {\mathbb C}^{Q\times Q}$, if $\tilde{\boldsymbol \Upsilon} \!=\! \big[\tilde{\boldsymbol \gamma}_1, \tilde{\boldsymbol \gamma}_2, \cdots\!, \tilde{\boldsymbol \gamma}_Q\big] \!\in\! {\mathbb C}^{Q\times Q}$ is unitary, then there holds
\begin{align}{\label{Lemma1}}
\max_{ \tilde{\boldsymbol\Upsilon} } & \sum_{q=1}^{Q} \frac{1}{ \tilde{\boldsymbol \gamma}_q^H \mathbf{A} \tilde{\boldsymbol \gamma}_q } = \sum_{q=1}^{Q} \frac{1}{\lambda_q (\mathbf{A}) } = \operatorname{tr} ( \mathbf{A}^{-1} ), \nonumber \\ & \hat{\boldsymbol \gamma}_q = \boldsymbol{\nu}_q(\mathbf{A}), q=1,2,\cdots\!, Q.
\end{align}
\end{lemma}
\begin{IEEEproof}
Refer to Appendix~\ref{Lemma_Derivation}.
\end{IEEEproof}

With Lemma 1, (\ref{max_SINR}) can be further decomposed into
\begin{align}{\label{CFO_Estimation}}
\hat\phi_k = \arg\max_{\tilde\phi_k} G_k\big(\tilde\phi_k\big),
\end{align}
where $G_k\big(\tilde\phi_k\big) = \operatorname{tr} \big( {\bf Q}_k^H {\bf\Xi}_k^{-1} \big(\tilde\phi_k\big) {\bf Q}_k \big)$, and
\begin{align}
\hat{\boldsymbol \gamma}_k^{(q)} & \!=\! {\bf Q}_k^{-H} \hat{\boldsymbol \beta}_k^{(q)} \!=\! {\bf Q}_k^{-H} \boldsymbol{\nu}_q \big( {\bf Q}_k^{-1} {\bf\Xi}_k \big(\hat\phi_k\big) {\bf Q}_k^{-H} \big), \nonumber \\
q & \!=\! 1,2,\cdots\!,Q_k.
\end{align}
Directly solving (\ref{CFO_Estimation}) necessitates a one-dimensional search that faces high complexity. However, since (\ref{CFO_Estimation}) falls into the scope of unconstraint maximization, we can solve it efficiently via Newton's method~\cite{Boyd_2004}, which ensures low complexity and guarantees the convergence to local optimum. The procedure of solving (\ref{CFO_Estimation}) via Newton's method is given as follows.

Given a coarse CFO estimate $\hat\phi_{k,0} $, we approximate ${\bf E}\big(\tilde\phi_k\big)$ with Taylor series expansion as
\begin{align}\label{E_approx}
{\bf E}\big(\tilde\phi_k\big) \approx \Big(  {\bf I}_N + \Delta {\tilde{\phi}}_k  {\bf D} + \frac{\Delta {\tilde{\phi}}_k^2}{2} {\bf D}^2 \Big) {\bf E}\big( \hat\phi_{k,0}  \big),
\end{align}
where $\Delta {\tilde{\phi}}_k \!=\! \tilde\phi_k \!-\! \hat\phi_{k,0}$ and ${\bf D} \!=\! {\mathrm j}\frac{2 {\mathrm{\pi}}  }{N} \operatorname{diag}\big([0,1,\cdots\!, N\!-\!1 ]\big)$.
Note that although the JSFA approach in~\cite{ZhangGLO16} is originally designed for separate users without mutual AoA-overlapping, it indeed can provide a valid coarse CFO estimate when the AoA regions of different users intersect.
Substituting (\ref{E_approx}) into (\ref{Xi_BDAF}), we can approximate ${\bf\Xi}_k\big(\tilde\phi_k\big)$ as
\begin{align}
{\bf\Xi}_k\big(\tilde\phi_k\big)
  \approx \mathbf{T}_k^{(0)} + \Delta {\tilde{\phi}}_k \mathbf{T}_k^{(1)} + \frac{ \Delta {\tilde{\phi}}_k^2 }{2} \mathbf{T}_k^{(2)},
\end{align}
where
\begin{align*}
\mathbf{T}_k^{(0)} \!=\! & \ {\mathcal U}_k^T  {\bf Y}^H {\bf E}\big( \hat\phi_{k,0} \big) {\bf P}_{{\bf B}_k}^\bot {\bf E}^H\big( \hat\phi_{k,0} \big) {\bf Y} {\mathcal U}_k^*, \nonumber \\
\mathbf{T}_k^{(1)} \!=\! & \ {\mathcal U}_k^T  {\bf Y}^H {\bf E}\big(  \hat\phi_{k,0}  \big) \big(  {\bf D} {\bf P}_{{\bf B}_k}^\bot \!+\! {\bf P}_{{\bf B}_k}^\bot {\bf D}^H  \big) {\bf E}^H\big( \hat\phi_{k,0} \big) {\bf Y} {\mathcal U}_k^*, \nonumber \\
\mathbf{T}_k^{(2)} \!=\! & \ {\mathcal U}_k^T  {\bf Y}^H {\bf E}\big(  \hat\phi_{k,0}  \big) \big( {\bf D}^2 {\bf P}_{{\bf B}_k}^\bot \!+\! {\bf P}_{{\bf B}_k}^\bot {\bf D}^{2H} \!+\! 2{\bf D} {\bf P}_{{\bf B}_k}^\bot  {\bf D}^H \big) \nonumber \\
 & \hspace{2em} \times {\bf E}^H\big( \hat\phi_{k,0} \big) {\bf Y} {\mathcal U}_k^*.
\end{align*}

Since $({\mathbf A} \!+\! {\mathbf B})^{-1} \!\approx\! {\mathbf A}^{-1} \!-\! {\mathbf A}^{-1} {\mathbf B} {\mathbf A}^{-1}$ holds when the entries of arbitrary invertible matrix ${\mathbf A}$ is much larger than those of ${\mathbf B}$, ${{\bf\Xi}_k^{-1}} \big(\tilde\phi_k\big)$ can be approximated by
\begin{align}{\label{inv_approx}}
{\bf\Xi}_k^{-1} \big(\tilde\phi_k\big)
  \approx & \ \! \big(\mathbf{T}_k^{(0)}\big)^{-1} - \Delta {\tilde{\phi}}_k \big(\mathbf{T}_k^{(0)}\big)^{-1} \mathbf{T}_k^{(1)} \big(\mathbf{T}_k^{(0)}\big)^{-1} \nonumber \\
  & - \frac{ \Delta {\tilde{\phi}}_k^2 }{2} \big(\mathbf{T}_k^{(0)}\big)^{-1} \mathbf{T}_k^{(2)} \big(\mathbf{T}_k^{(0)}\big)^{-1}.
\end{align}
Note that as long as $ {\tilde{\phi}}_k$ is sufficiently small, the approximation (\ref{inv_approx}) will be quite accurate. Combining (\ref{CFO_Estimation}) and (\ref{inv_approx}), we can obtain the following parabolic function of $\Delta \tilde{\phi}_k$
\begin{align}\label{equ5}
\Delta \hat\phi_k =  \arg\max_{\Delta \tilde\phi_k} \hspace{1mm}  \big\{ t_k^{(0)} - \Delta\tilde{\phi}_k t_k^{(1)} - \frac{\Delta \tilde{\phi}_k^2}{2} t_k^{(2)} \big\},
\end{align}
where
\begin{align*}
t_k^{(i)} =  \operatorname{tr} \big( {\bf Q}_k^{H} \big(\mathbf{T}_k^{(0)}\big)^{-1} \mathbf{T}_k^{(i)} \big(\mathbf{T}_k^{(0)}\big)^{-1} {\bf Q}_k \big), i=0,1,2.
\end{align*}

The optimal solution of $\tilde\phi_k$ for (\ref{equ5}) is thus given by
\begin{align}
\hat\phi_{k} = \hat\phi_{k,0} + \Delta \hat\phi_k = \hat\phi_{k,0} - t_k^{(1)} / t_k^{(2)}.
\end{align}

\subsection{Multi-Branch Beamforming, CFO Compensation and MRC-based Data Detection}
The achievable SINR of the MRC depends on whether the noise level inside each branch is equalized.
As a result, we perform the multi-branch beamforming with a set of adjusted adaptive beamformers given by
\begin{align}{\label{AdjustedBeamformer}}
\hat{\boldsymbol \omega}_k^{(q)} = \alpha_k^{(q)} \mathcal{U}_k^* \hat{\boldsymbol \gamma}_k^{(q)} = \alpha_k^{(q)} \mathcal{U}_k^* {\bf Q}_k^{-H} \hat{\boldsymbol \beta}_k^{(q)},
\end{align}
where $\alpha_k^{(q)} \!=\! \frac{1} {\sqrt{\lambda_k^{(q)}}}, q \!=\! 1,2,\cdots\!,Q_k$ corresponds to the noise equalization coefficients for the $k$th user. Since $\lambda_k^{(q)} \!=\! \lambda_q \big( {\bf Q}_k^{-1}  \\ {\bf\Xi}_k \big(\tilde\phi_k\big) {\bf Q}_k^{-H} \big)$, there is $\alpha_k^{(q)} \!=\! \sqrt{\lambda_q \big( {\bf Q}_k^{H} {\bf\Xi}_k^{-1} \big(\tilde\phi_k\big) {\bf Q}_k \big)}$.

The beamformer, $\hat{\boldsymbol \omega}_k^{(q)}$, in (\ref{AdjustedBeamformer}) can effectively mitigate the MUI originating from both overlapping and non-overlapping users for the $k$th user. Mathematically,
\begin{align}
{\bf Y}_i \hat{\boldsymbol \omega}_k^{(q)} = {\eta_i}(\phi_k)
{\bf E}(\phi_k) {\bf B}_i^{(k)} {\bf H}_k \hat{\boldsymbol \omega}_k^{(q)} + \hat{\bf N}_{i,k}^{(q)},
\end{align}
where $\hat{\bf N}_{i,k}^{(q)} \!=\! {\bf N}_i \hat{\boldsymbol \omega}_k^{(q)} \!+\! \sum_{k' \ne k}^K  {\eta_i}(\phi_k') {\bf E}(\phi_{k'}) {\bf B}_i^{(k')} {\bf H}_{k'} \hat{\boldsymbol \omega}_{k'}^{(q)} $ includes the noise and the mitigated MUI terms from the other users.
A total of $Q_k$ qualified beamforming branches can be obtained, immediately resulting in an approximate single-user transmission network with $Q_k$ equivalent receive antennas. Then, the conventional single-user CFO compensation, channel estimation, and MRC-based data detection can be performed.
Note that the multi-CFO estimation is performed at the BS within a single training block, and the CFO of each user is also compensated at the BS after user separation. Thus, no CFO feedback is required.

\begin{remark}
``Nonphysical models'' and ``physical models'' are two generic modeling approaches for MIMO channels~\cite{Yu_JWCMC2002}. The latter models the locations of scatterers/reflectors (equivalently, the angles and delays of multipath components) relative to the transmitter or the receiver. The physical model usually assumes that all scatterers are concentrated around the users and within certain region~\cite{Lee_TVT1973}. A more general ``multiple cluster'' model~\cite{Molisch_VTC2003} further takes into account the clusters of ``far scatterers'' from high-rise buildings (in urban environments) or mountains (rural environments)~\cite{Petrus_TC2002}.
Some experimental measurements validating the multi-cluster model can be found in~\cite{Yu_JWCMC2002, Spencer_JSAC2000}.
Though multi-cluster seldom occurs in practice, its existence degrades the performance of many existing multiuser frequency synchronization approaches, which (partially) depend on multiuser separation in the angle/spatial domain. In contrast, owing to the two-stage MUI suppression via the ACM and the ADAF vectors, the proposed ADAF approach need not distinguish the interference sources and can indiscriminately mitigate the MUI from the side clusters as effectively as that from overlapping or non-overlapping users. Thus, the proposed ADAF approach should be more robust to multi-cluster environments.

\begin{figure}[t]
\setlength{\abovecaptionskip}{-0.0cm}
\setlength{\belowcaptionskip}{-0.6cm}
\begin{center}
\includegraphics[width=65mm]{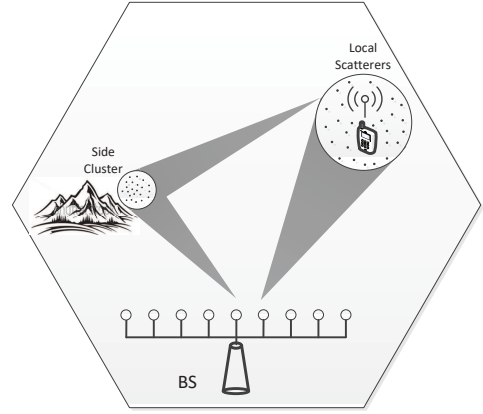}
\end{center}
\caption{ Illustration of the multi-cluster channel model, with local scatterers around the user and side cluster of far scatterers. }
\end{figure}

Without loss of generality, we assume that one cluster of ``far scatterers'' (referred to as side cluster) exists for each user, apart from the local scatterers (referred to as main cluster) centered around the user. The AS of the side cluster is defined as $\theta_{\mathrm{as}}^{(\mathrm{sc})}$ while the average AoA for the side cluster of the $k$th user is denoted as $\theta_k^{(\mathrm{sc})}$. In the presence of the side cluster, the received signal model in (\ref{ReceivedSignal}) can be modified as
\begin{align}{\label{ReceivedSignal_SideCluster}}
{\bf Y}_i = \sum\nolimits_{k=1}^K \eta_i(\phi_k) {\bf E}(\phi_k) \mathbf{B}_i^{(k)} \big(  {\bf H}_k + {\bf H}_k^{(\mathrm{sc})} \big) + {\bf N}_i,
\end{align}
where ${\bf H}_k^{(\mathrm{sc})} \!$ denotes the channel corresponding to the side clu-
ster and can be expressed as ${\bf H}_k^{(\mathrm{sc})} \!\!=\! \big[ {\bf h}^{(k)}_{1, \mathrm{sc}}, {\bf h}^{(k)}_{2, \mathrm{sc}}, \cdots\!, {\bf h}^{(k)}_{L, \mathrm{sc}} \big]\!^T$, \\ with ${\bf h}^{(k)}_{l, \mathrm{sc}} \!=\! \sum_{p=1}^{P^{(\mathrm{sc})}} g^{(k)}_{l,p,{\mathrm{sc}}} {\boldsymbol a}(\theta^{(k)}_{l,p,{\mathrm{sc}}} )$. Here, $P^{(\mathrm{sc})}$ is the number of subpaths under each delay and $\theta^{(k)}_{l,p,{\mathrm{sc}}}$ is uniformly distributed in the region $(\theta_k^{(\mathrm{sc})} \!-\! \theta_{\rm as}^{(\mathrm{sc})}, \theta_k^{(\mathrm{sc})} \!+\! \theta_{\rm as}^{(\mathrm{sc})} )$. Moreover, we assume $g^{(k)}_{l,p,{\mathrm{sc}}} \!\sim\! \mathcal{CN}\big( 0, {\sigma_{l, \mathrm{sc}}^{2}}/P^{(\mathrm{sc})} \big)$ and $\sum\nolimits_{l=1}^{L} {\sigma_{l, \mathrm{sc}}^{2}} \!=\! \eta$. Since the channel PDP of the main cluster is normalized, $\eta$ in fact represents the ratio between the signal strength within the side cluster and that within the main cluster and is referred to as side-to-main power ratio (SMPR) hereinbelow.

It should be clarified that the received signal model given by (\ref{ReceivedSignal}) only considers the main cluster, which will be adopted throughout the paper. Nevertheless, the robustness of the ADAF under the ``multi-cluster'' model (corresponding to (\ref{ReceivedSignal_SideCluster})) will also
be demonstrated by numerical simulations.
\end{remark}

\section{Performance Analysis of CFO Estimation}
The MSE of CFO estimation under high signal-to-noise ratio (SNR) is given by~\cite{W_Zhang2013TSP}
\begin{align}{\label{MSE_theo}}
\mathrm{MSE} \{ {{{{\phi }}}_{k}} \}={{\left. \frac{E\Big\{ {{\Big( \frac{\partial {{G}_{k}}\left( {{{\tilde{\phi }}}_{k}} \right)}{\partial {{{\tilde{\phi }}}_{k}}} \Big)}^{2}} \Big\}}{{{\Big[ E\Big\{ \frac{{{\partial }^{2}}{{G}_{k}}\left( {{{\tilde{\phi }}}_{k}} \right)}{\partial \tilde{\phi }_{k}^{2}} \Big\} \Big]}^{2}}} \right|}_{{{{\tilde{\phi }}}_{k}}={{\phi }_{k}}}},
\end{align}
where the cost function is $G_k\big(\tilde\phi_k\big) = \operatorname{tr} \big( {\bf Q}_k^{H} {\bf\Xi}_k^{-1}\big(\tilde\phi_k\big) {\bf Q}_k \big) = \operatorname{tr} \big( {\bf\Xi}_k^{-1}\big(\tilde\phi_k\big) \mathcal{U}_k^T {\bf Y}^H {\bf Y} \mathcal{U}_k^* \big)$.

From Appendix~\ref{MSE_Derivation}, the analytical MSE can be approximately expressed as
\begin{align}{\label{MSE_theo2}}
& \mathrm{MSE}\big\{ {{{\hat{\phi }}}_{k}} \big\} \approx \nonumber \\ & \frac{\frac{1}{{{L}^{3}}}{{\left[ {{g}_{1}}\left( k \right) \right]}^{2}}{{g}_{3}}\left( k \right){{d}_{k}}+\frac{\sigma _{\mathrm{n}}^{2}}{{{L}^{3}}}{{\left[ {{g}_{1}}\left( k \right) \right]}^{2}}{{g}_{2}}\left( k \right){{d}_{k}}}
{2{{\left[ \frac{\sigma _{\mathrm{n}}^{2}}{L}{{g}_{1}}\! \left( k \right){{g}_{2}}\! \left( k \right){{d}_{k}} \!+\! \frac{\sigma _{\mathrm{n}}^{2}}{{{L}^{2}}}{{g}_{0}}\! \left( k \right){{\left[ {{g}_{1}}\! \left( k \right) \right]}^{2}}{{d}_{k}} \!-\! \frac{1}{{{L}^{2}}}{{\left[ {{g}_{1}}\! \left( k \right) \right]}^{2}}{{d}_{k}} \right]}^{2}}},
\end{align}
where
\begin{align}{\label{gk_dk}}
{{g}_{0}}\left( k \right) & = \operatorname{tr}\left( {{\mathcal{R}}_{\mathcal{U}_{k}^{*}}} \right), \ {{g}_{3}}\left( k \right) = \frac{1}{N\!-\!L}{{g}_{1}}\left( k \right)-\sigma_{\mathrm n}^{2}{{g}_{2}}\left( k \right), \nonumber \\
{{g}_{1}}\left( k \right) & = \operatorname{tr}\left( {{\mathcal{R}}_{\mathcal{U}_{k}^{*}}}{{\mathbf{R}}_{k}} \right), \ {{g}_{2}}\left( k \right) = \operatorname{tr}\left( {{\mathcal{R}}_{\mathcal{U}_{k}^{*}}} {{\mathcal{R}}_{\mathcal{U}_{k}^{*}}} {{\mathbf{R}}_{k}} \right), \nonumber \\
d_k & = \operatorname{tr}\left( \mathbf{B}_{k}^{H}\mathbf{DP}_{{{\mathbf{B}}_{k}}}^{\bot }{{\mathbf{D}}^{H}}{{\mathbf{B}}_{k}} \right) = \left\| \mathbf{P}_{{{\mathbf{B}}_{k}}}^{\bot }{{\mathbf{D}}^{H}}{{\mathbf{B}}_{k}} \right\|^2.
\end{align}
Here, $ {{\mathcal{R}}_{\mathcal{U}_{k}^{*}}} \!=\! \mathcal{U}_{k}^{*} \! {{\big[ \mathcal{U}_{k}^{T}\mathbf{\Omega }_{k}^{\left( 1 \right)}\mathcal{U}_{k}^{*} \big]}\!^{-1}}\mathcal{U}_{k}^{T}$
with $\mathbf{\Omega }_{k}^{\left( 1 \right)} \!\approx\! \left( N\!\!-\!\!L \right)\! \big( \sigma_{\mathrm n}^{2}{{\mathbf{I}}_{M}} \\ \!+\!\sum\nolimits_{k'=1,k'\ne k}^{K}{\mathbf{H}_{k'}^{H}{{\mathbf{H}}_{k'}}} \big)$
and $
\! {{\mathbf{R}}_{k}} \!\approx\! \frac{1}{2{{\theta }_{\mathrm{as}}}}\! \int_{{{\theta }_{k}}-{{\theta }_{\mathrm{as}}}}^{{{\theta }_{k}}+{{\theta }_{\mathrm{as}}}}{{{\mathbf{a}}^{*}}\! \left( \theta  \right){{\mathbf{a}}^{T}}\! \left( \theta  \right)\! d\theta } $.
It should be indicated that the uniform channel PDP has been adopted to facilitate the MSE derivation, which can be readily extended to arbitrary channel PDPs.

Next, we attempt to simplify (\ref{MSE_theo2}) under two typical circumstances for massive MIMO channels with and without AoA-overlapping, respectively.

\vspace{-0.5em}
\subsection{Simplified MSE for Channels without AoA-Overlapping}
In this subsection, we will simplify the analytical MSE (\ref{MSE_theo2}) when the AoA regions of the channels of different users are non-overlapping. In such case, we have
\begin{align}
{{\mathcal{R}}_{\mathcal{U}_{k}^{*}}} \!=\! \mathcal{U}_{k}^{*}{{\left( \mathcal{U}_{k}^{T}\mathbf{\Omega }_{k}^{\left( 1 \right)}\mathcal{U}_{k}^{*} \right)}^{-1}}\mathcal{U}_{k}^{T} \!\approx\! \frac{1}{(N-L) \sigma_{\mathrm n}^{2} }\mathcal{U}_{k}^{*} \mathcal{U}_{k}^{T} \!=\! b \mathcal{U}_{k}^{*} \mathcal{U}_{k}^{T},
\end{align}
where $b \!=\! \frac{1}{(N-L) \sigma_{\mathrm n}^{2} }$. Denote ${{r}_{k}} \!=\! \operatorname{tr}\left( \mathcal{U}_{k}^{T}{{\mathbf{R}}_{k}}\mathcal{U}_{k}^{*} \right)$. Then, $g_{i}(k)$ for $i\!=\!0,1,2,3$ in (\ref{gk_dk}) can be simplified as
\begin{align}{\label{g0-g3}}
 {{g}_{0}}\left( k \right) & \!=\! \operatorname{tr}\left( {{\mathcal{R}}_{\mathcal{U}_{k}^{*}}} \right) \!=\! b\operatorname{tr}\left( \mathcal{U}_{k}^{T}\mathcal{U}_{k}^{*} \right) \!=\! b{{Q}_{k}}, \nonumber \\
{{g}_{1}}\left( k \right) & \!=\! \operatorname{tr}\left( {{\mathcal{R}}_{\mathcal{U}_{k}^{*}}}{{\mathbf{R}}_{k}} \right) \!=\! b\operatorname{tr}\left( \mathcal{U}_{k}^{T}{{\mathbf{R}}_{k}}\mathcal{U}_{k}^{*} \right) \!=\! b{{r}_{k}}, \nonumber \\
 {{g}_{2}}\left( k \right) & \!=\! \operatorname{tr}\left( {{\mathcal{R}}_{\mathcal{U}_{k}^{*}}} {{\mathcal{R}}_{\mathcal{U}_{k}^{*}}} {{\mathbf{R}}_{k}} \right) \!=\! {{b}^{2}}\operatorname{tr}\left( \mathcal{U}_{k}^{T}{{\mathbf{R}}_{k}}\mathcal{U}_{k}^{*} \right) \!=\! b{{g}_{1}}\left( k \right) \!=\! {{b}^{2}}{{r}_{k}}, \nonumber \\
 {{g}_{3}}\!\left( k \right) & \!=\!\! \frac{1}{N\!\!-\!L}{{g}_{1}}\!\left( k \right) \!-\! \sigma_{\mathrm n}^{2}{{g}_{2}}\!\left( k \right) \!=\! \sigma_{\mathrm n}^{2}\!\left[ b{{g}_{1}}\!\left( k \right) \!-\! {{g}_{2}}\!\left( k \right) \right] \!=\! 0.
\end{align}
Moreover, ${{r}_{k}}$ can be approximated as
\begin{align}{\label{rk}}
  {{r}_{k}}& =\frac{1}{M}\sum\limits_{q=1}^{{{Q}_{k}}}{{{\mathbf{a}}^{T}}\big( \vartheta _{k}^{\left( q \right)} \big){{\mathbf{R}}_{k}}{{\mathbf{a}}^{*}}\big( \vartheta _{k}^{\left( q \right)} \big)}, \nonumber \\
 & \approx \frac{1}{M}\sum\limits_{q=1}^{{{Q}_{k}}}{\frac{1}{2{{\theta }_{\mathrm{as}}}}\int_{{{\theta }_{k}}-{{\theta }_{\mathrm{as}}}}^{{{\theta }_{k}}+{{\theta }_{\mathrm{as}}}}{{{\left| {{\mathbf{a}}^{T}}\big( \vartheta _{k}^{\left( q \right)} \big){{\mathbf{a}}^{*}}\left( \theta  \right) \right|}^{2}}d\theta }}, \nonumber \\
 & =\frac{1}{M}\sum\limits_{q=1}^{{{Q}_{k}}}{\frac{1}{2{{\theta }_{\mathrm{as}}}}\int_{{{\theta }_{k}}-{{\theta }_{\mathrm{as}}}}^{{{\theta }_{k}}+{{\theta }_{\mathrm{as}}}}{\frac{{{\sin }^{2}}\big[ \chi M\big( \cos \vartheta _{k}^{\left( q \right)}-\cos \theta  \big) \big]}{{{\sin }^{2}}\big[ \chi \big( \cos \vartheta _{k}^{\left( q \right)}-\cos \theta  \big) \big]}d\theta }}, \nonumber \\
 & \ {\overset{*}{\mathop{\approx }}}\,\frac{1}{M{{\sin }}{{\theta }_{k}}}\sum\limits_{q=1}^{{{Q}_{k}}}{\frac{1}{2{{\theta }_{\mathrm{as}}}}\int_{\cos \vartheta _{k}^{\left( q \right)}-\cos \left( {{\theta }_{k}}-{{\theta }_{\mathrm{as}}} \right)}^{\cos \vartheta _{k}^{\left( q \right)}-\cos \left( {{\theta }_{k}}+{{\theta }_{\mathrm{as}}} \right)}{\frac{{{\sin }^{2}}\left( \chi Mx \right)}{{{\left( \chi x \right)}^{2}}}dx}}, \nonumber \\
 & \approx \frac{1}{ {\tilde d}^2 M{{\sin }}{{\theta }_{k}}}\sum\limits_{q=1}^{{{Q}_{k}}}{\frac{1}{2{{\theta }_{\mathrm{as}}}}{{\left. \int_{-\infty }^{\infty }{h\left( x \right)h\left( x \right){{\mathrm{e}}^{\mathrm{j}2 {\mathrm{\pi}} Fx}}dx} \right|}_{F=0}}}, \nonumber \\
 & \approx \frac{{{Q}_{k}}}{2 {\tilde d}^2 M{{\theta }_{\mathrm{as}}}{{\sin }}{{\theta }_{k}}}{{\left. \int_{-\infty }^{\infty }{H\left( f \right)H\left( F-f \right)df} \right|}_{F=0}}, \nonumber \\
 & =\frac{{{Q}_{k}}}{2 {\tilde d}^2 M{{\theta }_{\mathrm{as}}}{{\sin }}{{\theta }_{k}}}\int_{-\frac{1}{2}\tilde{d}M}^{\frac{1}{2}\tilde{d}M}{1df}, \nonumber \\
 & = \frac{{{Q}_{k}}}{2\tilde{d}{{\theta }_{\mathrm{as}}}{{\sin }}{{\theta }_{k}}}\overset{\tilde{d}=0.5}{\mathop{=}}\,\frac{{{Q}_{k}}}{{{\theta }_{\mathrm{as}}}{{\sin }}{{\theta }_{k}}},
\end{align}
where $h \! \left( x \right)\!=\!\tilde{d}M\! \operatorname{sinc} \! \big( \tilde{d}Mx \big)$ and its Fourier Transform $H\! \left( f \right) \\ \!=\! \mathcal{F}\! \left\{ h\left( x \right) \right\} \!=\! u\big( f \!+\! \frac{1}{2}\tilde{d}M \big) \!-\! u\big( f \!-\! \frac{1}{2}\tilde{d}M \big)$, with $u(\cdot)$ being the unit step function.
Moreover, for ${\overset{*}{\mathop{\approx }}}\,$, we have taken a variable substitution of ${x\!=\! \cos \vartheta_{k}^{\left( q \right)} \!-\! \cos \theta }$ before approximating $d\theta \!=\! \frac{dx}{\sin \theta }$ with $\frac{dx}{\sin {{\theta }_{k}}}$.
The approximation is due to the fact that for small AS, $\theta_{\mathrm{as}}$, and $\theta \!\in\! ({\theta }_{k} \!-\! {\theta }_{\mathrm{as}}, {\theta }_{k} \!+\! {\theta }_{\mathrm{as}})$, $\sin \theta$ varies slightly and can be approximated by the intermediate value $\sin {{\theta }_{k}}$.

Substituting (\ref{g0-g3}) and (\ref{rk}) into (\ref{MSE_theo2}) yields
\begin{align}{\label{MSE_theo3}}
 \mathrm{MSE} \{ {{{{\phi }}}_{k}} \} & \approx \frac{\frac{\sigma_{\mathrm n}^{2}}{{{L}^{3}}}b}{{2{\left[ \frac{\sigma_{\mathrm n}^{2}}{L}b+\frac{\sigma_{\mathrm n}^{2}}{{{L}^{2}}}{{g}_{0}}\left( k \right)-\frac{1}{{{L}^{2}}} \right]}^{2}}{{g}_{1}}\left( k \right){{d}_{k}}}, \nonumber \\
 &= \frac{L\sigma_{\mathrm n}^{2}}{2}\frac{\alpha_1}{{{r}_{k}}{{d}_{k}}}\approx {{\alpha }_{1}}\frac{\sigma_{\mathrm n}^{2}L{{\theta }_{\mathrm{as}}}{{\sin }}{{\theta }_{k}}}{2{{Q}_{k}}{{d}_{k}}},
\end{align}
where ${{\alpha }_{1}}\!=\!{{\big( \frac{N-L}{N-2L-{{Q}_{k}}} \big)}^{2}}$. We can further approximate $Q_k$ (integer) with the following continuous value $Q_k \!\approx\! \frac{\cos (\theta_k - \theta_{\mathrm{as}}) - \cos (\theta_k + \theta_{\mathrm{as}}) }{2}M \!=\! M \sin \theta_k \sin \theta_{\mathrm{as}}$. Moreover, $d_k$ only depends on the number of subcarriers, $N$, and the channel length, $L$. Thus, the MSE of CFO estimation decreases as expected with the increasing numbers of antennas, $M$, and subcarriers, $N$, and is independent of the average AoA, $\theta_k$.

\vspace{-0.5em}
\subsection{Asymptotic MSE for Channels with Very Severe AoA-Overlapping}
In this subsection, we will derive the asymptotic MSE in the presence of severe MUI due to mutual overlapping of the AoA regions among different users.

Denote ${{\mathcal{H}}_{k}} \!=\! \sum\nolimits_{k'=1,k'\ne k}^{K}{{{\mathbf{H}}_{k'}}}$. Besides, for $k' \!\ne\! k''$, there holds
\begin{align*}
& E\{\mathbf{H}_{k'}^{H}{{\mathbf{H}}_{k''}}\}= E\left\{ \sum_{l=1}^{L} {\bf h}^{(k')*}_l {\bf h}^{(k'')T}_l \right\}, \nonumber \\
=\ \! & \sum_{l=1}^{L} \sum_{p=1}^{P} \sum_{q=1}^{P} \underbrace{E\left\{ g_{l,p}^{(k')*} g_{l,q}^{(k'')} \right\}}_{0} {\boldsymbol a}^*(\theta_{l,p}^{(k')} ) {\boldsymbol a}^T(\theta_{l,q}^{(k'')} ) = \mathbf{0},
\end{align*}
which leads to $\mathcal{H}_{k}^{H}{{\mathcal{H}}_{k}} \!=\! \sum\nolimits_{k'=1,k'\ne k}^{K}\sum\nolimits_{k''=1,k''\ne k}^{K} {\mathbf{H}_{k'}^{H}{{\mathbf{H}}_{k''}}}$ $\approx \sum\nolimits_{k'=1,k'\ne k}^{K}{\mathbf{H}_{k'}^{H}{{\mathbf{H}}_{k'}}}$.
Thus, $\mathbf{\Omega }_{k}^{\left( 1 \right)}$ can be approximated as
\begin{align}
\mathbf{\Omega }_{k}^{\left( 1 \right)} & =\left( N-L \right)\Big( \sigma_{\mathrm n}^{2}{{\mathbf{I}}_{M}} + \sum\nolimits_{k'=1,k'\ne k}^{K}{\mathbf{H}_{k'}^{H}{{\mathbf{H}}_{k'}}} \Big), \nonumber \\
& \approx \left( N-L \right)\left( \sigma_{\mathrm n}^{2}{{\mathbf{I}}_{M}} + \mathcal{H}_{k}^{H}{{\mathcal{H}}_{k}} \right).
\end{align}
Besides, there holds
\begin{align}{\label{For_g0}}
& {{\mathcal{H}}_{k}}\mathcal{U}_{k}^{*}\mathcal{U}_{k}^{T}\mathcal{H}_{k}^{H} \!\approx\! \sum\limits_{k'=1,k'\ne k}^{K}{{{\mathbf{H}}_{k'}}\mathcal{U}_{k}^{*}\mathcal{U}_{k}^{T}\mathbf{H}_{k'}^{H}}, \nonumber \\
\!\approx\! & \sum\limits_{k'=1,k'\ne k}^{K}{\frac{1}{L} \operatorname{tr}\left( \mathcal{U}_{k}^{*}\mathcal{U}_{k}^{T}{{\mathbf{R}}_{k'}} \right){{\mathbf{I}}_{L}}} \!=\! \frac{1}{L}\operatorname{tr}\left( \mathcal{U}_{k}^{T} {{\mathcal{R}}_{k}} \mathcal{U}_{k}^{*} \right){{\mathbf{I}}_{L}},
\end{align}
where ${{\mathcal{R}}_{k}}=\sum\nolimits_{k'=1,k'\ne k}^{K}{{{\mathbf{R}}_{k'}}}$.

With the matrix inverse lemma~\cite{Petersen2012}, we can obtain
\begin{align}
  & {{\big( \mathcal{U}_{k}^{T}\mathbf{\Omega }_{k}^{\left( 1 \right)}\mathcal{U}_{k}^{*} \big)}^{-1}}=\frac{1}{N\!-\!L}{{\left( \sigma_{\mathrm n}^{2}{{\mathbf{I}}_{{{Q}_{k}}}} + \mathcal{U}_{k}^{T}\mathcal{H}_{k}^{H}{{\mathcal{H}}_{k}}\mathcal{U}_{k}^{*} \right)}^{-1}}, \nonumber \\
 = & \frac{1}{\left( N\!-\!L \right)\sigma_{\mathrm n}^{2}}\! \left( {{\mathbf{I}}_{{{Q}_{k}}}} \!\!-\! \mathcal{U}_{k}^{T}\mathcal{H}_{k}^{H}{{\left( \sigma_{\mathrm n}^{2}{{\mathbf{I}}_{L}} \!+\! {{\mathcal{H}}_{k}}\mathcal{U}_{k}^{*}\mathcal{U}_{k}^{T}\mathcal{H}_{k}^{H} \right)}^{-1}}{{\mathcal{H}}_{k}}\mathcal{U}_{k}^{*} \right)\!, \nonumber \\
 \approx &\ \! b\left( {{\mathbf{I}}_{{{Q}_{k}}}} \!-\! a\mathcal{U}_{k}^{T}\mathcal{H}_{k}^{H}{{\mathcal{H}}_{k}}\mathcal{U}_{k}^{*} \right),
\end{align}
where $a=\frac{1}{\sigma_{\mathrm n}^{2} + \frac{1}{L}\operatorname{tr}\left( \mathcal{U}_{k}^{T}{{\mathcal{R}}_{k}}\mathcal{U}_{k}^{*} \right)}$.
Therefore, we have
\begin{align}
{{\mathcal{R}}_{\mathcal{U}_{k}^{*}}} & = \mathcal{U}_{k}^{*}{{\big( \mathcal{U}_{k}^{T}\mathbf{\Omega }_{k}^{\left( 1 \right)}\mathcal{U}_{k}^{*} \big)}\!^{-1}}\mathcal{U}_{k}^{T} \nonumber \\
& \approx b\mathcal{U}_{k}^{*}\mathcal{U}_{k}^{T} - ab\mathcal{U}_{k}^{*}\mathcal{U}_{k}^{T}\mathcal{H}_{k}^{H} {{\mathcal{H}}_{k}}\mathcal{U}_{k}^{*}\mathcal{U}_{k}^{T}.
\end{align}

For severe AoA overlapping among channels of different users, there is $\frac{1}{L}\operatorname{tr}\left( \mathcal{U}_{k}^{T}{{\mathcal{R}}_{k}}\mathcal{U}_{k}^{*} \right) \!\gg\! \sigma_{\mathrm n}^{2}$ under moderate and high SNR conditions. In this way, we have $a \!\approx\! \frac{1}{ \frac{1}{L} \! \operatorname{tr}\left( \mathcal{U}_{k}^{T}{{\mathcal{R}}_{k}}\mathcal{U}_{k}^{*} \right)}$ or
$a \operatorname{tr}\! \left( \mathcal{U}_{k}^{T}{{\mathcal{R}}_{k}}\mathcal{U}_{k}^{*} \right) \!\approx\! L$.
Note that in order to deal with the multiuser uplink with significant AoA-overlapping and severe mutual MUI, the ADAF requires a relatively large number of antennas $M$. Furthermore, the number of pilots, $N$, should be increased accordingly with $M$ to avoid underdetermination.

As a result, for large $M$ and $N$, ${{{g}}_{0}}\left( k \right)\!=\! \operatorname{tr}\! \left( {{\mathcal{R}}_{\mathcal{U}_{k}^{*}}} \right)$ in (\ref{gk_dk}) can be approximated as
\begin{align}
  {{{g}}_{0}}\left( k \right) &  \!\approx\! b\operatorname{tr}\! \left( {{\mathcal{U}}^{T}}\mathcal{U}_{k}^{*} \right) \!-\! ab\operatorname{tr}\left( \mathcal{U}_{k}^{*}\mathcal{U}_{k}^{T}\mathcal{H}_{k}^{H}{{\mathcal{H}}_{k}}\mathcal{U}_{k}^{*}\mathcal{U}_{k}^{T} \right)\!, \nonumber \\
 & \!=\! b{{Q}_{k}} \!-\! ab\operatorname{tr}\! \left( {{\mathcal{H}}_{k}}\mathcal{U}_{k}^{*}\mathcal{U}_{k}^{T}\mathcal{H}_{k}^{H} \right).
\end{align}
Besides, we have $\operatorname{tr}\! \left( {{\mathcal{H}}_{k}}\mathcal{U}_{k}^{*}\mathcal{U}_{k}^{T}\mathcal{H}_{k}^{H} \right) \!\approx\! \operatorname{tr}\! \left( \frac{1}{L}\! \operatorname{tr}\! \left( \mathcal{U}_{k}^{T} {{\mathcal{R}}_{k}} \mathcal{U}_{k}^{*} \right) \! {{\mathbf{I}}_{L}} \right)$ $\!\approx\! \operatorname{tr}\! \left( \mathcal{U}_{k}^{T} {{\mathcal{R}}_{k}} \mathcal{U}_{k}^{*} \right)$ based on (\ref{For_g0}), which leads to
\begin{align}
{{{g}}_{0}}\left( k \right) \!\approx\! b{{Q}_{k}} \!-\! ba\operatorname{tr}\left( \mathcal{U}_{k}^{T}{{\mathcal{R}}_{k}}\mathcal{U}_{k}^{*} \right)
 \!\approx\! b\left( {{Q}_{k}}-L \right). \label{g0}
\end{align}
Similarly, ${{g}_{1}}\left( k \right)$ in (\ref{gk_dk}) can be approximated by
\begin{align}
 {{g}_{1}}\left( k \right) & =\operatorname{tr}\left( {{\mathcal{R}}_{\mathcal{U}_{k}^{*}}}{{\mathbf{R}}_{k}} \right), \nonumber \\ & =b\operatorname{tr}\left( \mathcal{U}_{k}^{T}{{\mathbf{R}}_{k}}\mathcal{U}_{k}^{*} \right)-ab\operatorname{tr}\left( {{\mathcal{H}}_{k}}\mathcal{U}_{k}^{*}\mathcal{U}_{k}^{T}{{\mathbf{R}}_{k}}\mathcal{U}_{k}^{*}\mathcal{U}_{k}^{T}\mathcal{H}_{k}^{H} \right), \nonumber \\
 & \approx b{{r}_{k}}-ab\operatorname{tr}\left( \mathcal{U}_{k}^{T}{{\mathbf{R}}_{k}}\mathcal{U}_{k}^{*}\mathcal{U}_{k}^{T}{{\mathcal{R}}_{k}}\mathcal{U}_{k}^{*} \right), \nonumber \\
 & \approx b{{r}_{k}}-ab\operatorname{tr}\left( \operatorname{diag}\left( \mathcal{U}_{k}^{T}{{\mathbf{R}}_{k}}\mathcal{U}_{k}^{*} \right)\mathcal{U}_{k}^{T}{{\mathcal{R}}_{k}}\mathcal{U}_{k}^{*} \right), \nonumber \\
 & \approx b{{r}_{k}}-ab\operatorname{tr}\Big( \frac{1}{{{Q}_{k}}}\operatorname{tr}\left( \mathcal{U}_{k}^{T}{{\mathbf{R}}_{k}}\mathcal{U}_{k}^{*} \right){{\mathbf{I}}_{{{Q}_{k}}}}\mathcal{U}_{k}^{T}{{\mathcal{R}}_{k}}\mathcal{U}_{k}^{*} \Big), \nonumber \\
 & \approx b{{r}_{k}}-\frac{bL}{{{Q}_{k}}}{{r}_{k}}=b\Big( 1-\frac{L}{{{Q}_{k}}} \Big){{r}_{k}}. \label{g1}
\end{align}
Moreover, since
\begin{align*}
  {{\mathcal{R}}_{\mathcal{U}_{k}^{*}}}{{\mathcal{R}}_{\mathcal{U}_{k}^{*}}} =
  & \ \! {{b}^{2}}\mathcal{U}_{k}^{*}\mathcal{U}_{k}^{T}-2a{{b}^{2}} \mathcal{U}_{k}^{*}\mathcal{U}_{k}^{T}\mathcal{H}_{k}^{H} {{\mathcal{H}}_{k}}\mathcal{U}_{k}^{*}\mathcal{U}_{k}^{T} \nonumber \\
  & +{{a}^{2}}{{b}^{2}}\mathcal{U}_{k}^{*} \mathcal{U}_{k}^{T}\mathcal{H}_{k}^{H} {{\mathcal{H}}_{k}}\mathcal{U}_{k}^{*}\mathcal{U}_{k}^{T}\mathcal{H}_{k}^{H} {{\mathcal{H}}_{k}}\mathcal{U}_{k}^{*}\mathcal{U}_{k}^{T},
\end{align*}
we can approximate ${{{g}}_{2}}\left( k \right)$ as
\begin{align}{\label{g2}}
 & {{g}_{2}}\left( k \right)=\operatorname{tr}\left( {{\mathcal{R}}_{\mathcal{U}_{k}^{*}}}{{\mathcal{R}}_{\mathcal{U}_{k}^{*}}}{{\mathbf{R}}_{k}} \right), \nonumber \\
 \approx & \ \! {{b}^{2}}\operatorname{tr} \! \left( \mathcal{U}_{k}^{T}{{\mathbf{R}}_{k}}\mathcal{U}_{k}^{*} \right) - 2a{{b}^{2}}\operatorname{tr} \! \left( {{\mathcal{H}}_{k}}\mathcal{U}_{k}^{*}\mathcal{U}_{k}^{T}{{\mathbf{R}}_{k}}\mathcal{U}_{k}^{*}\mathcal{U}_{k}^{T}\mathcal{H}_{k}^{H} \right) \nonumber \\
 & + {{a}^{2}}{{b}^{2}}\operatorname{tr} \! \left( {{\mathcal{H}}_{k}}\mathcal{U}_{k}^{*}\mathcal{U}_{k}^{T}\mathcal{H}_{k}^{H}{{\mathcal{H}}_{k}}\mathcal{U}_{k}^{*} \mathcal{U}_{k}^{T}{{\mathbf{R}}_{k}}\mathcal{U}_{k}^{*}\mathcal{U}_{k}^{T}\mathcal{H}_{k}^{H} \right)\!, \nonumber \\
 \approx & \ \! {{b}^{2}}{{r}_{k}}-2a{{b}^{2}} \operatorname{tr}\left( \mathcal{U}_{k}^{T} {{\mathbf{R}}_{k}}\mathcal{U}_{k}^{*}\mathcal{U}_{k}^{T} {{\mathcal{R}}_{k}}\mathcal{U}_{k}^{*} \right) \nonumber \\
 &+\frac{{{a}^{2}}{{b}^{2}}}{L}\operatorname{tr}\left( \mathcal{U}_{k}^{T}{{\mathcal{R}}_{k}}\mathcal{U}_{k}^{*} \right)\operatorname{tr}\left( \mathcal{U}_{k}^{T}{{\mathbf{R}}_{k}}\mathcal{U}_{k}^{*}\mathcal{U}_{k}^{T}{{\mathcal{R}}_{k}}\mathcal{U}_{k}^{*} \right), \nonumber \\
 \approx & \ \! {{b}^{2}}{{r}_{k}}-\frac{2{{b}^{2}}L}{{{Q}_{k}}}{{r}_{k}}+\frac{{{b}^{2}}L}{{{Q}_{k}}}{{r}_{k}}={{b}^{2}} \Big( 1-\frac{L}{{{Q}_{k}}} \Big){{r}_{k}}, \nonumber \\
 = & \ \! b{{g}_{1}}\left( k \right).
\end{align}
Combining (\ref{g1}) and (\ref{g2}) leads to
\begin{align}{\label{g3}}
{{g}_{3}}\left( k \right) & = \frac{1}{N\!-\!L}{{g}_{1}}\left( k \right) - \sigma_{\mathrm n}^{2}{{g}_{2}}\left( k \right) \nonumber \\
 & = \sigma_{\mathrm n}^{2}\left[ b{{g}_{1}}\left( k \right) \!-\! {{g}_{2}}\left( k \right) \right] = 0.
\end{align}

By substituting (\ref{g0}), (\ref{g1}), (\ref{g2}) and (\ref{g3}) into (\ref{MSE_theo2}), the asymptotic MSE can be expressed as
\begin{align}{\label{MSE_theo4}}
  \mathrm{MSE} \{ {{{{\phi }}}_{k}} \} & \!\approx\! \frac{\frac{\sigma_{\mathrm n}^{2}}{{{L}^{3}}}b}{{2{\left[ \frac{\sigma_{\mathrm n}^{2}}{L}b+\frac{\sigma_{\mathrm n}^{2}}{{{L}^{2}}}b\left( {{Q}_{k}} \!-\! L \right) \!-\! \frac{1}{{{L}^{2}}} \right]}^{2}}{{g}_{1}}\! \left( k \right) \! {{d}_{k}}}, \nonumber \\
  & \!=\! \frac{L\sigma_{\mathrm n}^{2}}{2} \frac{\alpha_2}{{{r}_{k}}{{d}_{k}}} \!\approx\! {{\alpha }_{2}}\frac{\sigma_{\mathrm n}^{2}L{{\theta }_{\mathrm{as}}}{{\sin }}{{\theta }_{k}}}{2{{Q}_{k}}{{d}_{k}}} \!\approx\! {{\alpha }_{1}}\frac{\sigma_{\mathrm n}^{2}L{{\theta }_{\mathrm{as}}}{{\sin }}{{\theta }_{k}}}{2{{Q}_{k}}{{d}_{k}}},
\end{align}
where ${{\alpha }_{2}} \!=\! {{\big( \frac{N-L}{N-L-{{Q}_{k}}} \big)}^{2}}  \frac{{{Q}_{k}}}{{{Q}_{k}}-L}$. Note that we have the approximation, $\alpha_2 \!\approx\! \alpha_1$, for the large number of antennas, $M$, since $Q_k$ increases linearly with $M$.

From the above discussion, the MSE with AoA-overlapping among users in (\ref{MSE_theo4}) asymptotically converges to the MSE without AoA intersection in (\ref{MSE_theo3}) for sufficiently large numbers of antennas and subcarriers. Such convergence of MSE analytically confirms the effectiveness of the ADAF in suppressing the MUI originating from the overlapping users.

Note that (\ref{MSE_theo4}) is obtained in the case of large numbers of antennas, $M$, and subcarriers (pilots), $N$. Moreover, as the number of pilots $N$ increases, there asymptotically holds
$d_k \!\approx\! \\ L\! \operatorname{tr}\! \left[ {{\mathbf{B}}_k^{H}} \mathbf{D}{{\mathbf{D}}^{H}}  \mathbf{B}_k \right] \!-\! \frac{L}{N}\! \operatorname{tr}\! \left[ {{\mathbf{B}}_k^{H}} {{\mathbf{D}}^{H}} \mathbf{B}_k{{\mathbf{B}}_k^{H}} \mathbf{DB}_k \right]
\!\approx\! L\! \operatorname{tr}\! \left( \mathbf{D}{{\mathbf{D}}^{H}} \right) \\ \!-\! \frac{L}{N}\operatorname{tr}\! \left( {{\mathbf{D}}^{H}} \right)\operatorname{tr}\! \left( \mathbf{D} \right) \!\approx\! \frac{{{ \rm{\pi} }^{2}}}{3}NL$. Hence, (\ref{MSE_theo4}) can be further simplified as
\begin{align}{\label{MSE_theo4bis}}
\mathrm{MSE} \{ {{{{\phi }}}_{k}} \} & \ \! \overset{\mathrm{large} \ \! M }{\mathop{\approx}}\,  {{\alpha }_{1}}\frac{\sigma_{\mathrm n}^{2}L{{\theta }_{\mathrm{as}}}{{\sin }}{{\theta }_{k}}}{2{{Q}_{k}}{{d}_{k}}}  \ \! \overset{\mathrm{large} \ \! N }{\mathop{\approx}}\, \alpha_1 \frac{3\sigma_{\mathrm n}^{2}} {2\mathrm{\pi}^2 MN} \frac{\theta_{\mathrm{as}}}{\sin\theta_{\mathrm{as}}}, \nonumber \\
& \! \overset{\mathrm{small} \ \theta_{\mathrm{as}} }{\mathop{\approx}}\, \alpha_1 \frac{3\sigma_{\mathrm n}^{2}} {2\mathrm{\pi}^2 MN}.
\end{align}

It should be pointed out that, we can also arrive at (\ref{MSE_theo4bis}) from (\ref{MSE_theo3}) for a large number of subcarriers, $N$, and small AS, $\theta_{\mathrm{as}}$. Consequently, for both separate and mutually overlapping users, we can draw the following conclusions:

1) The MSE of CFO estimation decreases as expected with the increasing numbers of antennas $M$ and subcarriers $N$;

2) The average AoA of the user does not affect the CFO estimation accuracy, neither the AS $\theta_{\mathrm{as}}$ if it is small;

3) For a small AS $\theta_{\mathrm{as}}$, the asymptotic MSE expression in (\ref{MSE_theo4bis}) of the proposed ADAF approach equals that of JSFA given by equation (31) in~\cite{Zhang_TWC2018}, except for the constant coefficient $\alpha_1$, which is slightly larger than $1$. This indicates that no matter the AoA regions of users are separate or mutually overlapping, the ADAF could in both cases achieve comparable CFO estimation performance as the JSFA (in the absence of AoA intersection).
Thus, the ADAF not only has wider applicability than the JSFA, but also can achieve almost the same performance as the JSFA under scenarios of separate users in the asymptotic sense.

\section{Simulation Results}
In this section, we demonstrate the effectiveness of the proposed ADAF approach through numerical examples. The total number of subcarriers is taken as $N\!=\!64$ and the channel length is $L\!=\!10$. The uniform channel PDP is assumed for simulation, i.e., $\sigma_l^2 \!=\! \frac{1}{L}, l\!=\!1,2,\cdots\!,L$. The training symbols are randomly drawn from QPSK constellations, while the data transmission employs 16-QAM. The average powers of both pilot and data symbols are normalized to 1, i.e., we have $\sigma_{\mathrm{s}}^2 \!=\! E\left\{ \frac{1}{N} \mathbf{x}_i^{(k)H} \mathbf{x}_i^{(k)} \right\} \!=\! 1, \forall i \!=\! 1,2,\cdots\!,N_b, \forall k \!=\! 1,2,\cdots\!,k$. Thus, the SNR is given by $\sigma_{\mathrm{s}}^2 / \sigma_{\mathrm{n}}^2 \!=\! 1/ \sigma_{\mathrm{n}}^2$.
Both $M\!=\!64$ and $128$ receive antennas are considered. The carrier frequency is fixed as $f_c \!=\! 2.4 \ \mathrm{GHz}$, while the block duration and subcarrier spacing are taken as $T_b \!=\! 66.7\ \mu\mathrm{s}$ and $f_s \!=\! 1.5 \!\times\! 10^{4} \ \mathrm{Hz}$, respectively. The normalized CFO is randomly generated from $-0.2$ to $0.2$.
The AS is fixed as $\theta_{\rm as} \!=\! 10^\circ$. Unless otherwise stated, both the AS and average AoAs of users are assumed perfectly known at BS.
The MSE of the normalized CFO and symbol error rate (SER) are adopted as the performance metrics. For comparison, we also include the existing JSFA scheme~\cite{ZhangGLO16}, which is only applicable to the case of separate users without AoA overlapping, and the UG scheme~\cite{Zhang_TWC2018}, which groups all the users with AoA-intersection to perform joint CFO estimation and subsequent data detection.

\begin{figure}[t]
\setlength{\abovecaptionskip}{-0.2cm}
\setlength{\belowcaptionskip}{-0.6cm}
\begin{center}
\includegraphics[width=80mm]{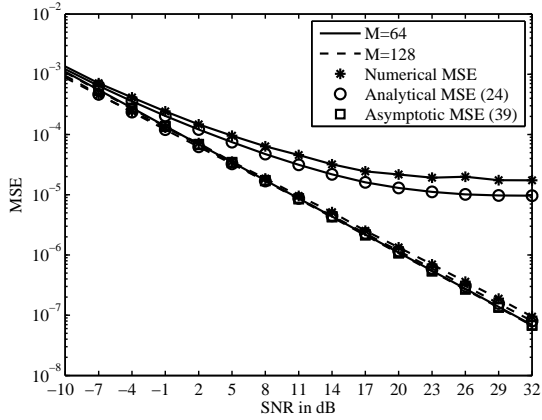}
\end{center}
\caption{ Comparison of analytical and numerical MSEs of the ADAF scheme, at $M=64, 128$, with $K=9$ users whose average AoAs are distributed as $\{ 38^\circ, 50^\circ, 70^\circ, 75^\circ, 95^\circ, 97^\circ, 117^\circ, 122^\circ, 142^\circ \}$. }
\end{figure}

In Fig. 3, we compare the analytical MSE and numerical MSE of the proposed ADAF scheme for $M \!=\! 64$ and $128$, respectively. The number of users is $K\!=\!9$ with average AoAs distributed as
$\{\underline{{{38}^{{}^\circ }},{{50}^{{}^\circ }}}, \underline{{{70}^{{}^\circ }},{{75}^{{}^\circ }}}, \underline{{{95}^{{}^\circ }},{{97}^{{}^\circ }}},\underline{{{117}^{{}^\circ }},{{122}^{{}^\circ }}}, {{142}^{{}^\circ }}\}$, which represents the scenario with rather severe AoA-overlapping among users. Note that the spatially proximate users are underlined.
From Fig. 3, we have the following observations:

1) For the ADAF approach, $M\!=\!64$ antennas is not sufficient to completely suppress the severe MUI stemming from the proximate users, which accounts for the obvious CFO estimation error floor. Nonetheless, such severe MUI could be effectively eliminated by increasing the number of antennas to $M\!=\!128$ and thereby the MSE performance floor is avoided.

2) The analytical MSE computed from (\ref{MSE_theo2}) tightly matches the numerical MSE, whether with performance floor ($M\!=\!64$) or not ($M\!=\!128$). This indicates that the severe AoA-overlapping and incomplete MUI suppression will not impact the accuracy of (\ref{MSE_theo2}).

3) The asymptotic MSE analysis (\ref{MSE_theo4}) at $M\!=\!64$ gradually deviates from its numerical counterpart as the SNR increases. However, the divergence between the MSE obtained by (\ref{MSE_theo4}) and the numerical MSE can be remarkably reduced and even disappears at $M\!=\!128$. This coincides with the fact that (\ref{MSE_theo4}) is an asymptotic result and thus is accurate only under a large number of antennas.

4) The analytical MSEs (\ref{MSE_theo4}) obtained at $M\!=\!64$ and $128$ coincide with each other, which is out of expectation yet can be explained as follows: On the one hand, the increasing number of antennas $M$ is favorable for MUI suppression and thereby enhances the CFO estimation accuracy. On the other hand, the increase of $M$ enlarges the dimension of ADAF vector ($Q_k$) and thus more parameters need to be estimated from the invariant $N$ pilots, which may compromise the MSE performance to some extent. These two contradictory factors offset each other, contributing to the coincidence of the MSE analysis results (\ref{MSE_theo4}) at $M\!=\!64$ and $128$. 

%
%

\begin{figure}[htbp]
\vspace{-0.5em}
\setlength{\abovecaptionskip}{-0.2cm}
\setlength{\belowcaptionskip}{-0.4cm}
\centering
\makeatletter\def\@captype{figure}\makeatother
  \subfigure{
    \includegraphics[width=80mm]{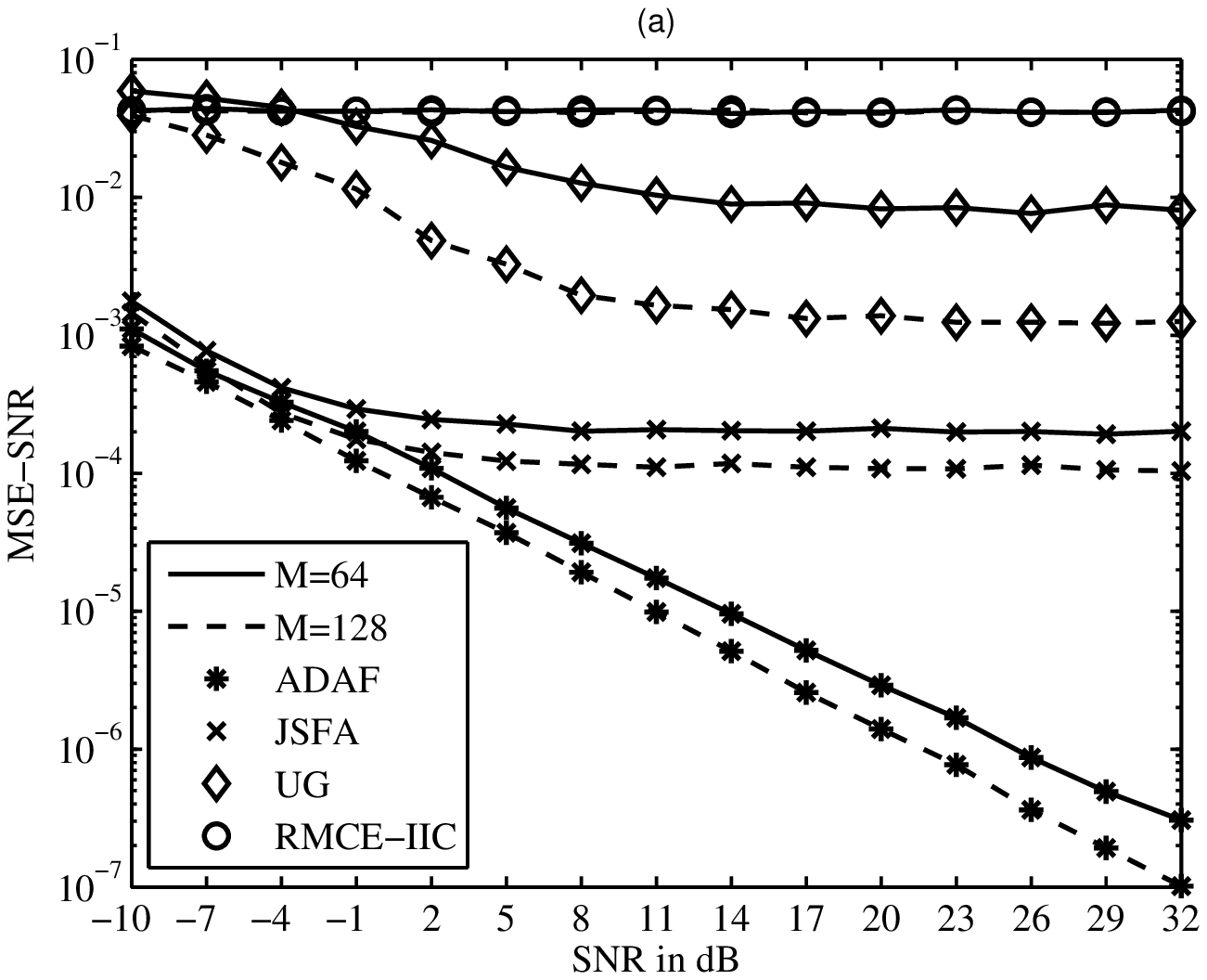}}
  \hspace{-0.2 in}
  \subfigure{
    \includegraphics[width=80mm]{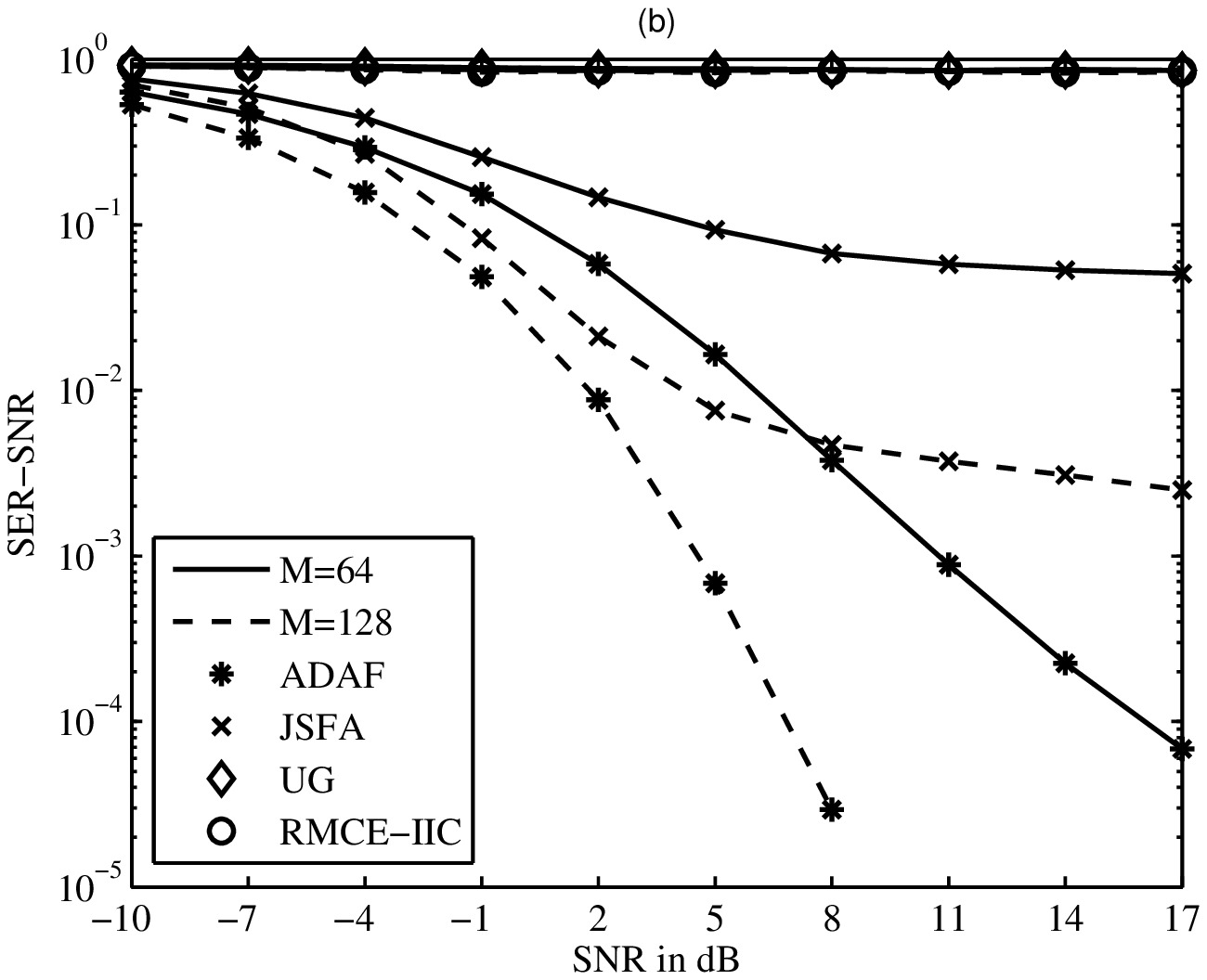}}
\caption{ (a) MSE and (b) SER performance comparison of ADAF, JSFA, UG and RMCE-IIC at $M \!=\! 64, 128$, with $K\!=\!9$ users whose average AoAs are $\{ 30^\circ, 45^\circ, 60^\circ, 75^\circ, 90^\circ, 105^\circ, 120^\circ, 135^\circ, 150^\circ\}$. }
\end{figure}


%
\setlength{\abovecaptionskip}{-0.1cm}
\setlength{\belowcaptionskip}{-0.8cm}
\begin{figure}[htbp]
\centering
\makeatletter\def\@captype{figure}\makeatother
  \subfigure{
    \includegraphics[width=80mm]{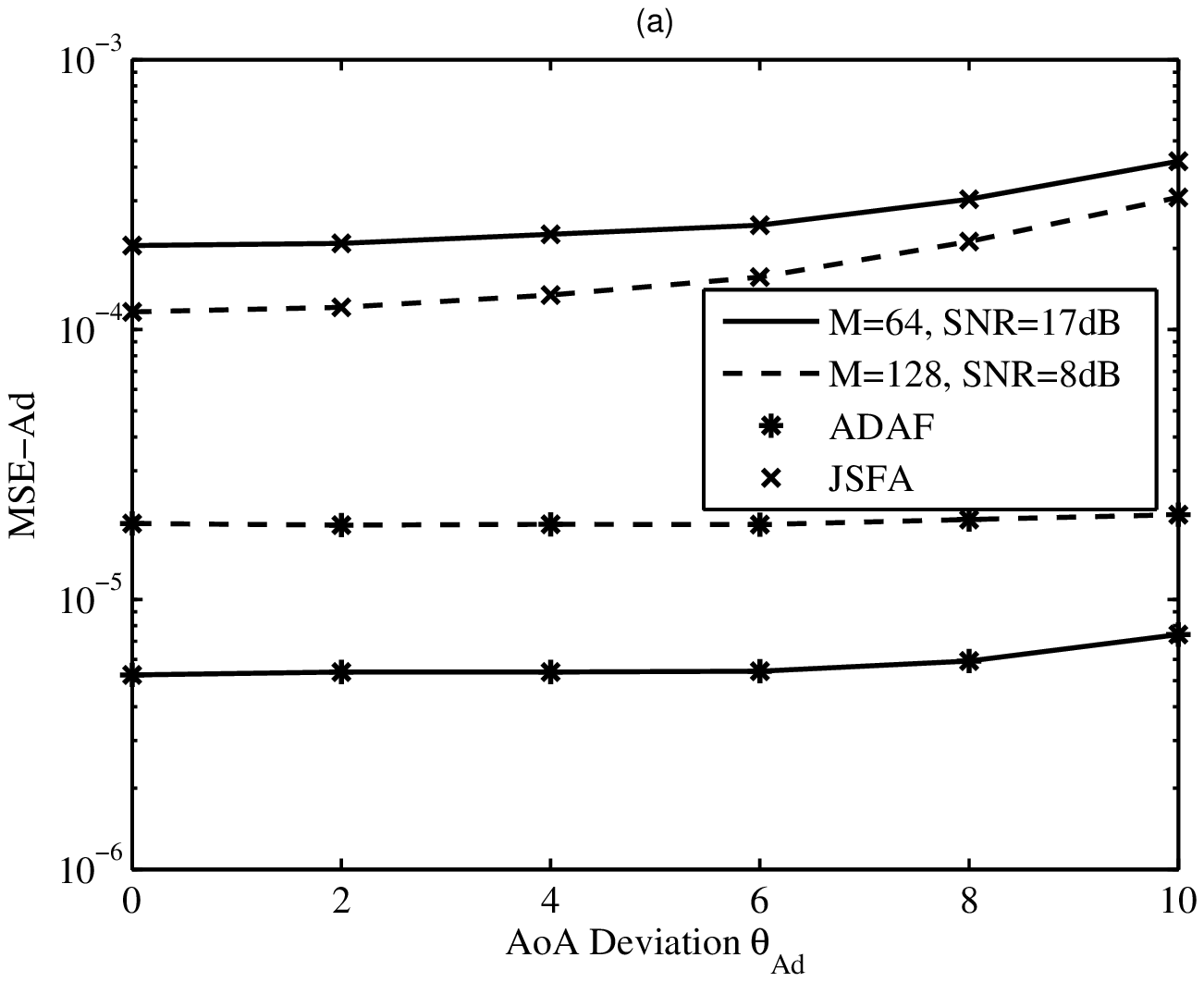}}
  \hspace{-0.2 in}
  \subfigure{
    \includegraphics[width=80mm]{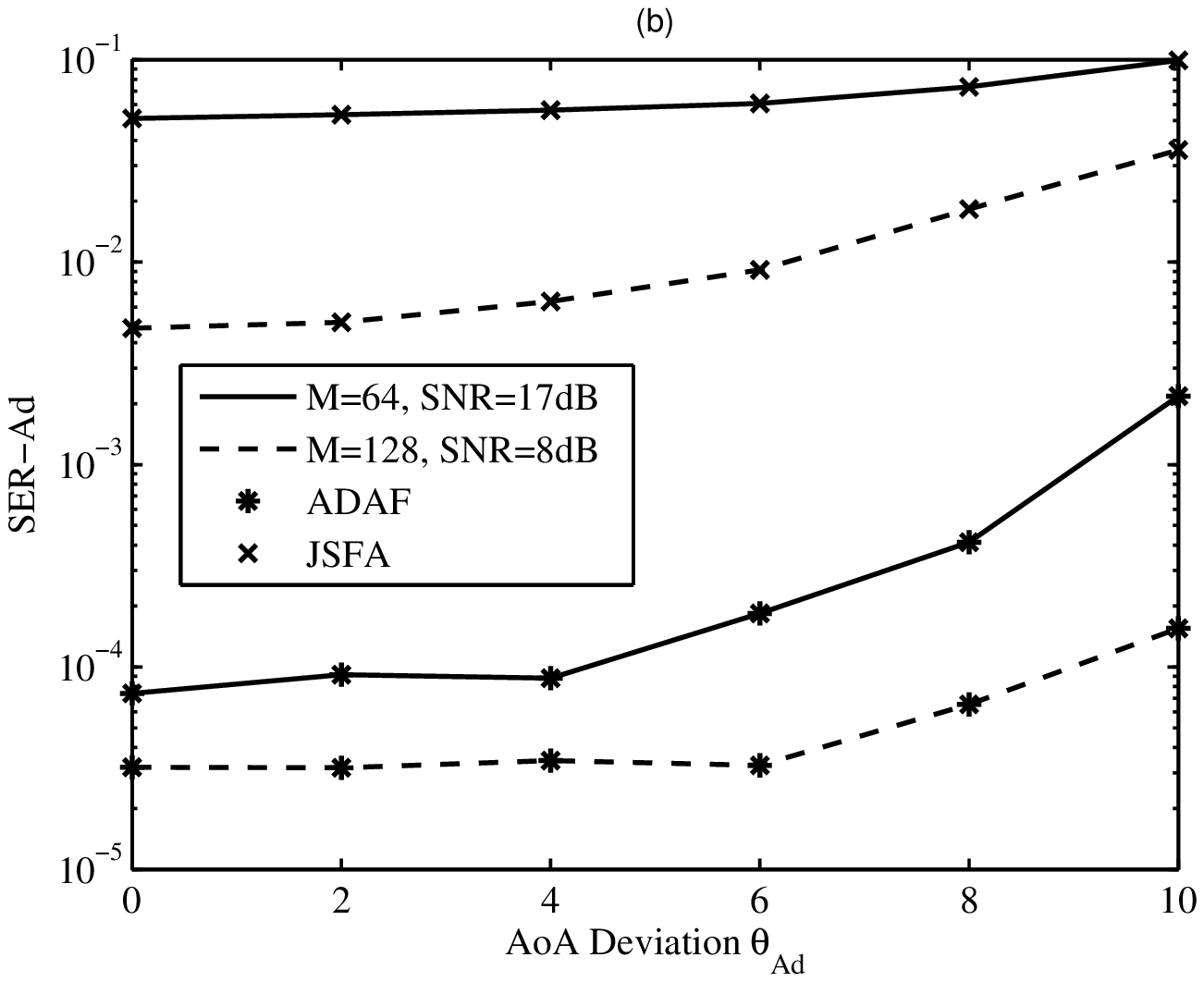}}
\caption{ (a) MSE and (b) SER performance comparison of ADAF and JSFA at $M \!=\! 64, 128$, with imperfect knowledge about average AoAs of $K\!=\!9$ users (same average AoAs as in Fig. 4). }
\end{figure}

Next, we compare the MSE and SER performances of the proposed ADAF method with the JSFA~\cite{ZhangGLO16}, the UG~\cite{Zhang_TWC2018} and the RMCE-IIC~\cite{Tsai-TWC13} in Fig. 4. Note that the RMCE-IIC in~\cite{Tsai-TWC13} was designed for single-antenna case and we extend it to the massive MIMO scenario. Besides, the transmitted data symbols are jointly detected with the detection method of the UG.
We consider $K\!=\!9$ users whose average AoAs are uniformly set as
$\{ 30^\circ, \!\ 45^\circ, \!\ 60^\circ, \!\ 75^\circ, \!\ 90^\circ, 105^\circ, 120^\circ, \!\ 135^\circ, \!\ 150^\circ\}$.
Here, the average AoAs are fixed instead of randomly generated to circumvent the complete superposition of the AoA regions among two or even more users. The scenario where several users completely superimpose will cause very severe MUI, making it impossible to separate the signals of different users simply in angle domain. As a result, we do not consider the scenario of complete superposition of users, which can be avoided via proper user scheduling~\cite{C_Sun2015TC} in practice.
From Fig. 4, the superiority of the ADAF algorithm over the other approaches is evident.

1) Due to the mutual overlapping among different users, the JSFA exploiting the MF beamformers cannot effectively eliminate the MUI caused by the adjacent overlapping users and thus suffers from high MSE and SER performance floor.

2) In contrast, the proposed ADAF scheme can successfully suppress the MUI originating from both the overlapping and non-overlapping users via the two-stage MUI suppression, contributing to the superior performance of the ADAF over the JSFA. Moreover, increasing the number of antennas $M$ provides higher spatial resolution and thus the potential for superior MUI suppression capability. Consequently, the performance of the ADAF can be significantly improved at $M\!=\!128$. However, increasing $M$ from 64 to 128 not only improves the CFO estimation performance, but also enhances the MUI suppression capacity and provides more diversities for MRC data detection among different branches. This accounts for the phenomenon that the SNR improvement is more significant than MSE improvement, by increasing the number of antennas.

3) With the considered average AoA distribution, the UG must jointly estimate the CFO and channel coefficients of all $K\!=\!9$ users, which amounts to determining $K(L+1)\!=\!99$ independent variables based on $N\!=\!64$ pilots. Thus, the under-determination problem occurs and it is expected that the UG fails to achieve frequency synchronization and data detection, while increasing the number of antennas $M$ would be in vain.

4) Since the IIC for the $k$th user requires the CFO and channel estimates of the other users at each iteration,
the CFOs and channels of all users are still jointly estimated, implying that the number of parameters to be estimated linearly increases with the number of users $K$. However, the number of pilots always remains $N\!=\!64$. Hence, the underdetermination problem occurs for large $K$, which accounts for the failure of the RMCE-IIC in achieving frequency synchronization and data detection in the case of $K\!=\!9$ coexisting users.

Note that the ADAF, the JSFA and the UG are all angle-domain approaches, while the RMCE-IIC~\cite{Tsai-TWC13} is code-domain approach. We then choose the ADAF and the RMCE-IIC as typical angle-domain and code-domain approaches to compare their computational complexities in terms of complex multiplications. The comparison is made for both the CFO estimation and data detection procedures. By taking the RMCE-IIC as example, the iterative CFO estimation is conducted in three stages: one-dimensional (1D) CFO search, channel estimation and IIC. Let $\eta$ and $\mu$ denote the number of iterations and the number of 1D searches required at each iteration, respectively. Then the complexities of the 1D CFO search, channel estimation and IIC for all users can be calculated to be $O\left( \mu \eta M\left( NL\!+\!N\!+\!L \right) \right)$, $O\left( \eta M\left( NL\!+\!N\!+\!L \right) \right)$ and $O\left(  \eta NL\left( M\!+\!1 \right) \right)$, respectively. Similarly, we can calculate the complexity for the ADAF. The complexities of the ADAF and the RMCE-IIC are summarized in Table I.

\begin{table}[!t]
\vspace{2em}
\caption{ Computational complexities of CFO estimation and data detection for the ADAF and the RMCE-IIC }\label{CC_3Algorithms}
\vspace{0.8em}
\centering
\begin{tabular}{|c||c|c|}
\hline
\!\!\!Algorithm\!\!\! & \multicolumn{2}{|c|}{Computational complexity} \\
\hline
\multirow{3}*{ADAF} & \!\!\!$\begin{matrix}
   \mathrm{CFO}  \\
   \mathrm{estimation} \\
\end{matrix}$\!\!\! & \!\!\!$O\!\left( \!K\!\left( \begin{matrix}
   L{{\left( N\!+\!L \right)}^{2}} \!+\! \eta N\left( M\!+\!3N\!+\!3Q_{k}^{2} \right) \!+  \\
   \eta {{Q}_{k}}\left( NM \!+\!3{{N}^{2}} \right) \!+\! Q_k^2 \left( N\!+\!4\eta Q_k \right)  \\
\end{matrix} \right) \!\right)$\!\!\!\! \\
\cline{2-3}
$~ $ & \!\!$\begin{matrix}
   \mathrm{Data}  \\
   \mathrm{detection} \\
\end{matrix}$\!\! & $O\left( \begin{matrix}
   KN{{Q}_{k}}{{N}_{b}}\left( {{\log }_{2}}N \!+\! 1 \right)  \!+  \\
   K{{Q}_{k}}\left( NL \!+\! N\!+\!2Q_{k}^{2} \right)   \\
\end{matrix} \right)$ \\
\hline
\multirow{3}*{\!\!\!RMCE-IIC\!\!\!} & \!\!\!$\begin{matrix}
   \mathrm{CFO}  \\
   \mathrm{estimation} \\
\end{matrix}$\!\!\! & \!\!\!$O\!\left(\! K\!\left( \begin{matrix}
   \mu \eta M\left( NL\!+\!N\!+\!L \right) \!+  \\
   \eta M\left( N\!+\!2NL\!+\!L \right) \!+\! \eta NL \\
\end{matrix} \right) \!\right)$\!\!\!\! \\
\cline{2-3}
$~ $ & \!\!$\begin{matrix}
   \mathrm{Data}  \\
   \mathrm{detection} \\
\end{matrix}$\!\! & $O\left( \begin{matrix}
   KMN{{\log }_{2}}N \!+\! K{{\left( M \!+\! K \right)}^{2}}{{N}^{3}} +  \\
   K\left( M\!+\!K\!+\!1 \right){{N}^{2}}\left( {{N}_{b}} \!-\!1 \right) \\
\end{matrix} \right)$ \\
\hline
\end{tabular}
\vspace{-0em}
\end{table}

\begin{table}[!t]
\vspace{2em}
\caption{ Comparison of the computational complexities between the ADAF and the RMCE-IIC }\label{CC_Comparison}
\vspace{0.5em}
\centering
\begin{tabular}{|c||c|c|c|}
\hline
\multirow{2}*{Algorithms} & \multicolumn{3}{|c|}{Computational complexity} \\
\cline{2-4}
$~ $ & CFO estimation & Data detection & Overall \\
\hline
ADAF & $6.90\times 10^6$ & $1.07\times 10^5$ & $7.01\times 10^6$ \\
\hline
RMCE-IIC & $1.11\times 10^8$ & $1.26\times 10^{10}$ & $1.27\times 10^{10}$ \\
\hline
\end{tabular}
\vspace{-2em}
\end{table}

Consider the scenario of $N\!=\!64, M\!=\!64, K \!=\!9, L \!=\!10, N_b \\ \!=\! 2, \eta\!=\!5, \theta_{\mathrm{as}} \!=\! 10^{\circ}$. Besides, $Q_k$ can be approximated by its expectation $E\{ Q_k \} \!=\! \frac{2\theta_{\mathrm{as}}}{\mathrm{\pi}} M \!\approx\! 7$. Moreover, we accomplish the 1D CFO search in two steps, with 0.01 and 0.001 being the search intervals. Such two-step search yields $\mu \!=\! 52$. Then, the orders of complexity required by the ADAF and the RMCE-IIC are provided in Table II. Obviously, the proposed ADAF algorithm profits from noticeably reduced complexity than the RMCE-IIC, especially for the data detection.

The sensitivity of the ADAF and the JSFA to the inaccurate AoA knowledge is assessed in Fig. 5. All the parameters remain the same as in Fig. 4, except that the knowledge about the average AoAs is assumed imperfect.
Suppose the BS takes $\hat \theta_k \!=\! \theta_k \!+\! \Delta \theta_k$ as the estimated average AoA of the $k$th user for ACM generation and CFO estimation. The AoA bias, $\Delta \theta_k$, is assumed to follow uniform distribution $\mathcal{U} \left(-\theta_{\mathrm{Ad}}, \theta_{\mathrm{Ad}} \right)$, where $\theta_{\mathrm{Ad}}$ is the maximum AoA deviation from the real average AoA. Note that in view of the AS $\theta_{\mathrm{as}} \!=\!\! 10^\circ$, the AoA bias $\Delta \theta_k$ generated with $\theta_{\mathrm{Ad}} \!=\!\! 10^\circ$ will be quite non-negligible.
Fig. 5 reveals that the MSE performance of the JSFA and especially that of the ADAF are very robust to the AoA imperfection. In contrast, the SER performance exhibits weak sensitivity to $\theta_{\mathrm{Ad}}$. Nevertheless, even at $\theta_{\mathrm{Ad}} \!=\! 10^\circ$, the SER performance exacerbation of the ADAF with $M\!=\!128$ antennas is still tolerable. Such insensitivity confirms that the AoA information is not prerequisite of the proposed ADAF scheme, although perfect knowledge about average AoAs is assumed available at the BS. An estimation of the average AoA, $\hat \theta_k$, with reasonable accuracy for the $k$th user, which can be easily obtained~\cite{Zhang_TWC2018}, is adequate for achieving satisfactory performance.

\begin{figure}[t]
\setlength{\abovecaptionskip}{-0.3cm}
\setlength{\belowcaptionskip}{-0.7cm}
\begin{center}
\includegraphics[width=90mm]{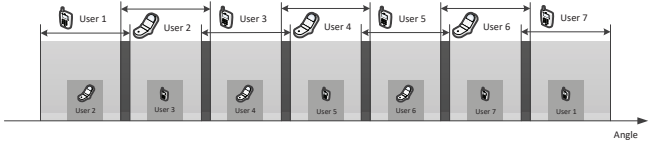}
\end{center}
\caption{ Illustration of the scenario with side cluster considered in Fig. 7 (overlapping users with AoA-intersection). }
\end{figure}

\setlength{\abovecaptionskip}{-0.2cm}
\setlength{\belowcaptionskip}{-0.6cm}
\begin{figure}[htbp]
\centering
\makeatletter\def\@captype{figure}\makeatother
  \subfigure{
    \includegraphics[width=81mm]{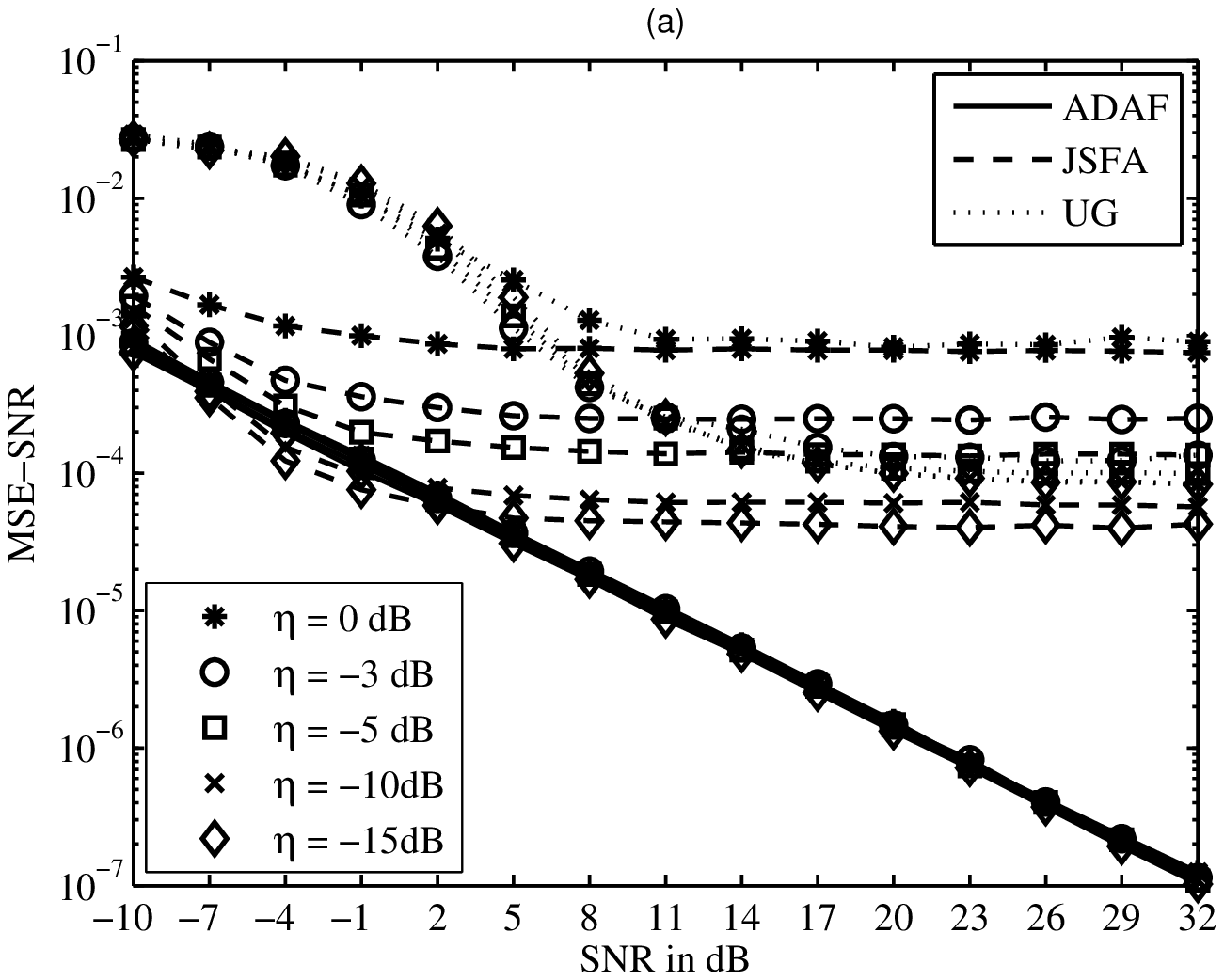}}
  \hspace{-0.2 in}
  \subfigure{
    \includegraphics[width=80mm]{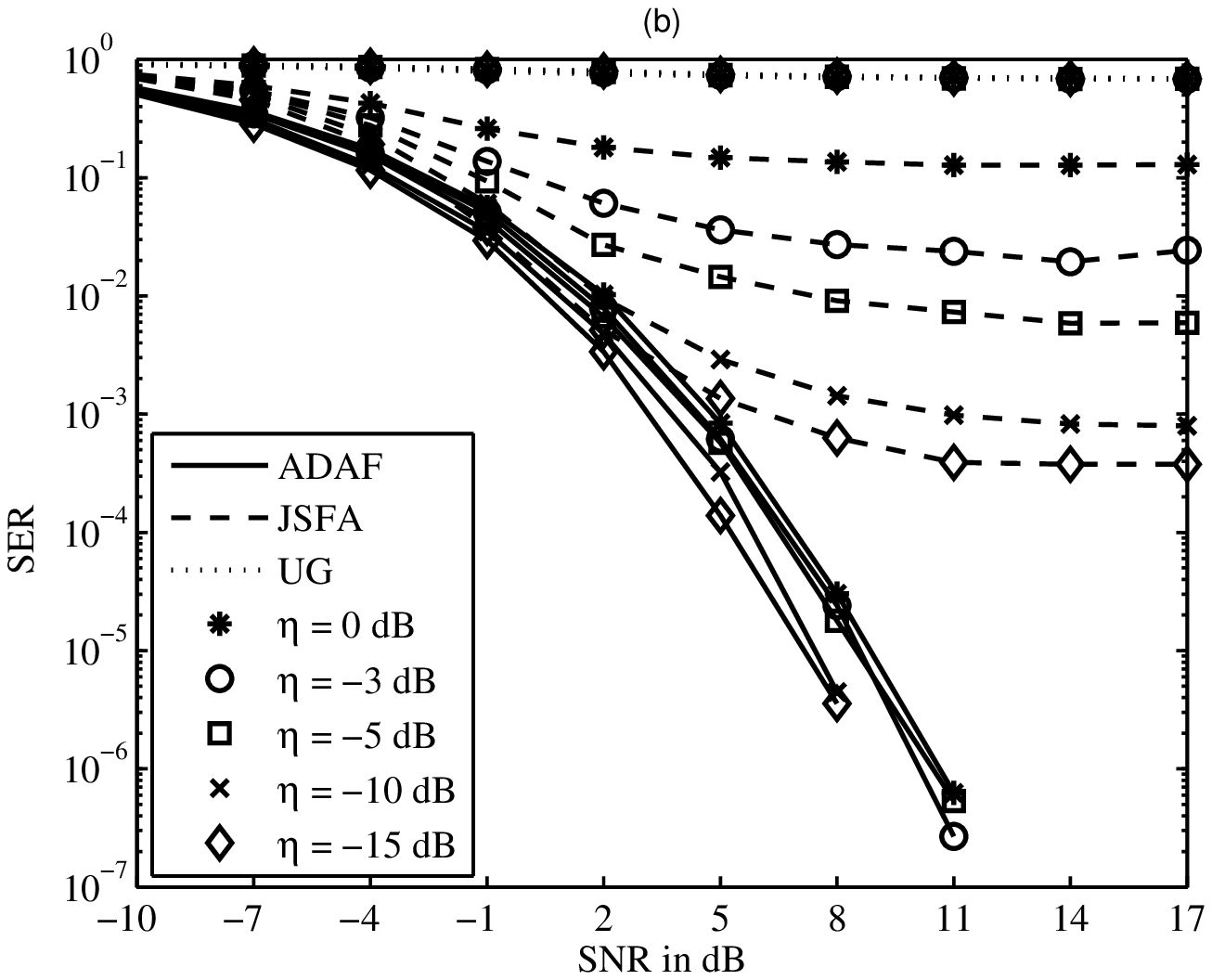}}
\caption{ Robustness of (a) MSE and (b) SER performance of ADAF, JSFA and UG to side cluster with $\eta \!=\! -\{ 0,3,5,10,15 \} \mathrm{dB}$, at $K\!=\!7$ (average AoAs: $\{ 36^\circ, 54^\circ, 72^\circ, 90^\circ, 108^\circ, 126^\circ, 144^\circ \}$) and $M\!=\!128$. }
\end{figure}

Then, the robustness of the ADAF, the JSFA and the UG against the side cluster is evaluated in Fig. 7 under the scenario with AoA-overlapping among $K\!=\!7$ users, whose average AoAs are distributed as
$\{ 36^\circ, 54^\circ, 72^\circ, 90^\circ, 108^\circ, 126^\circ, 144^\circ \}$, as illustrated in Fig. 6.
The AS of the side clusters is $\theta_{\mathrm{as}}^{(\mathrm{sc})} \!=\! 2^\circ$, while the average AoAs for the side clusters are fixed as
$\{ 54^\circ, 72^\circ, 90^\circ, 108^\circ, 126^\circ, 144^\circ, 36^\circ \}$.
Among all the possible distributions, such a distribution of the average AoAs tends to aggravate the MUI introduced by the side cluster and thus represents the relatively unfavorable circumstance detrimental to the system performance.
The average AoAs of the main clusters are assumed perfectly known and the signal strength within the side cluster is adjusted according to SMPR $\eta \!=\! \{ 0, -3, -5, -10, -15 \} \ \mathrm{dB} $. Besides, the number of antennas is fixed as $M\!=\!128$ in Fig. 7.

Fig. 7 confirms the robustness of the ADAF against the interfering side cluster. On the contrary, The performance of the JSFA degrades drastically with the increase of SMPR $\eta$ (equivalently, increase of signal strength within the side cluster and aggravation of corresponding MUI). Such divergence in the behavior of the ADAF and the JSFA is within expectation. In fact, the ADAF need not distinguish overlapping users from the interfering side clusters falling into the AoA region of the main cluster of the desired user. All the interference originating from overlapping or non-overlapping users and interfering side clusters could be effectively suppressed via the two-stage MUI suppression in an indiscriminating way.
Instead, the JSFA can only eliminate the MUI stemming from the non-overlapping users or remote side clusters and thus its performance accordingly degrades as the SMPR increases. Besides, although the joint estimation for intragroup users of the UG makes it unnecessary to combat the MUI from intragroup users, the UG is still incapable of dealing with the interfering side clusters related to users outside the group. Moreover, the underdetermination problem is unavoidable for the UG with a large number of users yet limited number of pilots $N$, leading to its complete failure to achieve frequency synchronization and data detection.


\section{Conclusions}
In this paper, we have developed an ADAF-based frequency synchronization scheme for multiuser MIMO uplink when the AoA regions of different users might intersect. By appropriately designing the ADAF vectors, we adaptively obtain the qualified beamformers that can substantially suppress the MUI from both overlapping and non-overlapping users. This conduces to an equivalent single-user transmission network. Thus, the CFO estimation and data detection can be carried out individually for each user, significantly reducing the computational burden.
Numerical results have corroborated the effectiveness of the ADAF algorithm and its superiority over existing competitors, \\ including the JSFA, the UG and the RMCE-IIC. Furthermore, the ADAF is quite robust to imperfect average AoAs knowledge about users and is almost insusceptible to the presence of the side clusters, even those with high SMPR.
All these rarely coexisting merits integrated into the ADAF enable it a promising technique for the uplink massive MIMO networks.

\vspace{-0.6em}
\appendices
\section{Derivation of Lemma 1}{\label{Lemma_Derivation}}
Denote the diagonal elements and the eigenvalues of $\mathbf{A}$ as $\mathbf{d}(\mathbf{A})$ and $\boldsymbol{\lambda}(\mathbf{A})$ (in descending order), respectively, and each element of $\mathbf{1}(\mathbf{A})$ is the average of all the diagonal elements of $\mathbf{A}$. Then, we have the following majorization inequality according to~\cite{DPPalomar_TSP2003} and~\cite{Marshall2009}
\begin{align}
\mathbf{1}\! \left( \mathbf{A} \right) \!=\! \mathbf{1}\! \left( \tilde{\boldsymbol \Upsilon}^{H} \mathbf{A} \tilde{\boldsymbol \Upsilon} \right) \!\prec\! \mathbf{d}\! \left( \tilde{\boldsymbol \Upsilon}^{H} \mathbf{A} \tilde{\boldsymbol \Upsilon} \right) \!\prec\! \boldsymbol{\lambda }\! \left( \tilde{\boldsymbol \Upsilon}^{H} \mathbf{A} \tilde{\boldsymbol \Upsilon} \right) \!=\! \boldsymbol{\lambda }\! \left( \mathbf{A} \right),
\end{align}
where $\mathbf{x} \prec \mathbf{y}$ represents that vector $\mathbf{x}$ is majorized by vector $\mathbf{y}$, mathematically, $\sum\nolimits_{q=1}^{{\bar{Q}}}{{{x}_{q}}}\le \sum\nolimits_{q=1}^{{\bar{Q}}}{{{y}_{q}}}$ for $ 1\le \bar{Q}\le Q-1$ and $\sum\nolimits_{q=1}^{Q}{{{x}_{q}}}=\sum\nolimits_{q=1}^{Q}{{{y}_{q}}}$.

Define $f\left( \mathbf{x} \right)=\sum\nolimits_{q=1}^{Q}{\frac{1}{{{x}_{q}}}} $ for $\mathbf{x}\in \mathcal{R}\subseteq \mathbb{R}_{+}^{Q}$. Then, $f\left( \mathbf{x} \right)$ is symmetric and there holds
\begin{align*}
& \left( {{x}_{i}}\!-\!{{x}_{j}} \right)\! \Big( \frac{\partial f}{\partial {{x}_{i}}}\!-\!\frac{\partial f}{\partial {{x}_{j}}} \Big) \!=\! \left( {{x}_{i}}\!-\!{{x}_{j}} \right) \Big( \frac{1}{x_{j}^{2}}\!-\!\frac{1}{x_{i}^{2}} \Big), \nonumber \\
 = & \frac{ {{\left( {{x}_{i}}\!-\!{{x}_{j}} \right)}^{2}}}{{{x}_{i}}{{x}_{j}}} \Big( \frac{1}{{{x}_{i}}} \!+\! \frac{1}{{{x}_{j}}} \Big)\! \ge 0, \forall \mathbf{x}\in {\mathcal{R}}, 1\le i\ne j\le d.
\end{align*}
Hence, $f\left( \mathbf{x} \right)$ is a Schur-convex function according to~\cite{Wayne1973} and $f\left( \mathbf{d}\left( {{\tilde{\boldsymbol \Upsilon}}^{H}} \mathbf{A} \tilde{\boldsymbol \Upsilon} \right) \right) \!=\!  \sum_{q=1}^{Q} \frac{1}{ \tilde{\boldsymbol \gamma}_q^H \mathbf{A} \tilde{\boldsymbol \gamma}_q }$.

Since $\mathbf{d}\left( {{\tilde{\boldsymbol \Upsilon}}^{H}} \mathbf{A} \tilde{\boldsymbol \Upsilon} \right) \!\prec\! \boldsymbol{\lambda }\left( \mathbf{A} \right)$ and $\mathbf{d} \left( {{\hat{\boldsymbol \Upsilon}}^{H}} \mathbf{A} \hat{\boldsymbol \Upsilon} \right) \!=\! \boldsymbol{\lambda }\left( \mathbf{A} \right)$ at ${\hat{\boldsymbol \gamma}}_q \!=\! \boldsymbol{\nu}_q (\mathbf{A}), q \!=\! 1,2, \cdots\!, Q$, then (\ref{Lemma1}) holds.

This completes the proof.

\section{Derivation of Analytical MSE}{\label{MSE_Derivation}}
The first-order derivative (FOD) and second-order derivative (SOD) of ${{G}_{k}}\big( {{{\tilde{\phi }}}_{k}} \big)$ with respect to ${{{\tilde{\phi }}}_{k}}$ are given by
\begin{align}{\label{gk_FOD}}
\frac{\partial {{G}_{k}}\big( {{{\tilde{\phi }}}_{k}} \big)}{\partial {{{\tilde{\phi }}}_{k}}} \!=\! -\operatorname{tr}\! \Big( \mathbf{\Xi }_{k}^{-1}\big( {{{\tilde{\phi }}}_{k}} \big)\frac{\partial {{\mathbf{\Xi }}_{k}}\big( {{{\tilde{\phi }}}_{k}} \big)}{\partial {{{\tilde{\phi }}}_{k}}}\mathbf{\Xi }_{k}^{-1}\big( {{{\tilde{\phi }}}_{k}} \big)\mathcal{U}_{k}^{T}{{\mathbf{Y}}^{H}}\mathbf{Y}\mathcal{U}_{k}^{*} \Big),
\end{align}
and
\begin{align}{\label{gk_SOD}}
\frac{{{\partial }^{2}}{{G}_{k}}\big( {{{\tilde{\phi }}}_{k}} \big)}{\partial \tilde{\phi }_{k}^{2}} \!=\! \operatorname{tr} & \bigg[ \mathbf{\Xi }_{k}^{-1}\big( {{{\tilde{\phi }}}_{k}} \big)\bigg( 2\frac{\partial {{\mathbf{\Xi }}_{k}}\big( {{{\tilde{\phi }}}_{k}} \big)}{\partial {{{\tilde{\phi }}}_{k}}}\mathbf{\Xi }_{k}^{-1}\big( {{{\tilde{\phi }}}_{k}} \big)\frac{\partial {{\mathbf{\Xi }}_{k}}\big( {{{\tilde{\phi }}}_{k}} \big)}{\partial {{{\tilde{\phi }}}_{k}}} \nonumber \\
& \!-\! \frac{{{\partial }^{2}}{{\mathbf{\Xi }}_{k}}\big( {{{\tilde{\phi }}}_{k}} \big)}{\partial \tilde{\phi }_{k}^{2}} \bigg)\mathbf{\Xi }_{k}^{-1}\big( {{{\tilde{\phi }}}_{k}} \big)\mathcal{U}_{k}^{T}{{\mathbf{Y}}^{H}}\mathbf{Y}\mathcal{U}_{k}^{*} \bigg].
\end{align}

1) Calculation of the derivatives of $\mathbf{\Xi }_{k}\big( {{{\tilde{\phi }}}_{k}} \big)$ with respect to $\tilde{\phi}_k$:

First, ${{\bf\Xi }_{k}} \big( {{\phi }_k} \big)$ can be approximately expressed as
\begin{align}\label{Xi_k}
 {{\left. E\left\{ {{\bf\Xi }_{k}} \big( {\tilde{\phi }_k} \big) \right\} \right|}_{\tilde{\phi }_k={{\phi }_{k}}}} & = \mathcal{U}_k^T {\bf Y}^H  {\bf E}(\phi_k)  {\bf P}_{{\bf B}_k}^\bot {\bf E}^H(\phi_k) {\bf Y} \mathcal{U}_k^*, \nonumber \\
 & = {{\mathcal{U}}^{T}} \mathbf{\Omega }_{k}^{\left( 1 \right)}{{\mathcal{U}}^{*}},
\end{align}
where
\begin{align}{\label{Omega_k1}}
 \mathbf{\Omega }_{k}^{\left( 1 \right)} \approx & \ \! {{\mathbf{N}}^{H}}\mathbf{E}\left( {{\phi }_{k}} \right)\mathbf{P}_{{{\mathbf{B}}_{k}}}^{\bot }{{\mathbf{E}}^{H}}\left( {{\phi }_{k}} \right)\mathbf{N} \nonumber \\
 & \hspace{-3.15em} +\! \sum\nolimits_{k'=1,k'\ne k}^{K} \! {\mathbf{H}_{k'}^{H}\mathbf{B}_{k'}^{H} \mathbf{E}\! \left( {{\phi }_{k}}\!-\!{{\phi }_{k'}} \right)\! \mathbf{P}_{{{\mathbf{B}}_{k}}}^{\bot } {{\mathbf{E}}^{H}}\! \left( {{\phi }_{k}}\!-\!{{\phi }_{k'}} \right) \! {{\mathbf{B}}_{k'}}{{\mathbf{H}}_{k'}}}\!, \nonumber \\
 \approx & \ \! \sigma_{\mathrm n}^{2} \operatorname{tr}\left( \mathbf{E}\left( {{\phi }_{k}} \right)\mathbf{P}_{{{\mathbf{B}}_{k}}}^{\bot }{{\mathbf{E}}^{H}}\left( {{\phi }_{k}} \right) \right){{\mathbf{I}}_{M}} \nonumber \\
 & \hspace{-3.15em} +\sum\nolimits_{k'=1,k'\ne k}^{K} {\operatorname{tr}\left( \mathbf{E}\left( {{\phi }_{k}}\!-\!{{\phi }_{k'}} \right) \mathbf{P}_{{{\mathbf{B}}_{k}}}^{\bot } {{\mathbf{E}}^{H}}\left( {{\phi }_{k}}\!-\!{{\phi }_{k'}} \right) \right)\mathbf{H}_{k'}^{H}{{\mathbf{H}}_{k'}}}, \nonumber \\
 =& \left( N-L \right)\big( \sigma_{\mathrm n}^{2}{{\mathbf{I}}_{M}} + \mathbf{\Pi }_{k}^{\left( 2 \right)} \big).
\end{align}
Here, $\mathbf{\Pi }_{k}^{\left( 2 \right)}=\sum\nolimits_{k'=1,k'\ne k}^{K}{\mathbf{H}_{k'}^{H}{{\mathbf{H}}_{k'}}}$.
From (\ref{Xi_k}), we can also obtain
\begin{align}{\label{inv_Xik}}
{{\left. E\left\{ {{\bf\Xi }_{k}^{-1}} \big( {\tilde{\phi }_k} \big) \right\} \right|}_{\tilde{\phi }_k={{\phi }_{k}}}} \approx {{\big( {{\mathcal{U}}_k^T} \mathbf{\Omega }_{k}^{\left( 1 \right)} {{\mathcal{U}}_k^*} \big)}^{-1}}.
\end{align}

Then, the FOD of ${{\bf\Xi }_{k}} \big( {\tilde{\phi }_k} \big)$ with respect to ${\tilde{\phi }_k} $ can be approximated as
\begin{align}{\label{Xik_FOD}}
& {{\left. E\left\{ \frac{\partial {{\mathbf{\Xi }}_{k}}\big( {{{\tilde{\phi }}}_{k}} \big)}{\partial {{{\tilde{\phi }}}_{k}}} \right\} \right|}_{{{{\tilde{\phi }}}_{k}}={{\phi }_{k}}}} \nonumber \\
=& \ \mathcal{U}_{k}^{T}{{\mathbf{Y}}^{H}}\mathbf{E}\left( {{\phi }_{k}} \right)\left( \mathbf{DP}_{{{\mathbf{B}}_{k}}}^{\bot }+\mathbf{P}_{{{\mathbf{B}}_{k}}}^{\bot }{{\mathbf{D}}^{H}} \right){{\mathbf{E}}^{H}}\left( {{\phi }_{k}} \right)\mathbf{Y}\mathcal{U}_{k}^{*}, \nonumber \\
\approx &\ \mathcal{U}_{k}^{T}\left[ {{\mathbf{N}}^{H}}\mathbf{\Gamma }_{k}^{\left( 1 \right)}+\mathbf{\Gamma }_{k}^{\left( 1 \right)H}\mathbf{N}+{{\mathbf{N}}^{H}}\mathbf{\Gamma }_{k}^{\left( 2 \right)}+\mathbf{\Gamma }_{k}^{\left( 2 \right)H}\mathbf{N} \right]\mathcal{U}_{k}^{*} \nonumber \\
& + \mathcal{U}_{k}^{T}\left[ \mathbf{\Phi }_{k}^{\left( 2 \right)}+\mathbf{\Omega }_{k}^{\left( 2 \right)}+\mathbf{\Psi }_{k}^{\left( 2 \right)}+\mathbf{\Psi }_{k}^{\left( 2 \right)H} \right]\mathcal{U}_{k}^{*}, \nonumber \\
\approx &\ \mathcal{U}_{k}^{T}\left[ {{\mathbf{N}}^{H}}\big( \mathbf{\Gamma }_{k}^{\left( 1 \right)}+\mathbf{\Gamma }_{k}^{\left( 2 \right)} \big)+\big( \mathbf{\Gamma }_{k}^{\left( 1 \right)H}+\mathbf{\Gamma }_{k}^{\left( 2 \right)H} \big)\mathbf{N} \right]\mathcal{U}_{k}^{*} \nonumber \\
& + \mathcal{U}_{k}^{T}\left[ \mathbf{\Psi }_{k}^{\left( 2 \right)}+ \mathbf{\Psi }_{k}^{\left( 2 \right)H} \right]\mathcal{U}_{k}^{*},
\end{align}
where
\begin{align}
 \mathbf{\Gamma }_{k}^{\left( 1 \right)} = & \ \! \mathbf{E}\left( {{\phi }_{k}} \right)\mathbf{P}_{{{\mathbf{B}}_{k}}}^{\bot }{{\mathbf{D}}^{H}}{{\mathbf{B}}_{k}}{{\mathbf{H}}_{k}}, \nonumber \\
\mathbf{\Gamma }_{k}^{\left( 2 \right)} = & \ \! \mathbf{E}\left( {{\phi }_{k}} \right)\big( \mathbf{DP}_{{{\mathbf{B}}_{k}}}^{\bot }+\mathbf{P}_{{{\mathbf{B}}_{k}}}^{\bot }{{\mathbf{D}}^{H}} \big) \nonumber \\
& \times \sum\nolimits_{k'=1,k'\ne k}^{K}{{{\mathbf{E}}^{H}}\left( {{\phi }_{k}}-{{\phi }_{k'}} \right){{\mathbf{B}}_{k'}}{{\mathbf{H}}_{k'}}}, \nonumber \\
\mathbf{\Phi }_{k}^{\left( 2 \right)} = & \ \! {{\mathbf{N}}^{H}}\mathbf{E}\left( {{\phi }_{k}} \right) \left( \mathbf{DP}_{{{\mathbf{B}}_{k}}}^{\bot } +\mathbf{P}_{{{\mathbf{B}}_{k}}}^{\bot } {{\mathbf{D}}^{H}} \right) {{\mathbf{E}}^{H}}\left( {{\phi }_{k}} \right)\mathbf{N}, \nonumber \\
\approx & \ \! \sigma_{\mathrm n}^{2} \operatorname{tr}\left( \mathbf{DP}_{{{\mathbf{B}}_{k}}}^{\bot }+\mathbf{P}_{{{\mathbf{B}}_{k}}}^{\bot }{{\mathbf{D}}^{H}} \right) {\mathbf I}_M = \mathbf{0}, \nonumber \\
\mathbf{\Omega }_{k}^{\left( 2 \right)} = & \sum\nolimits_{k'=1,k'\ne k}^{K} \mathbf{H}_{k'}^{H} \mathbf{B}_{k'}^{H} \mathbf{E}\left( {{\phi }_{k}}-{{\phi }_{k'}} \right) \nonumber \\
& \times \left( \mathbf{DP}_{{{\mathbf{B}}_{k}}}^{\bot }+\mathbf{P}_{{{\mathbf{B}}_{k}}}^{\bot }{{\mathbf{D}}^{H}} \right) {{\mathbf{E}}^{H}}\left( {{\phi }_{k}}-{{\phi }_{k'}} \right){{\mathbf{B}}_{k'}}{{\mathbf{H}}_{k'}}, \nonumber \\
 \approx & \operatorname{tr}\left( \mathbf{DP}_{{{\mathbf{B}}_{k}}}^{\bot }+\mathbf{P}_{{{\mathbf{B}}_{k}}}^{\bot }{{\mathbf{D}}^{H}} \right) \mathbf{\Pi }_{k}^{\left( 2 \right)} =\mathbf{0}, \nonumber \\
\mathbf{\Psi }_{k}^{\left( 2 \right)}
= & \ \! \mathbf{H}_{k}^{H}\mathbf{B}_{k}^{H} \mathbf{DP}_{{{\mathbf{B}}_{k}}}^{\bot } \sum\nolimits_{k'=1,k'\ne k}^{K} {{{\mathbf{E}}^{H}} \left( {{\phi }_{k}}-{{\phi }_{k'}} \right) {{\mathbf{B}}_{k'}}{{\mathbf{H}}_{k'}}}.
\end{align}

Next, the SOD of ${{\bf\Xi }_{k}} \big( {\tilde{\phi }_k} \big)$ with respect to ${\tilde{\phi }_k} $ can be approximated as
\begin{align}{\label{Xik_SOD}}
 & {{\left. E\left\{ \frac{{{\partial }^{2}}{{\mathbf{\Xi }}_{k}}\big( {{{\tilde{\phi }}}_{k}} \big)}{\partial \tilde{\phi }_{k}^{2}} \right\} \right|}_{{{{\tilde{\phi }}}_{k}}={{\phi }_{k}}}} \nonumber \\
 = & \ \!\! \mathcal{U}_{k}^{T}{{\mathbf{Y}}^{H}} \! \mathbf{E}\!\left( {{\phi }_{k}} \right)\!\! \Big( \! \underbrace{ {{\mathbf{D}}^{2}} \mathbf{P}_{{{\mathbf{B}}_{k}}}^{\bot } \!\!+\!\! \mathbf{P}_{{{\mathbf{B}}_{k}}}^{\bot } \! {{\mathbf{D}}^{2H}} \!\!+\!\! 2\mathbf{D\!P}_{{{\mathbf{B}}_{k}}}^{\bot } \! {{\mathbf{D}}^{H}} }_{{\mathcal{P}}_k} \! \Big) \! {{\mathbf{E}}^{H}}\!\! \left( {{\phi }_{k}} \right)\!\! \mathbf{Y}\mathcal{U}_{k}^{*}\!, \nonumber \\
 \approx & \ \! \mathcal{U}_{k}^{T} \left[ \mathbf{\Phi }_{k}^{\left( 3 \right)} + \mathbf{\Omega }_{k}^{\left( 3 \right)} + \mathbf{\Psi }_{k}^{\left( 3 \right)} + \mathbf{\Psi }_{k}^{\left( 3 \right)H} + 2{{\mathbf{\Phi }}_{k}} \right]{{\mathcal{U}}_k^{*}}, \nonumber \\
 \approx & \ \! {{\mathcal{U}}_k^{T}} \left[ \mathbf{\Psi }_{k}^{\left( 3 \right)} + \mathbf{\Psi }_{k}^{\left( 3 \right)H} + 2{{\mathbf{\Phi }}_{k}} \right]\mathcal{U}_{k}^{*},
\end{align}
where
\begin{align}
 \mathbf{\Phi }_{k}^{\left( 3 \right)} & \!=\!   {{\mathbf{N}}^{H}}\mathbf{E}\left( {{\phi }_{k}} \right) {{\mathcal{P}}_k} {{\mathbf{E}}^{H}}\left( {{\phi }_{k}} \right)\mathbf{N} \approx \sigma_{\mathrm n}^2 \operatorname{tr}\left( {{\mathcal{P}}_k} \right) {{\mathbf{I}}_{M}} = \mathbf{0}, \nonumber \\
 \mathbf{\Omega }_{k}^{\left( 3 \right)} &\!\!= \!\!\sum\nolimits_{k'=1,k'\ne k}^{K} \!\! {\mathbf{H}_{k'}^{H}\mathbf{B}_{k'}^{H}\mathbf{E} \! \left( {{\phi }_{k}} \!\!-\! {{\phi }_{k'}} \right)\! {{\mathcal{P}}_k} {{\mathbf{E}}^{H}} \!\! \left( {{\phi }_{k}} \!\!-\! {{\phi }_{k'}} \! \right) \! {{\mathbf{B}}_{k'}} \! {{\mathbf{H}}_{k'}}}\!, \nonumber \\
 & \!\approx \operatorname{tr} \left( {{\mathcal{P}}_k} \right) \mathbf{\Pi }_{k}^{\left( 2 \right)} =\mathbf{0}, \nonumber \\
 \mathbf{\Psi }_{k}^{\left( 3 \right)}
 & \!=\! \mathbf{H}_{k}^{H}\mathbf{B}_{k}^{H}\left( {{\mathbf{D}}^{2}} \mathbf{P}_{{{\mathbf{B}}_{k}}}^{\bot } + 2\mathbf{DP}_{{{\mathbf{B}}_{k}}}^{\bot } {{\mathbf{D}}^{H}} \right) \nonumber \\
 & \hspace{1em} \times \sum\nolimits_{k'=1,k'\ne k}^{K}{{{\mathbf{E}}^{H}}\left( {{\phi }_{k}}-{{\phi }_{k'}} \right){{\mathbf{B}}_{k'}}{{\mathbf{H}}_{k'}}}, \nonumber \\
 {{\mathbf{\Phi }}_{k}} & \!=\! \mathbf{H}_{k}^{H} \mathbf{B}_{k}^{H} \mathbf{DP}_{{{\mathbf{B}}_{k}}}^{\bot } {{\mathbf{D}}^{H}} {{\mathbf{B}}_{k}} {{\mathbf{H}}_{k}}.
\end{align}

In addition, we simplify ${{\mathbf{Y}}^{H}} \mathbf{Y} \!\!=\!\! {{\left( \! \sum\limits_{k'=1}^{K} \!\! {\mathbf{E}\! \left( {{\phi }_{k'}} \right)\! {{\mathbf{B}}_{k'}} {{\mathbf{H}}_{k'}}} \!\!+\!\! \mathbf{N} \! \right)}^{H}} \\ \left( \sum\limits_{k''=1}^{K}{\mathbf{E}\left( {{\phi }_{k''}} \right){{\mathbf{B}}_{k''}}{{\mathbf{H}}_{k''}}} \!+\! \mathbf{N} \right) $ as
\begin{align}{\label{YYH}}
& {{\mathbf{Y}}^{H}}\mathbf{Y}
 \approx \sum\limits_{k'=1}^{K}{\mathbf{H}_{k'}^{H}\mathbf{B}_{k'}^{H} {{\mathbf{B}}_{k'}}{{\mathbf{H}}_{k'}}} \nonumber \\
 & + \sum\limits_{k'=1}^{K}{\sum\limits_{k''=1,k''\ne k'}^{K}{\mathbf{H}_{k'}^{H} \mathbf{B}_{k'}^{H} \mathbf{E}\left( {{\phi }_{k''}} \!-\! {{\phi }_{k'}} \right) {{\mathbf{B}}_{k''}} {{\mathbf{H}}_{k''}}}} \!+\! {{\mathbf{N}}^{H}}\mathbf{N}, \nonumber \\
\approx & \sum\limits_{k'=1}^{K}{\mathbf{H}_{k'}^{H}{{\mathbf{H}}_{k'}}} +\sum\limits_{k'=1}^{K}{\sum\limits_{k''=1,k''\ne k'}^{K} {\operatorname{tr}\left( \mathbf{E}\left( {{\phi }_{k''}} \!-\! {{\phi }_{k'}} \right) \right)\mathbf{H}_{k'}^{H} {{\mathbf{H}}_{k''}}}} \nonumber \\
& + N\sigma_{\mathrm n}^{2}{{\mathbf{I}}_{M}}, \nonumber \\
\approx & \sum\limits_{k'=1}^{K}{\mathbf{H}_{k'}^{H}{{\mathbf{H}}_{k'}}}.
\end{align}

Denoting $\mathbf{\Pi }_{k}^{\left( 1 \right)} \!=\! \mathbf{H}_{k}^{H}{{\mathbf{H}}_{k}}$ and $\mathbf{\Pi } \!=\! \mathbf{\Pi }_{k}^{\left( 1 \right)} \!+\! \mathbf{\Pi }_{k}^{\left( 2 \right)}$, we arrive at ${{\mathcal{U}}_k^{T}} {{\mathbf{Y}}^{H}}\mathbf{Y}{{\mathcal{U}}_k^{*}} \approx {{\mathcal{U}}_k^{T}} \mathbf{\Pi }{{\mathcal{U}}_k^{*}}$.

2) Calculation of the derivatives of $G_k\big( {{{\tilde{\phi }}}_{k}} \big)$ with respect to $\tilde{\phi}_k$:

\newcounter{mytempeqncnt}
\begin{figure*}[ht]
\normalsize
\setcounter{mytempeqncnt}{\value{equation}}
\setcounter{equation}{56}
\begin{equation*}
\ \mathrm{MSE}\big\{ {{{\hat{\phi }}}_{k}} \big\} \!\approx\!
\frac{
   2\operatorname{tr}\big( {{\mathcal{R}}_{\mathcal{U}_{k}^{*}}}\mathbf{\Psi }_{k}^{\left( 2 \right)}{{\mathcal{R}}_{\mathcal{U}_{k}^{*}}}\mathbf{\Pi } \big) \operatorname{tr}\big( {{\mathcal{R}}_{\mathcal{U}_{k}^{*}}}\mathbf{\Psi }_{k}^{\left( 2 \right)H}{{\mathcal{R}}_{\mathcal{U}_{k}^{*}}}\mathbf{\Pi } \big)
   + 2\sigma _{\mathrm{n}}^{2}\operatorname{tr}\left( {{\mathcal{R}}_{\mathcal{U}_{k}^{*}}} {{\mathbf{\Phi }}_{k}} {{\mathcal{R}}_{\mathcal{U}_{k}^{*}}}\mathbf{\Pi }{{\mathcal{R}}_{\mathcal{U}_{k}^{*}}}{{\mathcal{R}}_{\mathcal{U}_{k}^{*}}}\mathbf{\Pi } \right) }
{\left[\begin{matrix}
  2\sigma_{\mathrm n}^{2}\operatorname{tr}\left( {{\mathcal{R}}_{\mathcal{U}_{k}^{*}}}
 {{\mathbf{\Phi }}_{k}} \right) \operatorname{tr}\left( {{\mathcal{R}}_{\mathcal{U}_{k}^{*}}}{{\mathcal{R}}_{\mathcal{U}_{k}^{*}}}\mathbf{\Pi } \right)  + 2\sigma_{\mathrm n}^{2}\operatorname{tr}\left( {{\mathcal{R}}_{\mathcal{U}_{k}^{*}}} \right)\operatorname{tr}\left( {{\mathcal{R}}_{\mathcal{U}_{k}^{*}}} {{\mathbf{\Phi }}_{k}} {{\mathcal{R}}_{\mathcal{U}_{k}^{*}}}\mathbf{\Pi } \right)  \\
  + 2\operatorname{tr}\big( {{\mathcal{R}}_{\mathcal{U}_{k}^{*}}}\mathbf{\Psi }_{k}^{\left( 2 \right)}{{\mathcal{R}}_{\mathcal{U}_{k}^{*}}}\mathbf{\Psi }_{k}^{\left( 2 \right)H}{{\mathcal{R}}_{\mathcal{U}_{k}^{*}}}\mathbf{\Pi } \big)+2\operatorname{tr}\big( {{\mathcal{R}}_{\mathcal{U}_{k}^{*}}}\mathbf{\Psi }_{k}^{\left( 2 \right)H}{{\mathcal{R}}_{\mathcal{U}_{k}^{*}}}\mathbf{\Psi }_{k}^{\left( 2 \right)}{{\mathcal{R}}_{\mathcal{U}_{k}^{*}}}\mathbf{\Pi } \big) \\
  - \operatorname{tr}\left[ {{\mathcal{R}}_{\mathcal{U}_{k}^{*}}}\big( \mathbf{\Psi }_{k}^{\left( 3 \right)} \!+\! \mathbf{\Psi }_{k}^{\left( 3 \right)H} \!+\! 2{{\mathbf{\Phi }}_{k}} \big) {{\mathcal{R}}_{\mathcal{U}_{k}^{*}}}\mathbf{\Pi } \right] \\
\end{matrix}\right]^2},
\end{equation*}
\begin{equation} {\label{MSE_theo1}}
\approx
\frac{\operatorname{tr}\big( {{\mathcal{R}}_{\mathcal{U}_{k}^{*}}} \mathbf{\Psi }_{k}^{\left( 2 \right)} {{\mathcal{R}}_{\mathcal{U}_{k}^{*}}} {\mathbf{\Pi }_k^{(1)}} \big) \operatorname{tr}\big( {{\mathcal{R}}_{\mathcal{U}_{k}^{*}}}\mathbf{\Psi }_{k}^{\left( 2 \right)H}{{\mathcal{R}}_{\mathcal{U}_{k}^{*}}} {\mathbf{\Pi }_k^{(1)}} \big)
   \!+\! \sigma _{\mathrm{n}}^{2}\operatorname{tr}\big( {{\mathcal{R}}_{\mathcal{U}_{k}^{*}}} {{\mathbf{\Phi }}_{k}} {{\mathcal{R}}_{\mathcal{U}_{k}^{*}}}  {\mathbf{\Pi }_k^{(1)}} {{\mathcal{R}}_{\mathcal{U}_{k}^{*}}}{{\mathcal{R}}_{\mathcal{U}_{k}^{*}}} {\mathbf{\Pi }_k^{(1)}} \big)  }
{2\left[
 \sigma_{\mathrm n}^{2}\operatorname{tr}\left( {{\mathcal{R}}_{\mathcal{U}_{k}^{*}}} {{\mathbf{\Phi }}_{k}} \right)\operatorname{tr}\big( {{\mathcal{R}}_{\mathcal{U}_{k}^{*}}}{{\mathcal{R}}_{\mathcal{U}_{k}^{*}}} {\mathbf{\Pi }_k^{(1)}} \big)
 \!+\! \sigma_{\mathrm n}^{2} {{g}_{0}}\left( k \right) \operatorname{tr}\big( {{\mathcal{R}}_{\mathcal{U}_{k}^{*}}} {{\mathbf{\Phi }}_{k}} {{\mathcal{R}}_{\mathcal{U}_{k}^{*}}} {\mathbf{\Pi }_k^{(1)}} \big)
 \!-\! \operatorname{tr}\big( {{\mathcal{R}}_{\mathcal{U}_{k}^{*}}} {{\mathbf{\Phi }}_{k}} {{\mathcal{R}}_{\mathcal{U}_{k}^{*}}} {\mathbf{\Pi }_k^{(1)}} \big)
\right]^2}.
\end{equation}
\setcounter{equation}{\value{mytempeqncnt}+1}
\hrulefill
\vspace*{-0.5em}
\end{figure*}
\setcounter{equation}{\value{mytempeqncnt}}

Denote ${{\mathcal{R}}_{\mathcal{U}_{k}^{*}}} \!=\! \mathcal{U}_{k}^{*}{{\left[ \mathcal{U}_{k}^{T}\mathbf{\Omega }_{k}^{\left( 1 \right)}\mathcal{U}_{k}^{*} \right]}^{-1}}\mathcal{U}_{k}^{T}$.
By combining (\ref{gk_SOD}), (\ref{inv_Xik}), (\ref{Xik_FOD}), (\ref{Xik_SOD}) and (\ref{YYH}), the SOD of ${{G}_{k}}\big( {{{\tilde{\phi }}}_{k}} \big)$ with respect to $\tilde{\phi}_k$ can be approximated as
\begin{align}{\label{Gk_SOD2}}
& {{\left. E\left\{ \frac{{{\partial }^{2}}{{G}_{k}}\big( {{{\tilde{\phi }}}_{k}} \big)}{\partial \tilde{\phi }_{k}^{2}} \right\} \right|}_{{{{\tilde{\phi }}}_{k}}={{\phi }_{k}}}} \nonumber \\
\approx & \ 2\operatorname{tr}\! \left[ {{\mathcal{R}}_{\mathcal{U}_{k}^{*}}}{{\mathbf{N}}^{H}}\big( \mathbf{\Gamma }_{k}^{\left( 1 \right)} \!+\! \mathbf{\Gamma }_{k}^{\left( 2 \right)} \big){{\mathcal{R}}_{\mathcal{U}_{k}^{*}}}\big( \mathbf{\Gamma }_{k}^{\left( 1 \right)H} \!+\! \mathbf{\Gamma }_{k}^{\left( 2 \right)H} \big)\mathbf{N}{{\mathcal{R}}_{\mathcal{U}_{k}^{*}}}\mathbf{\Pi } \right] \nonumber \\
& +\! 2\operatorname{tr} \! \left[ {{\mathcal{R}}_{\mathcal{U}_{k}^{*}}}\big( \mathbf{\Gamma }_{k}^{\left( 1 \right)H} \!\!+\! \mathbf{\Gamma }_{k}^{\left( 2 \right)H} \big)\mathbf{N}{{\mathcal{R}}_{{{\mathcal{U}}^{*}}}}{{\mathbf{N}}^{H}}\big( \mathbf{\Gamma }_{k}^{\left( 1 \right)} \!+\! \mathbf{\Gamma }_{k}^{\left( 2 \right)} \big){{\mathcal{R}}_{\mathcal{U}_{k}^{*}}}\mathbf{\Pi } \right] \nonumber \\
& +\! 2\operatorname{tr}\! \left[ {{\mathcal{R}}_{\mathcal{U}_{k}^{*}}}\big( \mathbf{\Psi }_{k}^{\left( 2 \right)} \!+\! \mathbf{\Psi }_{k}^{\left( 2 \right)H} \big){{\mathcal{R}}_{\mathcal{U}_{k}^{*}}}\big( \mathbf{\Psi }_{k}^{\left( 2 \right)} \!+\! \mathbf{\Psi }_{k}^{\left( 2 \right)H} \big){{\mathcal{R}}_{\mathcal{U}_{k}^{*}}}\mathbf{\Pi } \right] \nonumber \\
& -\! \operatorname{tr}\! \left[ {{\mathcal{R}}_{\mathcal{U}_{k}^{*}}}\big( \mathbf{\Psi }_{k}^{\left( 3 \right)} \!+\! \mathbf{\Psi }_{k}^{\left( 3 \right)H} \!+\! 2{{\mathbf{\Phi }}_{k}} \big) {{\mathcal{R}}_{\mathcal{U}_{k}^{*}}}\mathbf{\Pi } \right], \nonumber \\
\approx &\ 2\sigma_{\mathrm n}^{2} \operatorname{tr}\! \left[ {{\mathcal{R}}_{\mathcal{U}_{k}^{*}}}\big( \mathbf{\Gamma }_{k}^{\left( 1 \right)H} \!\!+\! \mathbf{\Gamma }_{k}^{\left( 2 \right)H} \big)\big( \mathbf{\Gamma }_{k}^{\left( 1 \right)} \!+\! \mathbf{\Gamma }_{k}^{\left( 2 \right)} \big) \right] \operatorname{tr}\! \left( {{\mathcal{R}}_{\mathcal{U}_{k}^{*}}}{{\mathcal{R}}_{\mathcal{U}_{k}^{*}}}\mathbf{\Pi } \right) \nonumber \\
& +\!2\sigma_{\mathrm n}^{2} \operatorname{tr}\! \left( {{\mathcal{R}}_{\mathcal{U}_{k}^{*}}} \right) \! \operatorname{tr}\! \left[ {{\mathcal{R}}_{\mathcal{U}_{k}^{*}}}\big( \mathbf{\Gamma }_{k}^{\left( 1 \right)H} \!\!+\! \mathbf{\Gamma }_{k}^{\left( 2 \right)H} \big) \! \big( \mathbf{\Gamma }_{k}^{\left( 1 \right)} \!\!+\! \mathbf{\Gamma }_{k}^{\left( 2 \right)} \big){{\mathcal{R}}_{\mathcal{U}_{k}^{*}}}\mathbf{\Pi } \right] \nonumber \\
& +\!2\operatorname{tr}\! \left( {{\mathcal{R}}_{\mathcal{U}_{k}^{*}}}\mathbf{\Psi }_{k}^{\left( 2 \right)} {{\mathcal{R}}_{\mathcal{U}_{k}^{*}}}\mathbf{\Psi }_{k}^{\left( 2 \right)H} {{\mathcal{R}}_{\mathcal{U}_{k}^{*}}}\mathbf{\Pi } \right) \nonumber \\
& +\! 2\operatorname{tr}\! \left( {{\mathcal{R}}_{\mathcal{U}_{k}^{*}}} \mathbf{\Psi }_{k}^{\left( 2 \right)H} {{\mathcal{R}}_{\mathcal{U}_{k}^{*}}} \mathbf{\Psi }_{k}^{\left( 2 \right)} {{\mathcal{R}}_{\mathcal{U}_{k}^{*}}} \mathbf{\Pi } \right) \nonumber \\
& -\! \operatorname{tr}\! \left[ {{\mathcal{R}}_{\mathcal{U}_{k}^{*}}}\big( \mathbf{\Psi }_{k}^{\left( 3 \right)} \!+\! \mathbf{\Psi }_{k}^{\left( 3 \right)H} \!+\! 2{{\mathbf{\Phi }}_{k}} \big) {{\mathcal{R}}_{\mathcal{U}_{k}^{*}}}\mathbf{\Pi } \right].
\end{align}

Besides, combining (\ref{gk_FOD}), (\ref{inv_Xik}), (\ref{Xik_FOD}) and (\ref{YYH}), we can approximate the FOD of ${{G}_{k}}\big( {{{\tilde{\phi }}}_{k}} \big)$ with respect to $\tilde{\phi}_k$ as
\begin{align}{\label{Gk_FOD}}
& {{\left. E\left\{ \frac{\partial {{G}_{k}}\big( {{{\tilde{\phi }}}_{k}} \big)}{\partial {{{\tilde{\phi }}}_{k}}} \right\} \right|}_{{{{\tilde{\phi }}}_{k}}={{\phi }_{k}}}} \nonumber \\
\approx & \!-\! \operatorname{tr}\! \left( {{\mathcal{R}}_{\mathcal{U}_{k}^{*}}} \! \left[ {{\mathbf{N}}^{H}}\big( \mathbf{\Gamma }_{k}^{\left( 1 \right)} \!+\! \mathbf{\Gamma }_{k}^{\left( 2 \right)} \big) \!+\! \big( \mathbf{\Gamma }_{k}^{\left( 1 \right)H} \!+\! \mathbf{\Gamma }_{k}^{\left( 2 \right)H} \big)\mathbf{N} \right] {{\mathcal{R}}_{\mathcal{U}_{k}^{*}}} \mathbf{\Pi } \right) \nonumber \\
& \ \ -\! \operatorname{tr}\! \left[ {{\mathcal{R}}_{\mathcal{U}_{k}^{*}}}\big( \mathbf{\Psi }_{k}^{\left( 2 \right)} \!+\! \mathbf{\Psi }_{k}^{\left( 2 \right)H} \big) {{\mathcal{R}}_{\mathcal{U}_{k}^{*}}}\mathbf{\Pi } \right]\!.
\end{align}

The following Lemma is introduced for further calculating the square of FOD based on (\ref{Gk_FOD}).
\begin{lemma}
For arbitrary matrices ${\mathbf{A}} \in {\mathbb C}^{R\times N}$ and ${\mathbf{C}} \in {\mathbb C}^{M\times R}$ which are uncorrelated with the noise matrix $\mathbf{N}$, there holds
\begin{align}
\operatorname{tr}\left( \mathbf{ANC} \right)\operatorname{tr}\left( {{\mathbf{C}}^{H}}{{\mathbf{N}}^{H}}{{\mathbf{A}}^{H}} \right)\approx \sigma_{\mathrm n}^{2} \operatorname{tr}\left( \mathbf{A}{{\mathbf{A}}^{H}}{{\mathbf{C}}^{H}}\mathbf{C} \right).
\end{align}
\end{lemma}
\begin{IEEEproof}
Suppose $\! \mathbf{A} \!\!=\!\! [ {\mathbf{a}_{1}}, {\mathbf{a}_{2}}, \cdots\!, {{\mathbf{a}}_{R}} ]^T\!$,
$\mathbf{C} \!\!=\!\! [ {\mathbf{c}_{1}}, {\mathbf{c}_{2}}, \cdots\!, {{\mathbf{c}}_{R}} ]$.
Then we have
\begin{align*}
 & \operatorname{tr}\left( \mathbf{ANC} \right) \operatorname{tr}\left( {{\mathbf{C}}^{H}}{{\mathbf{N}}^{H}}{{\mathbf{A}}^{H}} \right) \nonumber \\
=& \! \sum\limits_{r=1}^{R}\! {{{\left[ \mathbf{ANC} \right]}_{rr}}} \! \sum\limits_{s=1}^{R} \! {{{\left[ {{\mathbf{C}}^{H}} {{\mathbf{N}}^{H}} \! {{\mathbf{A}}^{H}} \right]\!}_{ss}}} \!\!=\! \sum\limits_{r=1}^{R} {\sum\limits_{s=1}^{R} {\mathbf{a}_{r}^{T}\mathbf{N} {{\mathbf{c}}_{r}} \mathbf{c}_{s}^{H}{{\mathbf{N}}^{H}}\mathbf{a}_{s}^{*}}}, \nonumber \\
 \approx & \ \! \sigma_{\mathrm n}^{2}\sum\limits_{r=1}^{R}{\sum\limits_{s=1}^{R}{\mathbf{a}_{r}^{T}\text{tr}\left( {{\mathbf{c}}_{r}}\mathbf{c}_{s}^{H} \right)\mathbf{a}_{s}^{*}}} = \sigma_{\mathrm n}^{2}\sum\limits_{r=1}^{R}{\sum\limits_{s=1}^{R} {\mathbf{a}_{r}^{T}\mathbf{a}_{s}^{*}\mathbf{c}_{s}^{H}{{\mathbf{c}}_{r}}}}, \nonumber \\
 = & \ \! \sigma_{\mathrm n}^{2} \operatorname{tr}\left( \mathbf{A}{{\mathbf{A}}^{H}}{{\mathbf{C}}^{H}}\mathbf{C} \right).
\end{align*}

This completes the proof.
\end{IEEEproof}

With Lemma 2, we arrive at
\begin{align}{\label{Gk_FOD2}}
 & {{\left. E\left\{ {{\bigg( \frac{\partial {{G}_{k}}\big( {{{\tilde{\phi }}}_{k}} \big)}{\partial {{{\tilde{\phi }}}_{k}}} \bigg)}^{2}} \right\} \right|}_{{{{\tilde{\phi }}}_{k}}={{\phi }_{k}}}} \nonumber \\
 \approx & {{\left( \operatorname{tr}\left[ {{\mathcal{R}}_{\mathcal{U}_{k}^{*}}} \big( \mathbf{\Psi }_{k}^{\left( 2 \right)} +\mathbf{\Psi }_{k}^{\left( 2 \right)H} \big) {{\mathcal{R}}_{\mathcal{U}_{k}^{*}}} \mathbf{\Pi } \right] \right)}^{2}} \nonumber \\
 & +2\operatorname{tr}\Big( \underbrace{{{\mathcal{R}}_{\mathcal{U}_{k}^{*}}}\big( \mathbf{\Gamma }_{k}^{\left( 1 \right)H}+\mathbf{\Gamma }_{k}^{\left( 2 \right)H} \big)}_{\mathbf{A}}\mathbf{N}\underbrace{{{\mathcal{R}}_{\mathcal{U}_{k}^{*}}}\mathbf{\Pi }}_{\mathbf{C}} \Big) \nonumber \\
 & \ \ \ \times \operatorname{tr}\Big( \underbrace{\mathbf{\Pi }{{\mathcal{R}}_{\mathcal{U}_{k}^{*}}}}_{{{\mathbf{C}}^{H}}}{{\mathbf{N}}^{H}}\underbrace{\big( \mathbf{\Gamma }_{k}^{\left( 1 \right)}+\mathbf{\Gamma }_{k}^{\left( 2 \right)} \big){{\mathcal{R}}_{\mathcal{U}_{k}^{*}}}}_{{{\mathbf{A}}^{H}}} \Big), \nonumber \\
 \approx &\ \! 2\operatorname{tr}\big( {{\mathcal{R}}_{\mathcal{U}_{k}^{*}}}\mathbf{\Psi }_{k}^{\left( 2 \right)}{{\mathcal{R}}_{\mathcal{U}_{k}^{*}}}\mathbf{\Pi } \big)\operatorname{tr}\big( {{\mathcal{R}}_{\mathcal{U}_{k}^{*}}}\mathbf{\Psi }_{k}^{\left( 2 \right)H}{{\mathcal{R}}_{\mathcal{U}_{k}^{*}}}\mathbf{\Pi } \big) \nonumber \\
 + & 2\sigma_{\mathrm{n}}^{2}\operatorname{tr} \!\! \left[ {{\mathcal{R}}_{\mathcal{U}_{k}^{*}}} \big( \mathbf{\Gamma }_{k}^{\left( 1 \right)H} \!\!+\! \mathbf{\Gamma }_{k}^{\left( 2 \right)H} \big)\! \big( \mathbf{\Gamma }_{k}^{\left( 1 \right)} \!\!+\! \mathbf{\Gamma }_{k}^{\left( 2 \right)} \big){{\mathcal{R}}_{\mathcal{U}_{k}^{*}}}\mathbf{\Pi }{{\mathcal{R}}_{\mathcal{U}_{k}^{*}}}{{\mathcal{R}}_{\mathcal{U}_{k}^{*}}}\mathbf{\Pi } \right]\!\!.
\end{align}

Besides,
\begin{align}
  & \big( \mathbf{\Gamma }_{k}^{\left( 1 \right)H} \!+\! \mathbf{\Gamma }_{k}^{\left( 2 \right)H} \big)\big( \mathbf{\Gamma }_{k}^{\left( 1 \right)} \!+\! \mathbf{\Gamma }_{k}^{\left( 2 \right)} \big) \approx \mathbf{\Gamma }_{k}^{\left( 1 \right)H}\mathbf{\Gamma }_{k}^{\left( 1 \right)} \!+\! \mathbf{\Gamma }_{k}^{\left( 2 \right)H}\mathbf{\Gamma }_{k}^{\left( 2 \right)}, \nonumber \\
 \approx &\ \! {{\mathbf{\Phi }}_{k}} +\sum\nolimits_{k'=1,k'\ne k}^{K} \mathbf{H}_{k'}^{H} \mathbf{B}_{k'}^{H}\mathbf{E} \left( {{\phi }_{k}}\!-\!{{\phi }_{k'}} \right) \nonumber \\
  & \hspace{2.5em} \times \left( \mathbf{DP}_{{{\mathbf{B}}_{k}}}^{\bot }\!+\! \mathbf{P}_{{{\mathbf{B}}_{k}}}^{\bot } {{\mathbf{D}}^{H}} \right)^2  {{\mathbf{E}}^{H}} \left( {{\phi }_{k}}\!-\!{{\phi }_{k'}} \right) {{\mathbf{B}}_{k'}}{{\mathbf{H}}_{k'}}, \nonumber \\
  \approx & \ \! {{\mathbf{\Phi }}_{k}}+2\operatorname{tr}\left( \mathbf{DP}_{{{\mathbf{B}}_{k}}}^{\bot }\mathbf{DP}_{{{\mathbf{B}}_{k}}}^{\bot }\!+\!\mathbf{DP}_{{{\mathbf{B}}_{k}}}^{\bot }{{\mathbf{D}}^{H}} \right)\mathbf{\Pi }_{k}^{\left( 2 \right)}.
\end{align}
Moreover, there holds ${{\mathcal{R}}_{{{\mathcal{U}}_k^{*}}}} \Big[ {{\mathbf{\Phi }}_{k}}+2\operatorname{tr}\Big( \mathbf{DP}_{{{\mathbf{B}}_{k}}}^{\bot }\mathbf{DP}_{{{\mathbf{B}}_{k}}}^{\bot } \!+\! \mathbf{DP}_{{{\mathbf{B}}_{k}}}^{\bot }{{\mathbf{D}}^{H}} \Big)\mathbf{\Pi }_{k}^{\left( 2 \right)} \Big]\approx {{\mathcal{R}}_{{{\mathcal{U}}_k^{*}}}} {{\mathbf{\Phi }}_{k}}$.

3) Approximate expression of the analytical MSE:

By substituting (\ref{Gk_FOD2}) and (\ref{Gk_SOD2}) into (\ref{MSE_theo}), we can express the analytical MSE as (\ref{MSE_theo1}), shown on the top of the page.
\\
Here, the approximation of the denominator originates from the fact that $\mathbf{\Psi }_{k}^{\left( 2 \right)}$ and $\mathbf{\Psi }_{k}^{\left( 3 \right)}$ reflect the interrelation between the channel of the desired $k$th user and other users, while ${{\mathbf{\Phi }}_{k}}$ reflects the autocorrelation of the channel of the $k$th user.

Next, we attempt to simplify (\ref{MSE_theo1}) term by term. Since ${\mathbf B}_{k'}, k' \!\ne\! k$ spans the signal subspace of users other than the desired $k$th user, they can be regarded as noise for the $k$th user. Thus, similar to Lemma 2, there holds $\operatorname{tr}\left( {{\mathbf{A}}^{H}}\mathbf{B}_{k'}^{H}{{\mathbf{C}}^{H}} \right) \\ \operatorname{tr}\left( \mathbf{C}{{\mathbf{B}}_{k'}}\mathbf{A} \right) \!\approx\! \operatorname{tr}\left( {{\mathbf{A}}^{H}}\mathbf{AC}{{\mathbf{C}}^{H}} \right)$ for $k' \!\ne\! k$.
With such approximation, the first term at the nominator of (\ref{MSE_theo1}) can be simplified as
\setcounter{equation}{57}
\begin{align}{\label{approx_term1}}
  & \operatorname{tr}\big( {{\mathcal{R}}_{\mathcal{U}_{k}^{*}}}\mathbf{\Psi }_{k}^{\left( 2 \right)H}{{\mathcal{R}}_{\mathcal{U}_{k}^{*}}}\mathbf{\Pi }_{k}^{\left( 1 \right)} \big)\operatorname{tr}\big( {{\mathcal{R}}_{\mathcal{U}_{k}^{*}}}\mathbf{\Psi }_{k}^{\left( 2 \right)}{{\mathcal{R}}_{\mathcal{U}_{k}^{*}}}\mathbf{\Pi }_{k}^{\left( 1 \right)} \big), \nonumber \\
\approx & \!\! \sum\limits_{k'=1,k'\ne k}^{K} \!\!
   \operatorname{tr} \! \left( \mathbf{\Pi }_{k}^{\left( 1 \right)} \! {{\mathcal{R}}_{\mathcal{U}_{k}^{*}}} \! \mathbf{H}_{k'}^{H} \mathbf{B}_{k'}^{H} \mathbf{E}\! \left( {{\phi }_{k}} \!\!-\! {{\phi }_{k'}} \right) \! \mathbf{P}_{{{\mathbf{B}}_{k}}}^{\bot } \! {{\mathbf{D}}^{H}} \! {{\mathbf{B}}_{k}}{{\mathbf{H}}_{k}} {{\mathcal{R}}_{\mathcal{U}_{k}^{*}}} \! \right) \nonumber \\
  & \!\times\! \operatorname{tr} \! \Big( \underbrace{{{\mathcal{R}}_{\mathcal{U}_{k}^{*}}} \mathbf{H}_{k}^{H}\mathbf{B}_{k}^{H} \mathbf{DP}_{{{\mathbf{B}}_{k}}}^{\bot } {{\mathbf{E}}^{H}} \! \left( {{\phi }_{k}} \!-\! {{\phi }_{k'}} \right)}_{\mathbf{C}} {{\mathbf{B}}_{k'}} \! \underbrace{{{\mathbf{H}}_{k'}} {{\mathcal{R}}_{\mathcal{U}_{k}^{*}}} \mathbf{\Pi }_{k}^{\left( 1 \right)}}_{\mathbf{A}} \Big)\!, \nonumber \\
 \approx & \sum\nolimits_{k'=1,k'\ne k}^{K} \operatorname{tr} \Big( \mathbf{\Pi }_{k}^{\left( 1 \right)} {{\mathcal{R}}_{\mathcal{U}_{k}^{*}}} \mathbf{H}_{k'}^{H}{{\mathbf{H}}_{k'}} {{\mathcal{R}}_{\mathcal{U}_{k}^{*}}} \mathbf{\Pi }_{k}^{\left( 1 \right)} {{\mathcal{R}}_{\mathcal{U}_{k}^{*}}} \mathbf{H}_{k}^{H} \nonumber \\
 & \hspace{2em} \times \mathbf{B}_{k}^{H}\mathbf{DP}_{{{\mathbf{B}}_{k}}}^{\bot } {{\mathbf{D}}^{H}} {{\mathbf{B}}_{k}} {{\mathbf{H}}_{k}} {{\mathcal{R}}_{\mathcal{U}_{k}^{*}}} \Big), \nonumber \\
 = &\operatorname{tr}\Big( {{\mathbf{H}}_{k}} {{\mathcal{R}}_{\mathcal{U}_{k}^{*}}} \mathbf{\Pi }_{k}^{\left( 2 \right)} {{\mathcal{R}}_{\mathcal{U}_{k}^{*}}} \mathbf{H}_{k}^{H} {{\mathbf{H}}_{k}} {{\mathcal{R}}_{\mathcal{U}_{k}^{*}}} \mathbf{H}_{k}^{H} \nonumber \\
 & \hspace{2em} \times \mathbf{B}_{k}^{H} \mathbf{DP}_{{{\mathbf{B}}_{k}}}^{\bot } {{\mathbf{D}}^{H}} {{\mathbf{B}}_{k}} {{\mathbf{H}}_{k}} {{\mathcal{R}}_{\mathcal{U}_{k}^{*}}} \mathbf{H}_{k}^{H} \Big), \nonumber \\
 \approx &\operatorname{tr} \! \Big[ \frac{1}{L}\! \operatorname{tr}\big( {{\mathcal{R}}_{\mathcal{U}_{k}^{*}}}\mathbf{\Pi }_{k}^{\left( 2 \right)}{{\mathcal{R}}_{\mathcal{U}_{k}^{*}}}{{\mathbf{R}}_{k}} \big)\frac{1}{L}\! \operatorname{tr}\left( {{\mathcal{R}}_{\mathcal{U}_{k}^{*}}}{{\mathbf{R}}_{k}} \right) \nonumber \\
 & \hspace{2em} \times \mathbf{B}_{k}^{H}\mathbf{DP}_{{{\mathbf{B}}_{k}}}^{\bot }{{\mathbf{D}}^{H}} {{\mathbf{B}}_{k}} \frac{1}{L} \!\operatorname{tr}\left( {{\mathcal{R}}_{\mathcal{U}_{k}^{*}}} {{\mathbf{R}}_{k}} \right) \! \Big], \nonumber \\
 =& \frac{1}{{{L}^{3}}}{{\left[ {{g}_{1}} \! \left( k \right) \right]}^{2}}{{g}_{3}}\! \left( k \right){{d}_{k}},
\end{align}
where the expressions for ${{g}_{1}}\! \left( k \right)$, ${{g}_{3}}\! \left( k \right)$ and $d_k$ have been given in (\ref{gk_dk}), and ${{\mathbf{R}}_{k}} $ is defined as
\begin{align}{\label{boldR_k}}
{{\mathbf{R}}_{k}} \!=\! LE\left\{ \mathbf{h}_{l,k}^{*}\mathbf{h}_{l,k}^{T} \right\} \!\approx\! \frac{1}{2{{\theta }_{\mathrm{as}}}} \! \int_{{{\theta }_{k}}-{{\theta }_{\mathrm{as}}}}^{{{\theta }_{k}}+{{\theta }_{\mathrm{as}}}}{{{\mathbf{a}}^{*}}\left( \theta  \right){{\mathbf{a}}^{T}}\left( \theta  \right)d\theta }.
\end{align}

The second term at the nominator of (\ref{MSE_theo1}) can be simplified as
\begin{align}{\label{approx_term2}}
& \operatorname{tr}\! \big( {{\mathcal{R}}_{\mathcal{U}_{k}^{*}}}{{\mathbf{\Phi }}_{k}}{{\mathcal{R}}_{\mathcal{U}_{k}^{*}}}\mathbf{\Pi }_{k}^{(1)}{{\mathcal{R}}_{\mathcal{U}_{k}^{*}}}{{\mathcal{R}}_{\mathcal{U}_{k}^{*}}}\mathbf{\Pi }_{k}^{(1)} \big) \nonumber \\
= & \operatorname{tr} \! \Big( {{\mathbf{H}}_{k}}{{\mathcal{R}}_{\mathcal{U}_{k}^{*}}} \mathbf{H}_{k}^{H}\mathbf{B}_{k}^{H}\mathbf{DP}_{{{\mathbf{B}}_{k}}}^{\bot } {{\mathbf{D}}^{H}} {{\mathbf{B}}_{k}}{{\mathbf{H}}_{k}} {{\mathcal{R}}_{\mathcal{U}_{k}^{*}}} \mathbf{H}_{k}^{H} \nonumber \\
& \hspace{3.5em} \times {{\mathbf{H}}_{k}} {{\mathcal{R}}_{\mathcal{U}_{k}^{*}}} {{\mathcal{R}}_{\mathcal{U}_{k}^{*}}} \mathbf{H}_{k}^{H} \Big), \nonumber \\
\approx & \operatorname{tr}\Big[ \frac{1}{L} \operatorname{tr}\left( {{\mathcal{R}}_{\mathcal{U}_{k}^{*}}} {{\mathbf{R}}_{k}} \right) \mathbf{B}_{k}^{H} \mathbf{DP}_{{{\mathbf{B}}_{k}}}^{\bot } {{\mathbf{D}}^{H}} {{\mathbf{B}}_{k}} \nonumber \\
& \hspace{3.5em} \times \frac{1}{L} \operatorname{tr}\left( {{\mathcal{R}}_{\mathcal{U}_{k}^{*}}} {{\mathbf{R}}_{k}} \right) \frac{1}{L}\operatorname{tr}\left( {{\mathcal{R}}_{\mathcal{U}_{k}^{*}}} {{\mathcal{R}}_{\mathcal{U}_{k}^{*}}} {{\mathbf{R}}_{k}} \right) \Big], \nonumber \\
= & \frac{1}{{{L}^{3}}}{{\left[ {{g}_{1}}\left( k \right) \right]}^{2}}{{g}_{2}}\left( k \right){{d}_{k}}.
\end{align}

The terms appeared at the denominator of (\ref{MSE_theo1}) can be further simplified as
\begin{align}{\label{approx_term3}}
& \operatorname{tr}\left( {{\mathcal{R}}_{\mathcal{U}_{k}^{*}}} {{\mathbf{\Phi }}_{k}} \right) \operatorname{tr}\big( {{\mathcal{R}}_{\mathcal{U}_{k}^{*}}} {{\mathcal{R}}_{\mathcal{U}_{k}^{*}}}\mathbf{\Pi }_{k}^{(1)} \big) \nonumber \\
= & \operatorname{tr}\left( {{\mathbf{H}}_{k}}{{\mathcal{R}}_{\mathcal{U}_{k}^{*}}}\mathbf{H}_{k}^{H}\mathbf{B}_{k}^{H}\mathbf{DP}_{{{\mathbf{B}}_{k}}}^{\bot }{{\mathbf{D}}^{H}}{{\mathbf{B}}_{k}} \right)\operatorname{tr}\left( {{\mathbf{H}}_{k}}{{\mathcal{R}}_{\mathcal{U}_{k}^{*}}}{{\mathcal{R}}_{\mathcal{U}_{k}^{*}}}\mathbf{H}_{k}^{H} \right), \nonumber \\
\approx & \operatorname{tr}\Big( \frac{1}{L} \operatorname{tr}\left( {{\mathcal{R}}_{\mathcal{U}_{k}^{*}}}{{\mathbf{R}}_{k}} \right) \mathbf{B}_{k}^{H}\mathbf{DP}_{{{\mathbf{B}}_{k}}}^{\bot } {{\mathbf{D}}^{H}}{{\mathbf{B}}_{k}} \Big) \nonumber \\
& \hspace{1.5em} \times \operatorname{tr}\Big( \frac{1}{L}\operatorname{tr}\left( {{\mathcal{R}}_{\mathcal{U}_{k}^{*}}} {{\mathcal{R}}_{\mathcal{U}_{k}^{*}}} {{\mathbf{R}}_{k}} \right) {{\mathbf{I}}_{L}} \Big), \nonumber \\
=& \frac{1}{L}{{g}_{1}}\left( k \right){{g}_{2}}\left( k \right){{d}_{k}},
\end{align}
and
\begin{align}{\label{approx_term4}}
& \operatorname{tr}\big( {{\mathcal{R}}_{\mathcal{U}_{k}^{*}}}{{\mathbf{\Phi }}_{k}}{{\mathcal{R}}_{\mathcal{U}_{k}^{*}}}\mathbf{\Pi }_{k}^{(1)} \big) \nonumber \\
= & \operatorname{tr}\left( {{\mathbf{H}}_{k}} {{\mathcal{R}}_{\mathcal{U}_{k}^{*}}} \mathbf{H}_{k}^{H} \mathbf{B}_{k}^{H} \mathbf{DP}_{{{\mathbf{B}}_{k}}}^{\bot } {{\mathbf{D}}^{H}} {{\mathbf{B}}_{k}} {{\mathbf{H}}_{k}} {{\mathcal{R}}_{\mathcal{U}_{k}^{*}}} \mathbf{H}_{k}^{H} \right), \nonumber \\
\approx & \frac{1}{{{L}^{2}}}{{\left[ {{g}_{1}}\left( k \right) \right]}^{2}}{{d}_{k}}.
\end{align}

Substituting (\ref{approx_term1}), (\ref{approx_term2}), (\ref{approx_term3}) and (\ref{approx_term4}) into (\ref{MSE_theo1}) leads to (\ref{MSE_theo2}).

Next, we demonstrate ${{g}_{3}}\! \left( k \right) \!=\! \frac{1}{N\!-\!L}{{g}_{1}}\! \left( k \right) \!-\! \sigma_{\mathrm n}^{2}{{g}_{2}}\! \left( k \right)$. Denoting ${{\boldsymbol \Upsilon }_{k}} \!=\! \mathcal{U}_{k}^{T} \mathbf{\Pi }_{k}^{\left( 2 \right)}\mathcal{U}_{k}^{*}$, we have
\begin{align}
{{\mathcal{R}}_{\mathcal{U}_{k}^{*}}} \!\approx \mathcal{U}_{k}^{*}{{\left( \mathcal{U}_{k}^{T}\mathbf{\Omega }_{k}^{\left( 1 \right)}\mathcal{U}_{k}^{*} \right)}^{-1}}\mathcal{U}_{k}^{T} \!=\! \frac{1}{N\!\!-\!\!L}\mathcal{U}_{k}^{*}{{\left( \sigma _{\mathrm n}^{2}{{\mathbf{I}}_{{{Q}_{k}}}} \!\!+\!\! {{\boldsymbol \Upsilon }_{k}} \right)}^{-1}}\mathcal{U}_{k}^{T}.
\end{align}

Considering that
\begin{align*}
& {{g}_{3}}\left( k \right) \!=\! \operatorname{tr}\big( {{\mathcal{R}}_{\mathcal{U}_{k}^{*}}}{{\mathbf{R}}_{k}}{{\mathcal{R}}_{\mathcal{U}_{k}^{*}}}\mathbf{\Pi }_{k}^{\left( 2 \right)} \big), \nonumber \\
= & \frac{1}{{{\left( N\!\!-\!\!L \right)}^{2}}}\! \operatorname{tr}\! \Big( \underbrace{{{\left( \sigma _{\mathrm n}^{2}{{\mathbf{I}}_{{{Q}_{k}}}} \!\!+\!\! {{\boldsymbol \Upsilon }_{k}} \right)}^{-1}} \mathcal{U}_{k}^{T} {{\mathbf{R}}_{k}} \mathcal{U}_{k}^{*}{{\left( \sigma _{\mathrm n}^{2}{{\mathbf{I}}_{{{Q}_{k}}}} \!\!+\!\! {{\boldsymbol \Upsilon }_{k}} \right)}^{-1}} {{\boldsymbol \Upsilon }_{k}}}_{\mathcal{A}} \Big)\!,
\end{align*}
there holds $\sigma _{\mathrm n}^{2} {{\mathbf{I}}_{{{Q}_{k}}}} \! {{\left( \sigma _{\mathrm n}^{2} {{\mathbf{I}}_{{{Q}_{k}}}} \!\!+\!\! {{\boldsymbol \Upsilon }_{k}} \right)}^{-1}} \mathcal{U}_{k}^{T} {{\mathbf{R}}_{k}} \mathcal{U}_{k}^{*} {{\left( \sigma _{\mathrm n}^{2} {{\mathbf{I}}_{{{Q}_{k}}}} \!\!+\!\! {{\boldsymbol \Upsilon }_{k}} \right)}^{-1}} \!+ \\ \mathcal{A} =\! {{\left( \sigma _{\mathrm n}^{2}{{\mathbf{I}}_{{{Q}_{k}}}} \!+\! {{\boldsymbol \Upsilon }_{k}} \right)}^{-1}}\mathcal{U}_{k}^{T}{{\mathbf{R}}_{k}}\mathcal{U}_{k}^{*}$. Thus,
\begin{align}
 {{g}_{3}}\! \left( k \right) = & \frac{1}{{{\left( N\!-\!L \right)}^{2}}} \operatorname{tr} \! \Big[ {{\left( \sigma _{\mathrm n}^{2}{{\mathbf{I}}_{{{Q}_{k}}}} \!\!+\!\! {{\boldsymbol \Upsilon }_{k}} \right)}^{-1}} \mathcal{U}_{k}^{T}{{\mathbf{R}}_{k}}\mathcal{U}_{k}^{*} \nonumber \\
 & \!\!-\! \sigma _{\mathrm n}^{2}\mathcal{U}_{k}^{T}\mathcal{U}_{k}^{*}{{\left( \sigma_{\mathrm n}^{2}{{\mathbf{I}}_{{{Q}_{k}}}} \!\!+\!\! {{\boldsymbol \Upsilon }_{k}} \right)}^{-1}} \mathcal{U}_{k}^{T}{{\mathbf{R}}_{k}} \mathcal{U}_{k}^{*} {{\left( \sigma_{\mathrm n}^{2}{{\mathbf{I}}_{{{Q}_{k}}}} \!\!+\!\! {{\boldsymbol \Upsilon }_{k}} \right)}^{-1}} \Big]\!, \nonumber \\
 = & \frac{1}{N-L}\operatorname{tr}\left( {{\mathcal{R}}_{\mathcal{U}_{k}^{*}}}{{\mathbf{R}}_{k}} \right)-\sigma_{\mathrm{n}}^{2}\operatorname{tr}\left( {{\mathcal{R}}_{\mathcal{U}_{k}^{*}}}{{\mathbf{R}}_{k}} {{\mathcal{R}}_{\mathcal{U}_{k}^{*}}} \right), \nonumber \\
 =& \frac{1}{N-L}{{g}_{1}}\left( k \right)- \sigma_{\mathrm n}^{2}{{g}_{2}}\left( k \right).
\end{align}



\begin{thebibliography}{99}

%
%
%

\bibitem{Marzetta10}
T. Marzetta, ``Noncooperative cellular wireless with unlimited numbers of base station antennas,'' \emph{IEEE Trans. Wirel. Commun.}, vol. 9, no. 11, pp. 3590--3600, Nov. 2010.

\bibitem{LuJSTSP2014}
L. Lu, G. Y. Li, A. L. Swindlehurst, A. Ashikhmin, and R. Zhang, ``An Overview of Massive MIMO: Benefits and Challenges,'' \emph{IEEE J. Sel. Topics Signal Process.}, vol. 8, no. 5, pp. 742--758, Oct. 2014.

\bibitem{GaoTVT16}
H. Xie, F. Gao, S. Zhang, and S. Jin, ``A unified transmission strategy for TDD/FDD massive MIMO systems with spatial basis expansion model,'' \emph{IEEE Trans. Veh. Technol.}, vol. 66, no. 4, pp. 3170--3184, Apr. 2017.

\bibitem{GaoAccess16}
H. Xie, F. Gao, and S. Jin, ``An overview of low-rank channel estimation for massive MIMO systems,'' \emph{IEEE Access}, vol. 4, no. 9, pp. 7313--7321, Nov. 2016.

\bibitem{GaoJSAC16}
 H. Xie, B. Wang, F. Gao, and S. Jin, ``A full-space spectrum-sharing strategy for massive MIMO cognitive radio,'' \emph{IEEE J. Select. Areas Commun.}, vol. 34, no. 10, pp. 2537--2549, Oct. 2016.

%
%

\bibitem{Larsson_CM2014}
E. Larsson, O. Edfors, F. Tufvesson, and T. Marzetta, ``Massive MIMO for next generation wireless systems,'' \emph{IEEE Commun. Mag.,} vol. 52, no. 2, pp. 186--195, Feb. 2014.

\bibitem{Morelli_TWC2012}
M. Morelli, and M. Moretti, ``Carrier Frequency Offset Estimation for OFDM Direct-Conversion Receivers,'' \emph{IEEE Trans. Wirel. Commun.,} vol. 11, no. 7, pp. 2670--2679, Jul. 2012.

\bibitem{Morelli_WCL2013}
M. Morelli, and M. Moretti, ``Joint Maximum Likelihood Estimation of CFO, Noise Power, and SNR in OFDM Systems,'' \emph{IEEE Wirel. Commun. Lett.,} vol. 2, no. 1, pp. 42--45, Feb. 2013.



\bibitem{Tsai-TWC13}
Y. R. Tasi, H. Y. Huang, Y. C. Chen, and K. J. Yang, ``Simultaneous multiple carrier frequency offset estimation for coordinated multi-point transmission in OFDM systems,'' \emph{IEEE Trans. Wirel. Commun.}, vol. 12, no. 9, pp. 4558--4568, Sep. 2013.

\bibitem{Chen_TSP2008}
J. Chen, Y. C. Wu, S. Ma, and T. S. Ng, ``Joint CFO and channel estimation for multiuser MIMO-OFDM systems with optimal training sequences,'' \emph{IEEE Trans. Signal Process.}, vol. 56, no. 8, pp. 4008--4019, Aug. 2008.

\bibitem{Wu_EUJWCN2011}
Y. Wu, J. W. M. Bergmans, and S. Attallah, ``Carrier Frequency offset estimation for multi-user MIMO OFDM uplink using CAZAC sequences: Performance and sequence optimization,'' \emph{EURASIP J. Wirel. Commun. Netw.}, 2011, Article ID: 570680.

\bibitem{Zhang_TWC2016}
W. Zhang and F. Gao, ``Computationally efficient blind estimator of carrier frequency offset for MIMO-OFDM systems,'' \emph{IEEE Trans. Wirel. Commun.}, vol. 15, no. 11, pp. 7644--7656, Nov. 2016.

\bibitem{Cheng13}
H. V. Cheng and E. G. Larsson, ``Some fundamental limits on frequency synchronization in massive MIMO,'' in Proc. \emph{IEEE Asilomar}, Nov. 2013, pp. 1213--1217.

\bibitem{Mukherjee_TVT2016-2}
S. Mukherjee, and S. K. Mohammed, ``Impact of frequency selectivity on the information rate performance of CFO impaired single-carrier massive MU-MIMO uplink,'' \emph{IEEE Wirel. Commun. Lett.}, vol. 5, no. 6, pp. 648¨C651, Dec. 2016.

\bibitem{Mukherjee_TVT2016-1}
S. Mukherjee, S. K. Mohammed, and I. Bhushan, ``Impact of CFO estimation on the performance of ZF reciever in massive MU-MIMO systems,'' \emph{IEEE Trans. Veh. Technol.}, vol. 65, no. 11, pp. 9430--9436, Nov. 2016.

\bibitem{Mukherjee_GLOBECOM2015}
S. Mukherjee and S. K. Mohammed, ``Low-complexity CFO estimation for multi-user massive MIMO systems,'' in \emph{Proc. IEEE GLOBECOM}, Dec. 2015, pp. 1--7.

\bibitem{Zhang_TSP2016}
W. Zhang and F. Gao, ``Blind frequency synchronization for multiuser OFDM uplink with large number of receive antennas,'' \emph{IEEE Trans. Signal Process.}, vol. 64, no. 9, pp. 2255--2268, May 2016.


\bibitem{ZhangGLO16}
W. Zhang, F. Gao, and H. Wang, ``Frequency synchronization for massive MIMO multi-user uplink,'' in Proc. \emph{IEEE GLOBECOM}, Dec. 2016, pp. 1--6.

\bibitem{GE_SPAWC2017}
Y. Ge, W. Zhang, F. Gao, and L. Xing, ``Frequency synchronization for uplink massive MIMO with partly calibrated subarrays,'' in Proc. \emph{IEEE SPAWC}, Jul. 2017, pp. 1--5.

\bibitem{Zhang_TWC2018}
W. Zhang, F. Gao, S. Jin, and H. Lin, ``Frequency Synchronization for Uplink Massive MIMO Systems,'' \emph{IEEE Trans. Wirel. Commun.}, vol. 17, no. 15, pp. 235--249, Jan. 2018.

\bibitem{You15}
L. You, X. Gao, X. Xia, N. Ma, and Y. Peng, ``Pilot reuse for massive MIMO transmission over spatially correleated Rayleigh fading channels,'' \emph{IEEE Trans. Wirel. Commun.}, vol. 14, no. 6, pp. 3352--3366, Jun. 2015.

\bibitem{Sun15}
C. Sun, X. Gao, S. Jin, M. Matthaiou, Z. Ding, and C. Xiao, ``Beam division multiple access transmission for massive MIMO communications,'' \emph{IEEE Trans. Commun.}, vol. 63, no. 6, pp. 2170--2184, Jun. 2015.

\bibitem{Zhang_DSP2017}
W. Zhang, H. Li, P. Mu, and H. Wang, ``Robust Multi-Branch Space-Time Beamforming for OFDM System with Interference,'' \emph{Digital Signal Process.}, vol. 65, no. 8, pp. 63--70, Jun. 2017.

\bibitem{Meyer_2001}
C. D. Meyer, \emph{Matrix Analysis and Applied Linear Algebra}. Society for Industrial and Applied Mathematics, 2000.

\bibitem{Boyd_2004}
S. Boyd, and L. Vandenberghe, \emph{Convex Optimization}. Cambridge University Press, 2004.

\bibitem{Yu_JWCMC2002}
K. Yu, and B. Ottersten, ``Models for MIMO Propagation Channels, A Review,'' \emph{Wiley J. Wireless Communication and Mobile Computing, special issue on Smart antenna and MIMO systems}, Oct. 2002.

\bibitem{Lee_TVT1973}
W. C. Y. Lee, ``Effects on correlation between two mobile radio base-station antennas,'' \emph{IEEE Trans. Veh. Technol.}, vol. 22, no. 4, pp. 130--140, Nov. 1973.

\bibitem{Molisch_VTC2003}
A. F. Molisch, ``Effect of far scatterer clusters in MIMO outdoor channel models,'' in Proc. \emph{IEEE VTC-Spring}, Apr. 2003, pp. 534--538.

\bibitem{Petrus_TC2002}
P. Petrus, J. H. Reed, and T. S. Rappaport, ``Geometrical-based statistical macrocell channel model for mobile enviornments,'' \emph{IEEE Trans. Commun.}, vol. 50, no. 3, pp. 495--502, Mar. 2002.

\bibitem{Spencer_JSAC2000}
Q. H. Spencer, B. D. Jeffs, M. A. Jensen, and A. L. Swindlehurst, ``Modeling the statistical time and angle of arrival characteristics of an indoor multipath channel,'' \emph{IEEE J. Sel. Areas Commun.}, vol. 18, no. 3, pp. 347--360, Mar. 2000.

\bibitem{W_Zhang2013TSP}
W.~Zhang, and Q.~Yin, ``Blind maximum likelihood carrier frequency offset
  estimation for OFDM with multi-antenna receiver,'' \emph{IEEE Trans.
  Signal Process.}, vol.~61, no.~9, pp. 2295--2307, May 2013.


\bibitem{Petersen2012}
K. B. Petersen, and M. S. Pedersen, \emph{The matrix cookbook}. Available
online http://matrixcookbook.com, Nov. 2012.

\bibitem{C_Sun2015TC}
C. Sun, X. Gao, S. Jin, M. Matthaiou, Z. Ding, and C. Xiao, ``Beam Division Multiple Access Transmission for Massive MIMO Communications,'' \emph{IEEE Trans. Commun.}, vol. 63, no. 6, pp. 2170--2184, Jun. 2015.

\bibitem{DPPalomar_TSP2003}
D. P. Palomar, J. M. Cioffi, and M. A. Lagunas, ``Joint Tx-Rx beamforming design for multicarrier MIMO channels: a unified framework for convex optimization,'' \emph{IEEE Trans. Signal Process.}, vol. 51, no. 9, pp. 2381--2401, Sep. 2003.

\bibitem{Marshall2009}
A. W. Marshall, I. Olkin, and B. Arnold, \emph{Inequalities: Theory of Majorization
and Its Applications}, Springer, U.S., New York, 2009.

\bibitem{Wayne1973}
A. W. Roberts, and D. E. Varberg, \emph{Convex function}, Academic Press, U.S., New York, 1973.



\end{thebibliography}
\end{document}